\documentclass[9pt]{elife}

\usepackage{lipsum} 
\usepackage[version=4]{mhchem}
\usepackage{siunitx}

\usepackage{amsfonts}
\usepackage{caption}
\usepackage{amssymb}
\usepackage{graphicx}
\usepackage{mathrsfs}
\usepackage{array}
\usepackage{amsmath}
\usepackage{lscape}
\usepackage{bbold}
\usepackage{ucs}
\usepackage{color}
\usepackage{ulem}
\usepackage{ragged2e} 
\usepackage{color}

\newcommand{\esp}[1]{\left\langle #1\right\rangle}
\newcommand{\var}[1]{\text{Var}[#1]}
\newcommand{\rev}[1]{ \textcolor{black}{#1} }

\title{Learning protein constitutive motifs from sequence data}

\author{J\'er\^ome Tubiana}
\author{Simona Cocco}
\author{R\'emi Monasson}

\affil[]{Laboratory of Physics of the Ecole Normale Sup\'erieure, CNRS \& PSL Research, 24 rue Lhomond, 75005 Paris, France}
\corr{monasson@lpt.ens.fr}{RM}

\begin{document}

\maketitle

\begin{abstract}
Statistical analysis of evolutionary-related protein sequences provides insights about their structure, function, and history. We show that Restricted Boltzmann Machines (RBM), designed to learn complex high-dimensional data and their statistical features, can efficiently model protein families from sequence information. We here apply RBM to twenty protein families, and present detailed results for two short protein domains, Kunitz and WW, one long chaperone protein, Hsp70, and synthetic lattice proteins for benchmarking.  The features inferred by the RBM are biologically interpretable: they are related to structure (such as residue-residue tertiary contacts, extended secondary motifs ($\alpha$-helix and $\beta$-sheet) and intrinsically disordered regions), to function (such as activity and ligand specificity), or to phylogenetic identity. In addition, we use RBM to design new protein sequences with putative properties by composing and turning up or down the different modes at will. Our work therefore shows that RBM are a versatile and practical tool to unveil and exploit the genotype-phenotype relationship for protein families.
\end{abstract}

Sequencing of many organism genomes has led over the recent years to the collection of a huge number of protein sequences, gathered in databases such as UniProt or PFAM \cite{finn2013pfam}. Sequences with a common ancestral origin, defining a family (Fig.~1A), are likely to code for proteins with similar functions and structures, hence providing a unique window into the relationship between genotype (sequence content) and phenotype (biological features). In this context, various approaches have been introduced to infer protein properties from sequence data statistics, in particular amino-acid conservation and coevolution (correlation) \cite{teppa2012disentangling,de2013emerging}. 

A major objective of these approaches is to identify positions carrying amino acids with critical impact on the protein function, such as catalytic or binding sites, or specificity-determining sites controlling ligand specificity.  Principal Component Analysis (PCA) of the sequence data can be used to unveil groups of coevolving sites with specific functional role \cite{russ2005natural,rausell2010protein,halabi2009protein}. Other methods rely on phylogeny \cite{rojas2012ras}, entropy  (variability in amino-acid content) \cite{reva2011predicting,reva2007determinants}, or hybrid combination of both  \cite{mihalek2004family,ashkenazy2016consurf}.

Another objective is to extract structural information, such as the contact map of the three-dimensional fold. Considerable progress was brought by maximum-entropy methods, which rely on the computation of direct couplings between sites reproducing the pairwise coevolution statistics in the sequence data \cite{lapedes1999,weigt2009mp,jones2011psicov,cocco2018inverse}. Direct couplings provide very good estimators of contacts \cite{morcos2011direct,hopf2012three,kamisetty2013assessing,ekeberg2014fast}, and capture pairwise epistasis effects necessary to model fitness changes resulting from mutations \cite{chakraborty2014hiv,figliuzzi2016,hopf2017mutation}. 

Despite these successes, a unique, accurate framework capable of extracting the structural and functional features common to a protein family, as well as the phylogenetic variations specific to sub-families is still missing. Hereafter, we consider Restricted Boltzmann Machines (RBM) for this purpose. RBM are a powerful concept coming from machine learning \cite{ackley1987learning,hinton2012practical}; they are unsupervised (sequence data need not be annotated) and generative (able to generate new data). Informally speaking, RBM learn complex data distributions through their statistical features (Fig.~1B). 

In the present work, we have developed a method to learn efficiently RBM from protein sequence data. To illustrate the power and versatility of RBM, we have applied our approach to the sequence alignments of twenty different protein families. We report the results of our approach, with special emphasis on four families: the Kunitz domain, a protease inhibitor, historically important for protein structure determination \cite{ascenzi2003bovine}, the WW domain, a short module binding different classes of ligands \cite{sudol1995characterization}, Hsp70, a large chaperone protein \cite{bukau1998hsp70}, and  lattice-protein \textit {in silico} data \cite{shakhnovich1990enumeration,mirny2001} to benchmark our approach on exactly solvable models \cite{jacquin2016benchmarking}. Our study shows that RBM are able to capture (1) structure-related features, either local, such as tertiary contacts, or extended, such as secondary structure motifs  ($\alpha$-helix and $\beta$-sheet) or characteristic of intrinsically disordered regions; (2) functional features, i.e. groups of amino acids controlling specificity or activity; (3) phylogenetic features, related to sub-families sharing evolutionary determinants. Some of these features involves two residues only (as direct pairwise couplings do), others extend over large and not necessarily contiguous portions of the sequence (as in collective modes extracted with PCA). The pattern of similarities of each sequence with the inferred features defines a multi-dimensional representation of this sequence, highly informative about the biological properties of the corresponding protein (Fig.~1C). Focusing on representations of interest allows us, in turn, to design new sequences with putative functional properties.  In summary, our work shows that RBM offer  an effective computational tool to characterize and exploit quantitatively the genotype-phenotype relationship specific to a protein family.

\newpage
\begin{figure}[t]
\begin{fullwidth}
\centering
\includegraphics[width=17.8cm,angle=0]{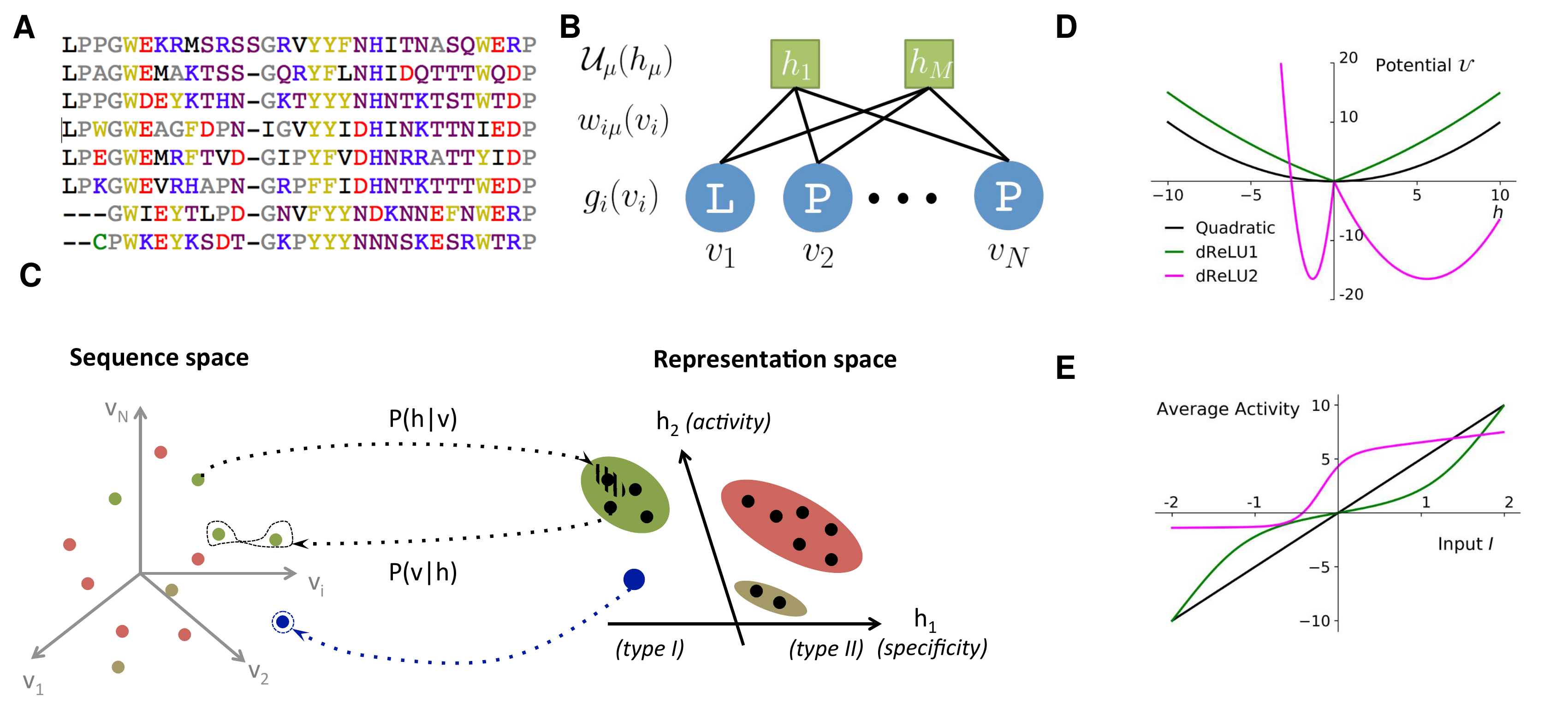}
\caption{ {\bf Reverse and forward modeling of proteins.} {\bf A.} Example of Multiple-Sequence Alignment (MSA), here of the WW domain (PF00397). Each column $i=1,...,N$ corresponds to a site on the protein, and each line to a different sequence in the family. Color code for amino acids: red = negative charge (E,D), blue = positive charge  (H, K, R), purple =  non-charged polar (hydrophilic) (N, T, S, Q), yellow = aromatic (F, W, Y), black = aliphatic hydrophobic  (I, L, M, V), green = cysteine (C), grey = other, small (A, G, P). {\bf B.} In a Restricted Boltzmann Machine (RBM), weights $w_{i\mu}$ connect the visible layer (carrying protein sequences $\bf v$) to the hidden layer (carrying representations $\bf h$). Biases on the visible and hidden units are introduced by the local potentials $g_i(v_i)$ and ${\cal U}_\mu(h_\mu)$. Due to the bipartite nature of the weight graph, hidden units are independent conditioned to a visible configuration, and vice versa. {\bf C.} Sequences $\bf v$ in the MSA (dots in sequence space, left) code for proteins with different phenotypes (dot colors). RBM define a probabilistic mapping from sequences $\bf v$ onto the representation space $\bf h$ (right), indicative of the phenotype of the corresponding protein, and encoded in the conditional distribution $P({\bf h}|{\bf v})$, Eqn.~(3) (black arrow).  The reverse mapping from representations to sequences is $P({\bf v}|{\bf h})$, Eqn.~(4) (black arrow). Sampling a subspace in the representation space (colored domains) defines in turn a complex subset of the sequence space, and allows one to design sequences with putative phenotypic properties, either found in the MSA (green circled dots) or not encountered in Nature (arrow out of blue domain). {\bf D.} Three examples of potentials ${\cal U}$ defining the hidden-unit type in RBM, see Eqn.~(1) and Panel B: quadratic  (black, $\gamma=0.2$,$\theta=0$) and double Rectified Linear Unit (dReLU) (Green: $\gamma_+=\gamma_-=0.1$, $\theta_+=-\theta_-=1$; Purple: $\gamma_+=1$, $\gamma_-=20$, $\theta_+=-6$, $\theta_-=25$) potentials. In practice, the parameters of the hidden unit potentials are fixed through learning of the sequence data. {\bf E.}  Average activity of hidden unit $h$, calculated from Eqn.~(3), as a function of the input $I$ defined in Eqn.~(2). The three curves correspond to the three choices of potentials in panel A. For the quadratic potential (Black), the average activity is a linear function of $I$. For dReLU1 (Green), small inputs $I$ barely activate the hidden unit, whereas dReLU2 (Purple) essentially binarizes the inputs $I$.
}
\end{fullwidth}
\label{fig1}
\end{figure}
\clearpage

\section*{Results}
\subsection*{Restricted Boltzmann Machines}
 
\subsubsection{Definition} 

A Restricted Boltzmann Machine (RBM) is a joint probabilistic model for sequences and representations, see Fig.~1C. It is formally defined on a bipartite, two-layer graph (Fig.~1B). Protein sequences ${\bf v} = (v_1,v_2,...,v_N)$ are displayed on the Visible layer, and representations  ${\bf h} = (h_1,h_2,...,h_M)$ on the Hidden layer. Each visible unit takes one out of $21$ values (20 amino acids + 1 alignment gap). Hidden-layer unit  values $h_\mu$ are real. The joint probability distribution of ${\bf v}, {\bf h}$ is 
\begin{equation}\label{Energy}
P({\bf v},{\bf h}) \propto \exp \bigg( \sum_{i=1}^N g_i(v_i) - \sum_{\mu =1}^M \mathcal{U}_\mu(h_\mu)  + \sum_{i,\mu} h_\mu\, w_{i\mu} (v_i) \bigg) \ ,
\end{equation}
up to a normalization constant. Here,  the weight matrix $w_{i\mu}$ couples the visible and the hidden layers, and $g_i(v_i)$ and $\mathcal{U}_\mu(h_\mu)$ are local potentials biasing the values of, respectively, the visible and the hidden variables (Figs.~1B,D).

\subsubsection{From sequence to representation, and back} 

Given a sequence $\bf v$ on the visible layer,  the hidden unit $\mu$ receives the input
\begin{equation}\label{inputmu}
I_\mu ({\bf v}) = \sum_{i} w_{i\mu} (v_i) \ .
\end{equation}
This expression is analogous to the score of a sequence with a position-specific weight matrix. \rev{Large positive or negative} $I_\mu$ \rev{values signal a good match between the sequence and, respectively, the positive and the negative components of the weights attached to unit} $\mu$, \rev{whereas small} $|I_\mu|$ \rev{correspond to a bad match.}

The input $I_\mu$ determines, in turn, the conditional probability of the activity $h_\mu$ of the hidden unit,
\begin{equation}
\label{condmu}
P( h_\mu | {\bf v}) \propto \exp \big(-\mathcal{U}_\mu(h_\mu)  + h_\mu\, I_\mu({\bf v}) \big) \ ,
\end{equation}
up to a normalization constant. The nature of the potential ${\cal U}$ is crucial to determine how the average activity $h$ varies with the input $I$, see  Fig.~1E and below. 

In turn, given a representation (set of activities) $\bf h$ on the hidden layer, the residues on site $i$ are distributed according to
\begin{equation}
\label{condv}
P( v_i | {\bf h}) \propto \exp \bigg( g_i(v_i)  + \sum _\mu h_\mu\, w_{i\mu}(v_i) \bigg)\ .
\end{equation}
Hidden units with large activities $h_\mu$ strongly bias this probability, and favor values of $v_i$ corresponding to large weights $w_{i\mu}(v_i)$. 

Use of Eqn.~(3) allows us to sample the representation space given a sequence, while Eqn.~(4) defines the sampling of sequences given a representation, see both directions in Fig.~1C. Iterating this process generates high-probability representations, which, in turn produce very likely sequences, and so on.

\subsubsection{Probability of a sequence}
\rev{The probability of a sequence}, $P({\bf v})$, \rev{is obtained by summing (integrating)} $P({\bf v},{\bf h})$ \rev{over all its possible representations} $\bf h$.
\begin{equation} \label{marginal}
P({\bf v}) = \int \prod_{\mu=1}^M dh_\mu P({\bf v}, {\bf h}) \propto \exp \bigg[ \sum_{i=1}^N g_i(v_i) + \sum_{\mu=1}^M \Gamma_\mu\big (I_\mu ({\bf v})\big)  \bigg] \ ,
\end{equation}
\rev{where} $\Gamma_\mu(I) = \log  \int dh \, e^{-U_\mu(h) + h \,I} $ \rev{is the cumulant-generating function associated to the potential} ${\cal U}_\mu$ \rev{and is a function of the input to hidden unit} $\mu$, see Eqn.~(\ref{inputmu}).

\rev{For quadratic potentials} ${\cal U}_\mu(h)=\frac {\gamma_\mu}{ 2} h^2 + \theta_\mu h$ (Fig.~1E), \rev{the conditional probability $P( h_\mu | {\bf v})$ is Gaussian, and the RBM is said to be Gaussian. The cumulant-generating functions }$\Gamma_\mu(I) = \frac{1}{2 \gamma_\mu}(I-\theta_\mu)^2$ \rev{are quadratic, and their sum in }Eqn.~(\ref{marginal}) \rev{gives rise to effective pairwise couplings between the visible units,} $J_{ij} (v_i,v_j) =\sum _\mu \frac{1}{\gamma_\mu} w_{i\mu} (v_i) w_{j\mu}(v_j)$. \rev{Hence, a Gaussian RBM is equivalent to a Hopfield-Potts model} \cite{cocco2013principal}, \rev{where the number} $M$ \rev{of hidden units plays the role of the number of Hopfield-Potts `patterns'.  }

\rev{Non-quadratic potentials }${\cal U}_\mu$, \rev{and, hence, non-quadratic} $\Gamma(I)$, \rev{introduce couplings to}  \textit{all orders} \rev{between the visible units, all generated from the weights} $w_{i\mu}$. \rev{RBM thus offer a practical way to go beyond pairwise models, and express complex, high-order dependencies between residues, based on the inference of a limited number of interaction parameters (controlled by }$M$\rev{). In practice, for each hidden unit, we consider the class of 4-parameter potentials, }
\begin{equation}
\mathcal{U}_\mu(h) = \frac12 \gamma_{\mu,+} h_+^2 + \frac12 \gamma_{\mu,-} h_-^2 + \theta_{\mu,+} h_+ + \theta_{\mu,-} h_-\ ,\quad \text{where} \quad h_+ = \max(h,0)\ , \quad h_- = \min(h,0) \ ,
\end{equation}
\rev{hereafter called double Rectified Linear Units (dReLU) potentials (Fig.~1E). Varying the parameters allows us to span a wide class of behaviors, including quadratic potentials, double-well potentials (leading to bimodal distributions for }$h_\mu$\rev{) and hard constraints (e.g. preventing }$h_\mu$ \rev{from being negative).}

\rev{RBM can thus be thought of both as a framework to extract representations from sequences through Eqn.~(3), and as a way to model complex interactions between residues in sequences through Eqn.~(5). They constitute a natural candidate to unify (and improve) PCA-based and Direct-Coupling-based approaches to protein modeling.}

\subsubsection{Learning}
The weights $w_{i\mu}$ and the defining parameters of the potentials $g_i$ and ${\cal U}_\mu$ are learned by maximizing the average log-probability $\esp{\log P(\bf v) }_{MSA}$ of the  sequences $\bf v$  in the Multiple Sequence Alignment (MSA). In practice, estimating the gradients of the average log-probability with respect to these parameters requires to sample from the model distribution $P({\bf v})$, which is done through Monte Carlo simulation of the RBM, see Methods. 

We also introduce penalty terms over the weights $w_{i\mu}(v)$  (and the local potentials $g_i(v)$ on visible units) to avoid overfitting, and to promote sparse weights. Sparsity facilitates the biological interpretation of weights and thus, emphasizes the correspondence between representation and phenotypic spaces (Fig.~1C). Crucially, imposing sparsity also forces the RBM to learn a so-called compositional representation, in which each sequence is characterized by a subset of strongly activated hidden units, of size large compared to 1 but small compared to $M$ \cite{tubiana2017emergence}. All technical details about the learning procedure are reported in Methods. 

In the next sections, we present results for selected values of the number of hidden units and of the regularization penalty. The values of these (hyper-)parameters are justified afterwards.

\subsection*{Kunitz domain}

\noindent
\subsubsection*{Description}
The majority of natural proteins are obtained by concatenating functional building blocks, called protein domains. The Kunitz domain, with a length of about 50-60 residues (protein family PF00014 \cite{finn2013pfam}) is present in several genes and its main function is to inhibit serine protease such as trypsin. Kunitz domains play a key role in the regulation of many important processes in the body such as tissue growth and remodeling, inflammation, body coagulation and fibrinolysis. They are implicated in several diseases such as tumor growth,  Alzheimer,  cardiovascular and inflammatory diseases and, therefore, have been largely studied and shown to have a large potential in drug design \cite{shigetomi2010anti,bajaj2001structure}.

Some examples of Kunitz domain-containing proteins include the Basic Pancreatic Trypsin Inhibitor (BPTI, 1 Kunitz domain), the Bikunin (2 domains) \cite{fries2000bikunin}, Hepatocyte growth factor activator inhibitor (HAI, 2 domains) and tissue  factor pathway inhibitor (TFPI, 3 domains) \cite{shigetomi2010anti,bajaj2001structure}.

Figure~2A shows the MSA sequence logo and the secondary structure of the Kunitz domain. It is characterized by two $\alpha$ helices  and two $\beta$ strands; Cystein-Cystein disulfide bridges largely contribute to the thermodynamic stability of the domain, as frequently observed in small proteins. BPTI structure was the first one ever resolved \cite{ascenzi2003bovine}, and is often used to benchmark folding predictions based on simulations \cite{levitt1975computer}  and coevolutionary approaches \cite{morcos2011direct,hopf2012three,kamisetty2013assessing,cocco2013principal,haldane2018coevolutionary}. We train a RBM with $M=100$ dReLU on the MSA of PF00014, constituted by $B=8062$ sequences with $N=53$ consensus sites.

\subsubsection*{Inferred weights and interpretations.}

Figure~2B shows the weights $w_{i\mu}(v)$ attached to 5 selected hidden units. Each logo identifies the amino-acid motifs in the sequences $\bf v$ giving rise to large (positive or negative) inputs $I$ onto the associated hidden unit, see Eqn.~(2). 

Weight 1 in Fig.~2B has large components on sites 45 and 49, in contact in the final $\alpha_2$ helix (Figs.~2A\&D). The distribution of the inputs $I_1$ partitions the MSA in three subfamilies (Fig.~2C, top panel, dark blue histogram). The two peaks in $I_1\simeq -2.5$ and $I_1\simeq 1.5$ identify sequences  in which the contact is due to an electrostatic interaction with, respectively, $(+,-)$ and $(-,+)$ charged amino acid on sites 45 and 49; the other peak in $I_1\simeq 0$  identify sequences realizing the contact differently, e.g. with an aromatic amino acid on site 45. Weight 1 shows also a weaker electrostatic component on site 53 in Fig.~2B; the 4-site separation between sites 45--49--53 fits well with the average helix turn of 3.6 amino acids (Fig.~2D).

Weight 2 focuses on the contact between residues 11-35, realized in most sequences by a C-C disulfide bridge (Fig.~2B and negative $I_2$ peak in Fig.~2C, top). A minority of sequences in the MSA, corresponding to $I_2>0$ and mostly coming from nematode organisms (Appendix 1, Fig.~19), do not show the C-C bridge. A subset of these sequences strongly and positively activate hidden unit~3 (Appendix 1, Fig.~19 and $I_3>0$ peak in Fig.~2C). Positive components in weight~3 logo suggest that these proteins stabilize their structure through electrostatic interactions between sites 10 ($-$ charge) and 33-36 ($+$ charges both), see Figs.~2B\&D, to compensate the absence of C-C bridge on the neighbouring sites 11-35. 

Weight 4 describes a feature mostly localized on the loop preceding the $\beta_1$-$\beta_2$ strands  (sites 7 to 16), see Figs.~2B\&D. Structural studies of the  trypsin-trypsin inhibitor complex have shown that this loop binds to proteases  \cite{marquart1983geometry}; site 12 is in contact with the active site of the protease and is therefore key to the inhibitory activity of the Kunitz domain. The two amino acids (R, K) having a large positive contribution to weight 4  in position 12  are basic and bind to negatively charged residues (D, E) on the active site of trypsin-like serine proteases. While several Kunitz domains with known trypsin inhibitory activity, such as BPTI, TFPI, TPPI-2,... give rise to large and positive inputs $I_4$, Kunitz domains with no trypsin/chymotrypsin inhibition activity, e.g. associated to COL7A1 and COL6A3 genes \cite{chen2001carboxyl,kohfeldt1996conversion}, correspond to negative or vanishing values of $I_4$. Hence, hidden unit 4 possibly separates the Kunitz domains having trypsin-like protease inhibitory activity from the others. 

This interpretation is also in agreement with mutagenesis experiments carried out on sites 7 to 16 to test the inhibitory effects of Kunitz domains BPT1,  HAI-1, and TFP1 against trypsine-like proteases  \cite{bajaj2001structure,kirchhofer2003tissue,shigetomi2010anti,grzesiak2000inhibition,chand2004structure}.  In \cite{kirchhofer2003tissue} it was shown that mutation R12A on the first domain (out of two) of HAI-1 destroyed its inhibitory activity; a similar effect was observed with R12X, with X a non basic residue, in the first two domains (out of three) of TFP1 as discussed in \cite{bajaj2001structure}. The affinity between human serine proteases and the mutants G9F, G9S, G9P of bovine BPTI was shown to decrease in \cite{grzesiak2000inhibition}. Conversely, in \cite{kohfeldt1996conversion} it was shown that the set of mutations P10R, D13A, F14R could convert the COL6A3 domain into a Trypsin inhibitor. All these results are in agreement with the above interpretation and the logo of weight 4. Note that, though several sequences have large $I_4$ (top histogram in Fig.~2C), many correspond to small or negative values. This may be explained by the facts that (i) many of the Kunitz domains analyzed are present in two or more copies, and as such, are not all required to strongly bind trypsin \cite{bajaj2001structure} and (ii) Kunitz domain may have other specificities encoded by other hidden units. In particular, weight 34 in Supporting Information, displays on site 12 large components associated to medium to large size hydrophobic residues (L, M, Y), and is possibly related to other serine protease specificity classes such as chymotrypsin \cite{appel1986chymotrypsin}.

Weight~5 codes for a complex extended mode. To interpret this feature, we display in Fig.~2C (bottom histogram) the distributions of Hamming distances between all pairs of sequences in the MSA (gray histograms) and between the 100 sequences $\bf v$ in the MSA with largest inputs $|I_\mu({\bf v})|$ to the corresponding hidden unit (light blue histograms).  For hidden unit 5, the distances between those top-input sequences are smaller than between random sequences in the MSA, suggesting that weight 5 is characteristic of a  cluster of closely related sequences. Here, these sequences correspond to the protein Bikunin present in most mammals and some other vertebrates \cite{shigetomi2010anti}. Conversely, for other hidden units (e.g. 1,2), both histograms are quite similar, showing that the corresponding weight motifs are found in evolutionary distant sequences.

\rev{The five weights above were chosen based on several criteria:  (i) Weight norm, which is a proxy for the relevance of the hidden unit. Hidden units with larger weight norms contribute more to the likelihood, whereas weight with low norms may arise from noise/overfitting. (ii) Weight sparsity. Hidden units with sparse weights are more easily interpretable in terms of structural/functional constraints. (iii) Shape of input distributions. Hidden units with multimodal input distributions separate the family in subfamilies, and are therefore potentially interesting. (iv) Comparison with available literature. (v) Diversity.} The remaining 95 inferred weights are shown in Supporting Information. We find a variety of structural features, \textit{e.g.} pairwise contacts as in weights 1 and 2, \rev{also reminiscent of the localized, low-eigenvalue modes of the Hopfield-Potts model }\cite{cocco2013principal},  and phylogenetic features (activated by evolutionary related sequences as hidden unit $5$); the latter include in particular stretches of gaps, mostly located at the extremities of the sequence \cite{cocco2013principal}. Several weights have strong components on the same sites as weight 4, showing the complex pattern of amino acids controlling binding affinity.

\begin{figure}
\begin{fullwidth}
\centering
\includegraphics[scale=0.95,angle=90]{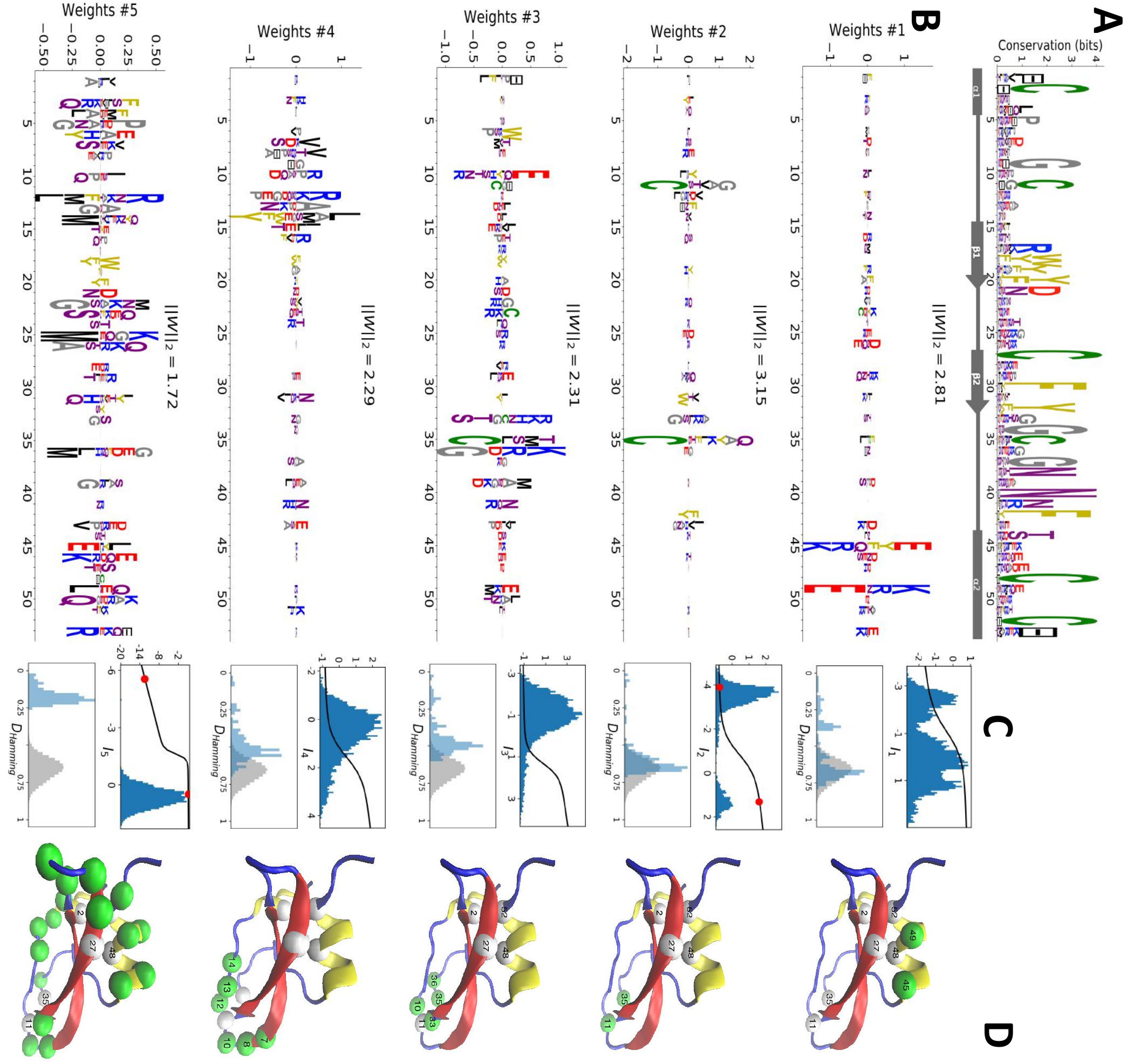}
\caption{ {\bf Modeling Kunitz Domain with RBM.} {\bf A.} Sequence logo and secondary structure of the Kunitz domain (PF00014), showing two  $\alpha$-helices and two $\beta$-strands. Note the presence of the three  C-C disulfide bridges between 11-35, 2-52, 27-48. {\bf B.} Weight logos for five hidden units, see text. Positive and negative weights are shown by letters located, respectively, above and below the zero axis. Values of the norms $\|W_\mu\|_2 = \sqrt{\sum _{i,v} w_{i\mu}(v)^2}$ are given. Same color code for the amino acids as in Fig.~1A. {\bf C.} Top: Distribution of inputs $I_\mu({\bf v})$ over the sequences $\bf v$ in the MSA (dark blue), and average activity vs. input function (full line, left scale); red points correspond to activity levels used for design in Fig.~5. Bottom: Histograms of Hamming distances between sequences in the MSA (grey) and between the 20 sequences (light blue) with largest (for unit 2,3,4) or smallest (1,5) $I_\mu$. {\bf D.} 3D visualization of the weights, shown on PDB structure 2knt \cite{merigeau19981} using VMD \cite{humphrey1996vmd}. White spheres denote the positions of the 3 disulfide bridges in the wild type sequence. Green spheres locate residues $i$ such that $\sum_v |w_{i\mu}(v) |>S$, with $S= 1.5$ for hidden units $\mu=1,2,3$, $S=1.25$ for $\mu=4$, and $S=0.5$ for $\mu=5$. }
\label{fig2}
\end{fullwidth}
\end{figure}
\clearpage
\subsection*{WW domain}

\subsubsection{Description}
WW is a protein-protein interaction domain found in many eukaryotes and human signalling proteins, involved in essential cellular processes such as transcription, RNA processing, protein trafficking, receptor signalling. WW is a short domain of length 30-40 amino-acids (Fig.~3A, PFAM PF00397, $B=7503$ sequences, $N=31$ consensus sites), which folds into a three-stranded antiparallel $\beta$-sheet. The domain name stems from the two conserved tryptophans (W) at positions 5-28 (Fig.~3A), which serve as anchoring sites for the ligands. WW domains bind to a variety of proline (P)-rich peptide ligands, and can be divided into four groups, based on their preferential binding affinity \cite{sudol2000new}. Group I binds specifically to PPXY motif - where X is any amino acid; Group II to PPLP motifs; Group III to proline-arginine containing sequences (PR); Group IV to phosphorylated serine/threonine-proline sites [p(S/T)P]. Modulation of binding properties allow hundreds of WW domain to specifically interact with hundreds of putative ligands in mammalian proteomes.

\subsubsection{Inferred weights and interpretation} 
Four weight logos of the inferred RBM are shown in Fig.~3B; the remaining 96 weights are given in Supporting Information. Weight 1 codes for a contact between sites 4-22 realized either by two amino acids with oppositive charges ($I_1<0$), or by one \rev{small} and one negatively charged amino acid ($I_1>0$). Weight 2 shows a $\beta$-sheet--related feature, with large entries defining a set of mostly hydrophobic ($I_2>0$) or hydrophilic ($I_2<0$) residues localized on the $\beta_1$ and $\beta_2$ strands (Fig.~3B) and in contact on the 3D fold, see Fig.~3D. The activation histogram in FIg.~3C, with a large peak on negative $I_2$, suggest that this part of the WW domain is exposed to the solvent in most, but not all, natural sequences.

Weights 3 and 4 are supported by sites on the $\beta_2$-$\beta_3$ binding pocket and on the $\beta_1$-$\beta_2$ loop of the WW domain. The distributions of activities in Fig.~3C highlight different groups of sequences in the MSA that strongly correlate with experimental ligand-type identification, see Fig.~3E. We find that
(i) Type I domains are characterized by $I_3<0$ and $I_4>0$;
(ii) Type II/III domains are characterized by $I_3 >0 $ and $I_4>0$;
(iii) There is no clear distinction between Type II and Type III domains;
(iv) Type IV domains are characterized by $I_3>0$ and $I_4<0$.
These findings are in good agreement with various  studies:

(i) Mutagenesis experiment  have shown the importance of sites 19, 21, 24, 26 for binding specificity \cite{espanel1999single,fowler2010high}. For the YAP1 WW domain, as confirmed by various studies  (see Table~2 in \cite{fowler2010high}), the mutations H21X and T26X  reduce the binding affinity to Type I ligands, while Q24R increases it and S12X has no effect. This is in agreement with  the negative components of weight 3 (Fig.~3B): $I_3$ increases upon mutations H21X and T26X, decreases upon  Q24R and is unaffected by S12X. Moreover the mutation L19W alone, or combined with
H21[D/G/K/R/S] could switch the specificity from Type I to Type II/III \cite{espanel1999single}. These results are consistent with Fig. 3E: YAP1 (blue cross) is of Type I but one or two mutations move it to the right side, closer to the other cluster (orange crosses).  Espanel and Sudol \cite{espanel1999single} also proposed that Type II/III specifity required the presence of an aromatic amino acid (W/F/Y) on site 19, in good agreement with weight 3. 

(ii) The distinction between Types II and III is unclear in the literature, because WW domains often have high affinity with both ligand types. 
  
 (iii) Several studies \cite{russ2005natural,kato2002determinants,jager2006structure} have demonstrated the importance of the $\beta_1$-$\beta_2$ loop for achieving Type IV specificity, which requires a longer, more flexible loop, as opposed to short rigid loop for other types. The length of the loop is encoded in weight 4 through the gap symbol on site 13: short and long loops correspond to, respectively, positive and negative $I_4$. The importance of residues R11 and R13 was shown in \cite{kato2002determinants} and \cite{russ2005natural}, where removing R13 of Type IV hPin1 WW domain reduced its binding affinity to [p(S/T)P] ligands.  These observations agree with the logo of weight 4, whoch authorizes substitutions between K and R on sites 11 and 13. 
  
 (iv) A specificity-related sector of eight sites was identified in \cite{russ2005natural}, five of which carry the top entries of weight~3 (green balls in Fig.~3D). Our approach not only provides another specificity-related feature (weight 4) but also the motifs of amino acids affecting Type I \& IV specificity, in good agreement with the experimental findings of \cite{russ2005natural}.

\newpage
\begin{figure}
\begin{fullwidth}
\centering
\includegraphics[scale=0.95,angle=90]{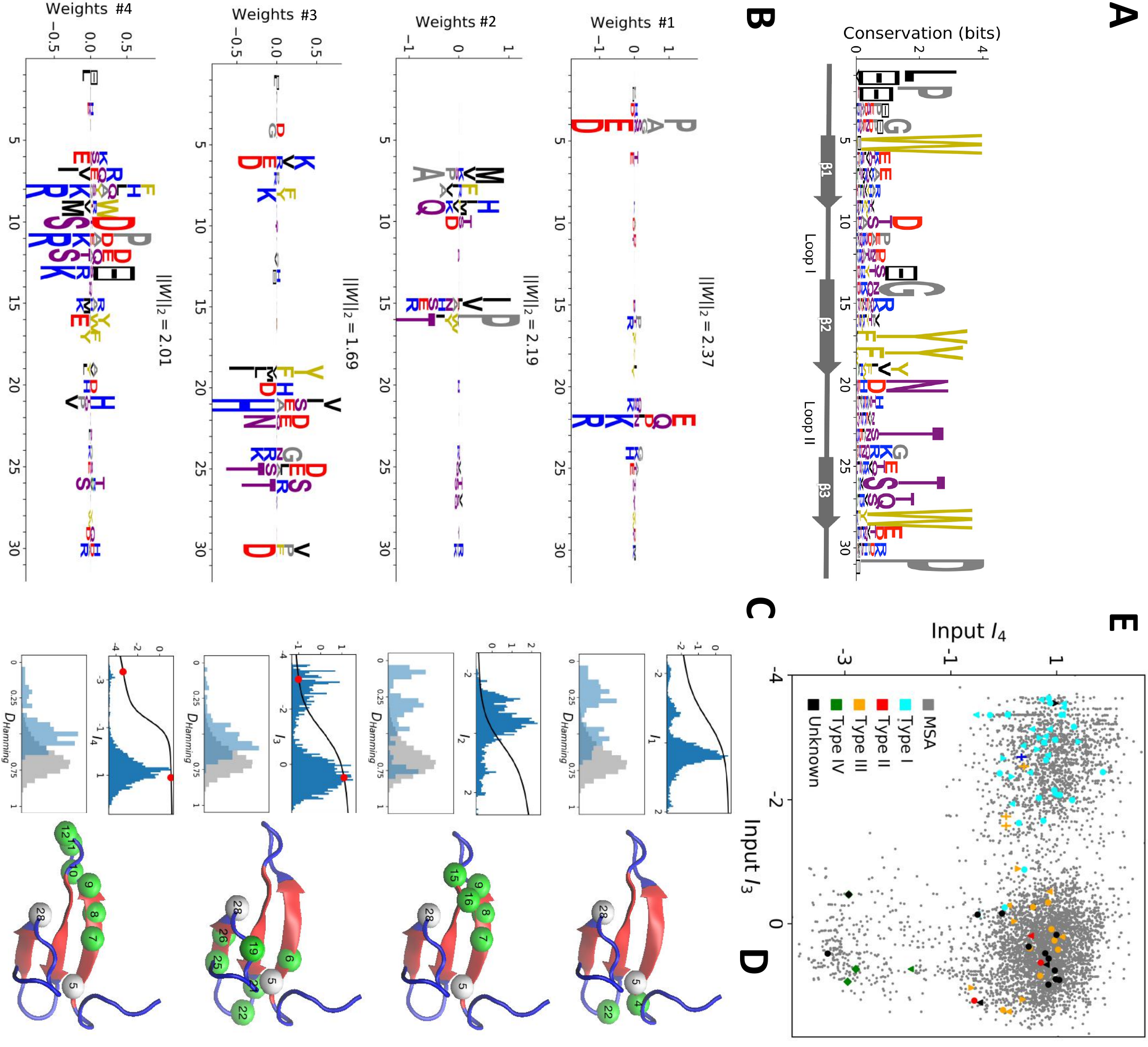}
\caption{{\bf Modeling WW Domain with RBM.} {\bf A.} Sequence logo and secondary structure of the WW domain (PF00397), with three $\beta$-strands. Note the two conserved W in positions 5 and 28. {\bf B.} Weight logos for four representative hidden units, same as Fig.~3B. {\bf C.} Corresponding inputs, average activities and distances between top-20 feature activating sequences, same as Fig.~3C. {\bf D.} 3D visualization of the features, shown on the PDB structure 1e0m \cite{macias2000structural}. White spheres locate the two W. Green spheres locate residues $i$ such that $\sum_v |w_{i\mu}(v)|>0.7$ for each hidden unit $\mu$. {\bf E.} Scatter plot of inputs $I_3$ vs. $I_4$. Gray dots represent the sequences in the MSA; they cluster into three main groups. Colored dots show artificial or natural sequences whose specificities, given in the legend, were tested experimentally. Upper triangle: natural, from \cite{russ2005natural}. Lower triangle: artificial, from \cite{russ2005natural}. Diamond: natural, from \cite{otte2003ww}. Crosses: YAP1 (0) and variants (1 and 2 mutations from YAP1), from \cite{espanel1999single}. The three clusters match the standard ligand type classification. }
\label{fig3}
\end{fullwidth}
\end{figure}
\clearpage

\subsection*{Hsp70 Protein}

\subsubsection{Description}
70-kDa heat shock proteins (Hsp70) form a highly-conserved family represented in essentially all organisms. Hsp70, together with other chaperone proteins, perform a variety of essential functions in the cell: they can assist folding and assembly of newly synthetized proteins, trigger refolding cycles of misfolded proteins, transport unfolded proteins through organelle membranes, and when necessary, deliver non-functional proteins to the proteasome, endosome or lysosome for recycling \cite{bukau1998hsp70,young2004pathways,zuiderweg2017remarkable}. There are 13 HSP70s protein-encoding genes in humans, differing by where (nucleus/cytoplasm, mitochondria, endoplasmic reticulum) and when they are expressed. Some, such as HSPA8 (Hsc70) are constitutively expressed whereas others such as HSPA1 and HSPA5 are stress-induced (respectively by heat shock and glucose deprivation). Notably, Hsc70 can make up to 3\% of the total total mass of proteins within the cell, and is thus one of its most important housekeeping genes.
Structurally, Hsp70 are multi-domain proteins of length of 600-670 sites (631 for E-Coli DNaK gene). They consist of 
\begin{itemize}
\item A Nucleotide Binding Domain (NBD, ~400 sites) that can bind and hydrolyse ATP. 
\item A Substrate Binding Domain (SBD sites), folded in a beta-sandwich structure, which binds to the target peptide or protein.
\item A flexible, hydrophobic interdomain-linker linking the NBD and the SBD.
\item A LID domain, constituted by several (up to 5) $\alpha$ helices, which encapsulates the target protein and blocks its release.
\item An unstructured C-terminal tail of variable length, important for detection and interaction with other co-chaperones, such as Hop proteins \cite{scheufler2000structure}.
\end{itemize}

Hsp70 functions by adopting two different conformations, see Figs.~4A\&B. When the NBD is bound to ATP, the NBD and the SBD are held together and the LID is open, such that the protein has low binding affinity to substrate peptides. After hydrolysis of ATP to ADP, the NBD and the SBD detach from one another, and the LID is closed, yielding high binding affinity and effectively trapping the peptides between the SBD and the LID. By cycling between both conformations, Hsp70 can bind to misfolded proteins, unfold them by stretching (e.g. with two Hsp70 bound at two ends of the protein) and release them for refold cycles. Since Hsp70 alone have low ATPase activity, this cycle requires another type of co-chaperone, J-protein, which simultaneously binds to the target protein and the Hsp70 to stimulate its ATPase activity, as well as a Nucleotide Exchange Factor (NEF) that favors swaps of the ADP back to ATP and hence release of the target protein, see Fig.~1 in \cite{zuiderweg2017remarkable}. 

We have constructed a multiple sequence alignment for HSP70 with $N=675$ consensus sites and $B=32,170$ sequences, starting from the seeds of \cite{malinverni2015large}, and queried SwissProt and Trembl UniprotKB databases using HMMER3 \cite{eddy2011accelerated}. Annotated sequences were grouped based on their phylogenetic origin and functional role. Prokaryotes mainly express two Hsp70 proteins: DnaK ($B=17,118$ sequences in the alignment), which are the prototype Hsp70, and HscA ($B=3,897$), which are specialized in chaperoning of Iron-Sulfur cluster containing proteins. Eukaryotes Hsp70 were grouped by location of expression (Mitochondria: $B=851$, Chloroplaste: $B=416$, Endoplasmic reticulum: $B=433$, Nucleus/Cytoplasm and others: $B=1,452$). We also singled out Hsp110 sequences, which, despite the high homology with Hsp70, correspond to non-allosteric proteins ($B=294$). We have then trained a dReLU RBM over the full MSA with $M=200$ hidden units. We show below the weight logos, structures and input distributions for ten selected hidden units, see Fig. 4 and Appendix 1, Figs.~21-26.

\subsubsection{Inferred weights and interpretation}

Weight 1 encodes a variability of the length of the loop within the IIB subdomain of the NBD, see stretch of gaps from sites 301 to 306. As shown in Fig.~4D (projection along x axis), it separates prokaryotic DNaK proteins - for which the loop is 4-5 sites longer - from most Eukaryotic Hsp70 proteins and prokaryotic HscA. An additional hidden unit (Weight 6 in Appendix 1, Fig.~21) further separates Eukaryotic Hsp70 from HscA proteins, whose loops are 4-5 sites shorter (distribution of inputs $I_6$ in Appendix 1, Fig.~26). This structural difference between the three families was previously reported and is of high functional importance to the NBD \cite{buchberger1994conserved,brehmer2001tuning}. Shorter loops increase the nucleotide exchange rates (and thus the release of target protein) in the absence of NEF, and the loop size controls interactions with NEF proteins \cite{brehmer2001tuning,briknarova2001structural,sondermann2001structure}. Hsp70 proteins having long and intermediate loop size interact specifically with respectively GrpE and Bag-1 NEF proteins, whereas short, HscA-like loops do not interact with any of them. This cochaperone specificity allows for functional diversification within the cell; for instance, Eukaryotic Hsp70 proteins expressed within mitochondria and chloroplasta, such as the human gene HSPA9 and the Chlamydomonas reinhardtii HSP70B share the long loop with prokaryotic DNaK proteins, and therefore do not interact with Bag proteins. Within the DNaK subfamily, two main variants of the loop can be isolated as well (Weight 7 in Appendix 1, Fig.~22), hinting at more NEF-protein specificities.

Weight 2 encodes a small collective mode localized on $\beta_4-\beta_5$ strands, at the edge of the $\beta$ sandwich within the SBD. Weight are quite large ($w\sim2$), and the input distribution is bimodal, separating notably HscA and chloroplastal Hsp70 ($I_2>0$) from mitochondrial Hsp70 and the other Eukaryotic Hsp70 ($I_2<0$). We note also a similarity in structural location and amino-acid content with weight 3 of the WW--domain, which controls binding specificity (Fig.~3B). Though we have not found trace of this motif in the literature, this suggests that it could be important for binding substrate specificity. Endoplasmic reticulum-specific Hsp70 proteins can also be separated from the other Eukaryotic proteins by looking at appropriate hidden units, see Weight 8 in Appendix 1, Fig.~22 and distribution of input $I_8$ in Appendix 1, Fig.~26.

RBM can also extract collective modes of coevolution spanning multiple domains, as shown by Weight 3 (Appendix 1, Fig.~21). The residues supporting Weight 3 (green spheres in Figs.~4A\&B) are physically contiguous in the ADP conformation, but not in the ATP conformation. Hence, Weight 3 captures inter-domain coevolution between the SBD and the LID domains. 

Weight 4 (sequence logo in Appendix 1, Fig.~21) also codes for a wide, inter--domain collective mode, localized at the interface between the SBD and the NBD domains. When the Hsp70 protein is in the ATP conformation, the sites carrying weight 4 are physically contiguous, whereas in the ADP state they are far apart, see yellow spheres in Fig.~4A\&B. Moreover, its input distribution, shown in Fig.~4E, separates the non-allosteric Hsp110 subfamily ($I_4 \sim 0$) from the other subfamilies ($I_4 \sim 40$), suggesting that this motif is important for allostery. Several mutational studies have highlighted 21 important sites for allostery within E-Coli DNaK \cite{smock2010interdomain}; 7 of these positions carries the top entries of Weight 3, 4 appear in another Hsp110-specific hidden unit (Weight 9 in Appendix 1, Fig.~22), and several others are highly conserved and do not coevolve at all. 

Lastly, Weight 5 (Fig. 4C) codes for a collective mode located mainly on the unstructured C-terminal tail, with few sites on the LID domain. Its amino-acid content is strikingly similar across all sites: positive weights for hydrophilic residues (in particular, lysine), and negative weights for tiny, hydrophobic residues. Hydrophobic-rich or hydrophilic-rich sequences are found in the MSA, see Appendix 1, Fig.~28. This motif is consistent with the role of the tail for cochaperone interaction: hydrophobic residues are important for formation of Hsp70-Hsp110 complexes via the Hop protein \cite{scheufler2000structure}. High-charge content is also frequently encountered and at the basis of recognition mechanism in intrinsically disordered protein regions \cite{oldfield2014intrinsically}, which could suggest the existence of different protein partners.

Some of the results presented here were previously obtained with others coevolutionary methods. In \cite{malinverni2015large}, the authors showed that Direct Coupling Analysis could detect conformation-specific contacts; this is similar to hidden units, respectively, 3 and 4 presented here, located on contiguous sites in the, respectively, ADP-bound and ATP-bound conformations. In \cite{smock2010interdomain}, an inter-domain sector of sites  discriminating between allosteric and non-allosteric sequences was found. This sector share many sites with our weight 4, and is also localized at the SBD/NBD edge. However, only a sector could be retrieved with sector analysis, whereas many other meaningful collective modes could be extracted using RBM.

\begin{figure}
\begin{fullwidth}
\centering
\includegraphics[scale=0.9,angle=90]{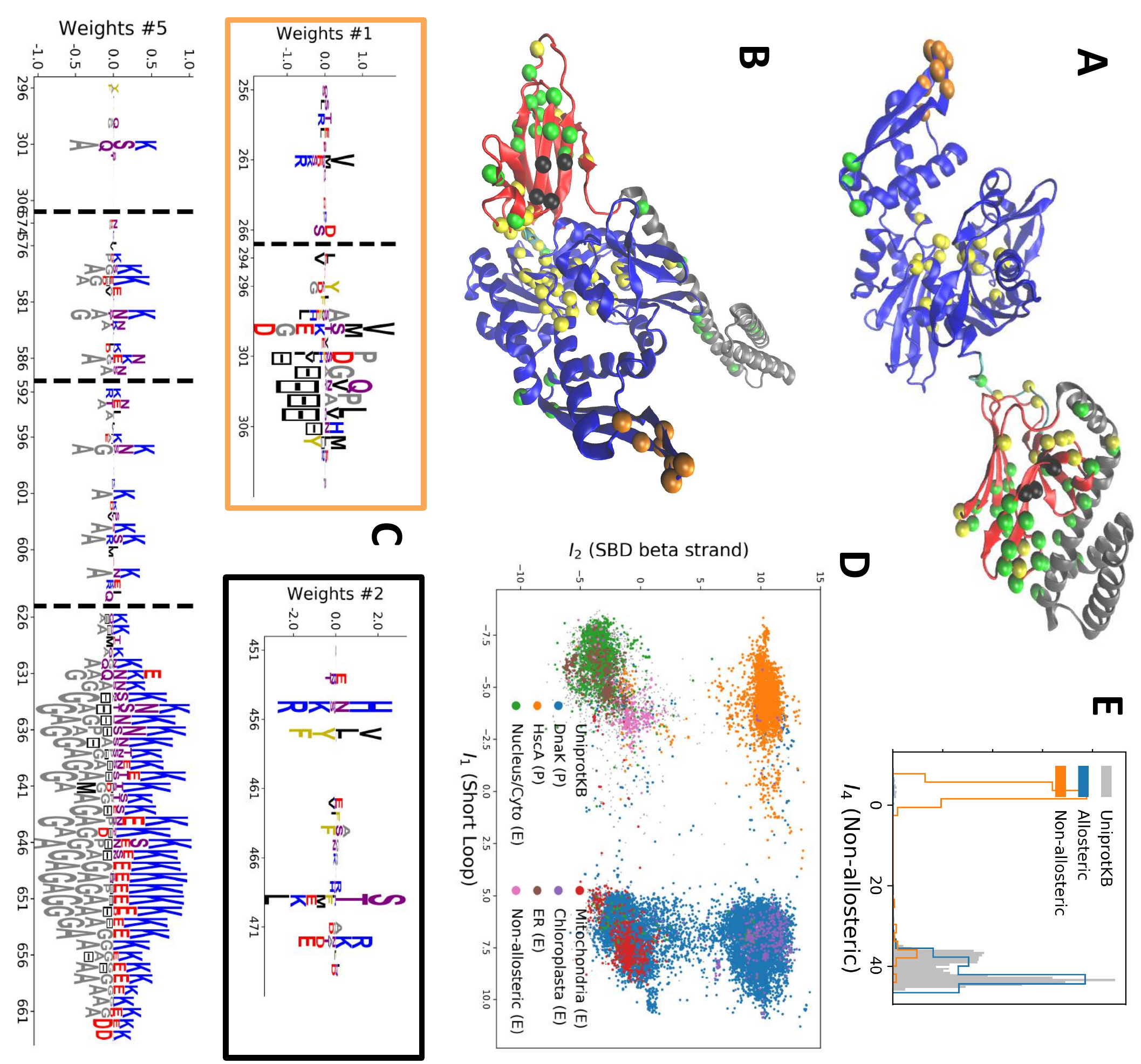}
\caption{{\bf Modeling HSP70 with RBM.}  {\bf A, B.} 3D structures of the DNaK E-coli HSP70 protein in the ADP-bound ({\bf A}: PDB: 2kho \cite{bertelsen2009solution}) and ATP-bound ({\bf B}: PDB: 4jne \cite{qi2013allosteric}) conformations. The colored spheres show the sites carrying the largest entries in the weights in panel C. {\bf C.} Weight logos for hidden units $\mu=1$, 2 and 5; see Appendix 1, Fig.~21 for the other hidden units. Due to the large protein length, we show only weights for positions $i$ with large weights ($\sum_v |w_{i\mu}(v)| > 0.4\times \max_i \sum_v |w_{i\mu}(v)|$), with surrounding positions up to $\pm 5$ sites away; dashed lines vertical locate the left edges of the intervals.  Protein backbone colors: Blue=NBD, Cyan=Linker, Red=SBD, Gray=LID. Colors: Orange=Unit 1 (NBD loop), black = Unit 2 (SBD $\beta$ strand), green= Unit 3 (SBD/LID), yellow = Unit 4 (Allosteric). {\bf D.} Scatter plot of inputs $I_1$ vs. $I_2$. Gray dots represent the sequences in the MSA, and cluster into four main groups. Colored dots represent the main sequence categories based on gene phylogeny, function and expression. {\bf E.} Histogram of input $I_4$, showing separation between allosteric and non-allosteric protein sequences in the MSA.}
\label{fig4}
\end{fullwidth}
\end{figure}
\clearpage
\begin{figure}
\begin{fullwidth}
\centering
\includegraphics[width = 1.2\columnwidth,angle=0]{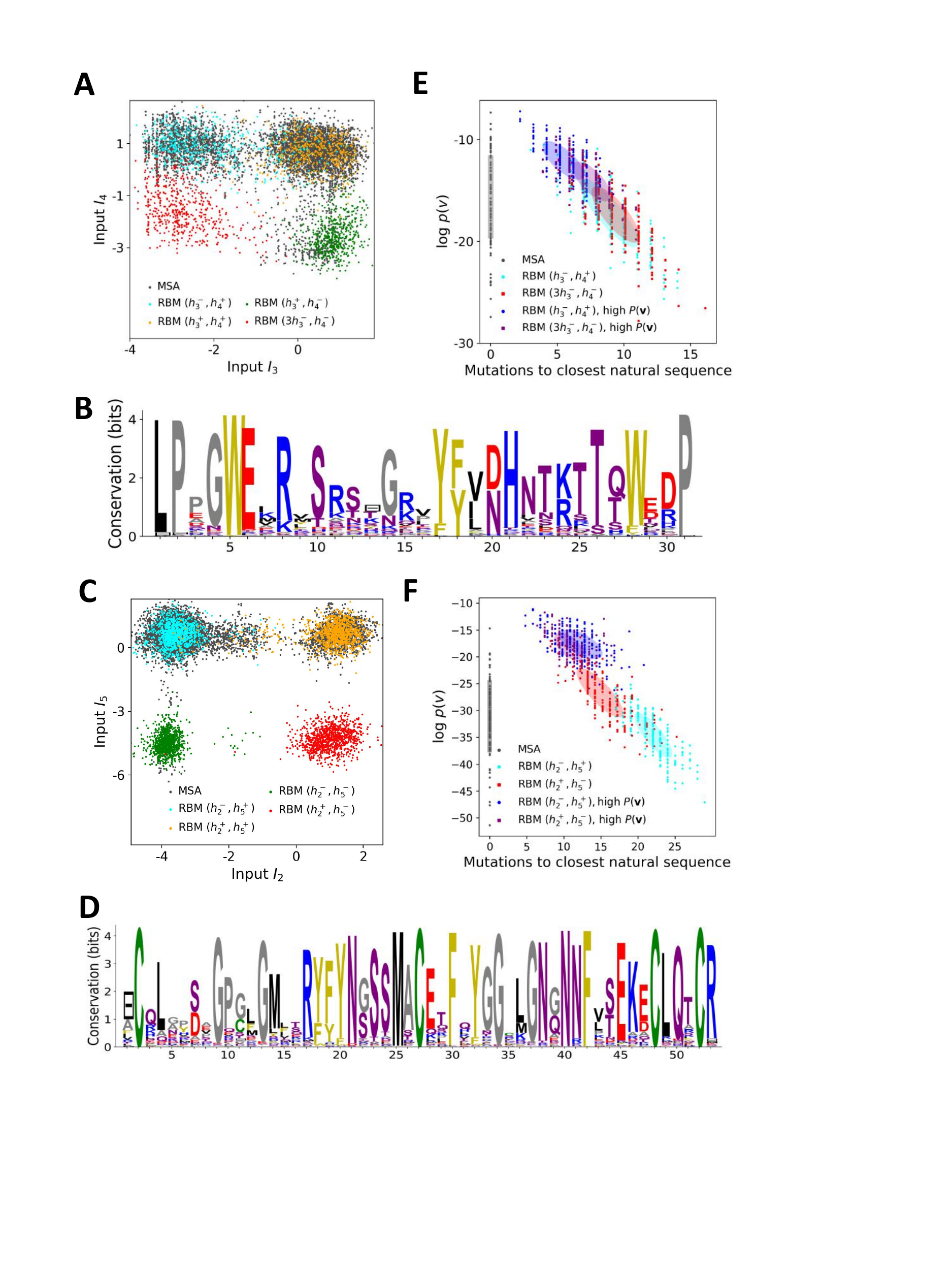}
\vskip -4cm
\caption{{\bf Sequence design with RBM.} {\bf A.} Conditional sampling of WW domain-modeling RBM. Sequences are drawn according to Eqn.~(3), with activities $(h_3,h_4)$ fixed to $(h_4^-,h_4^+)$, $(h_3^+,h_4^-)$, $(h_3^+,h_4^+)$ and $(3 h_3^-,h_4^-)$, see red points indicating the values of $h_3^\pm,h_4^\pm$ in Fig.~3C. Natural sequences in the MSA are shown with gray dots, and generated sequences with colored dots. Four clusters of sequences are obtained; the first three are putatively associated to, respectively, ligand-specific groups I, II/III and IV. The sequences in the bottom left cluster, obtained through very strong conditioning, do not resemble any of the natural sequences in the MSA; their binding specificity is unknown. {\bf B.} Sequence logo of the red sequences in panel {\bf A}, with  `long $\beta_1$-$\beta_2$ loop' and `type I' features. {\bf C.} Conditional sampling of Kunitz domain-modeling RBM, with activities $(h_2,h_5)$ fixed to $(h_2^\pm,h_5^\pm)$, see red dots indicating $h_2^\pm,h_5^\pm$ in Fig.~2C. Red sequences combine the absence of the 11-35 disulfide bridge and a strong activation of the Bikunin-AMBP feature, though these two phenotypes are never found together in natural sequences. {\bf D.} Sequence logo of the red sequences in panel {\bf }C, with  `no disulfide bridge' and `bikunin' features. {\bf E.} Scatter plot of the number of mutations to the closest natural sequence vs log-probability, for natural (gray) and artificial (colored) WW domain sequences. Same color code as panel {\bf A}; dark dots were generated with the high-probability trick, based on duplicated RBM (Methods). Note the existence of many high-probability artificial sequences far away from the natural ones. {\bf F.} Same scatter plot as in panel {\bf E} for natural and artificial Kunitz domain sequences. }
\label{fig6}
\end{fullwidth}
\end{figure}
\clearpage
\subsection*{Sequence Design}

The biological interpretation of the features inferred by the RBM guides us to sample new sequences $\bf v$ with putative functionalities. In practice, we sample from the conditional distribution $P( {\bf v} | {\bf h} )$, Eqn.~(3), where a few hidden-unit activities in the representation $\bf h$ are fixed to desired values, while the others are sampled from Eqn.~(4).
For WW domains, we condition on the activities of hidden units 3 and 4, related to binding specificity. Fixing $h_3$ and $h_4$ to levels corresponding to the peaks in the histograms of inputs in Fig.~3C allows us to generate sequences belonging specifically to each one of the three ligand-specificity clusters, see Fig.~5A. 

In addition, sequences with combinations of activities that are not encountered in the natural MSA can be engineered. As an illustration, we generate by conditional sampling hybrid WW-domain sequences with strongly negative values of $h_3$ and $h_4$, corresponding to a Type I-like $\beta_2$-$\beta_3$ binding pocket and a long, Type IV-like $\beta_1$-$\beta_2$ loop, see Fig.~5A\&B. 

For Kunitz domains, the property `no 11-35 disulfide bond' holds only for some sequences of nematode organisms, whereas the Bikunin-AMBP gene is present only in vertebrates; they are thus never observed simultaneously in natural sequences. Sampling our RBM conditioned to appropriate levels of $h_2$ and $h_5$ allows us to generate sequences with both features activated, see Figs.~5C\&D. 

The sequences designed by RBM are far away from all natural sequences in the MSA, but have comparable probabilities, see Figs.~5E (WW) and 5F (Kunitz). Their probabilities estimated with pairwise direct-coupling models (trained on the same data), whose ability to identify functional and artificial sequences has already been tested \cite{balakrishnan2011learning,cocco2018inverse}, are also large, see Appendix 1, Fig.~7. 

Our RBM framework can also be modified to design sequences with very high probabilities, even larger than in the MSA, by appropriate duplication of the hidden units (Methods). This trick can be combined with conditional sampling, see Fig.~5E\&F.

\subsection*{Contact Predictions}

As illustrated above, co-occurrence of large weight components in highly sparse features often corresponds to nearby sites on the 3D fold. To extract structural information in a systematic way, we use our RBM to derive effective pairwise interactions between sites, which can then serve as estimators for contacts as in direct-coupling based approaches \cite{cocco2018inverse}. The derivation is sketched in Fig.~6A. We consider a sequence ${\bf v}^{a,b}$ with residues $a$ and $b$ on, respectively, sites $i$ and $j$. Single mutations $a\to a'$ or $b\to b'$ on, respectively, site $i$ or $j$ are accompanied by changes in the log probability of the sequence indicated by the full arrows in Fig.~6A. Comparing the change resulting from the double mutation with the sum of the changes resulting from the two single mutations provides our RBM-based estimate of the epistatic interaction, see Eqns.~(10,11) in Methods. These interactions are well correlated with the outcomes of the Direct-Coupling Analysis, see Appendix 1, Fig.~9.

Figure~6 shows that the quality of prediction of the contact maps of the Kunitz  (panel B) and the WW (panel C) domains with RBM is comparable to state-of-the-art methods based on direct couplings \cite{morcos2011direct}; predictions for long-range contacts are reported in Appendix 1, Fig.~10. The quality of contact prediction with RBM 
\begin{itemize}
\item does not seem to depend much on the choice of the hidden-unit potential, compare the Gaussian and dReLU PPV performances in Figs.~6B,C\&D, though the latter have better performance in terms of sequence scoring than the former, see Appendix 1, Figures~1, 2\&5.
\item strongly increases with the number of hidden units, see Appendix 1, Fig.~11,12. This dependence is not surprising, as the number $M$ of hidden units acts in practice as a regularizor over the effective coupling matrix between residues. In the case of Gaussian RBM, the value of $M$ fixes the maximal rank of the matrix $J_{ij}(v_i,v_j)$, see Methods. The value $M=100$ of the number of hidden units is small compared to the maximal ranks $R=20\times N$ of the couplings matrices of the Kunitz ($R=1060$) and WW ($R=620$) domains, and explains why Direct-Coupling Analysis gives slightly better performance than RBM in the contact predictions of Figs.~6B\&C.
\item worsens with stronger weight-sparsifying regularizations, see Appendix 1, Fig.~12, as expected.
\end{itemize}

\rev{We further tested RBM distant contact predictions in a fully blind setting on the 17 protein families (the Kunitz domain plus 16 other domains) that were used for benchmarking plmDCA} \cite{ekeberg2014fast}\rev{, a state-of-the-art procedure for inferring pairwise couplings in Direct-Coupling Analysis. The number of hidden units was fixed to} $M = 0.3\, R$, \textit{i.e.} \rev{proportionally to the domain lengths, and the regularization strength was fixed to} $\lambda_1^2=0.1$. \rev{Contact predictions averaged over all families are reported in Fig.~6D for different choices of the hidden-unit potentials (Gaussian and dReLU). We find that performances are comparable to the ones of plmDCA, but the computational cost of training RBM is substantially higher.}

\clearpage
\begin{figure}
\begin{fullwidth}
\centering
\includegraphics[scale=.95,angle=0]{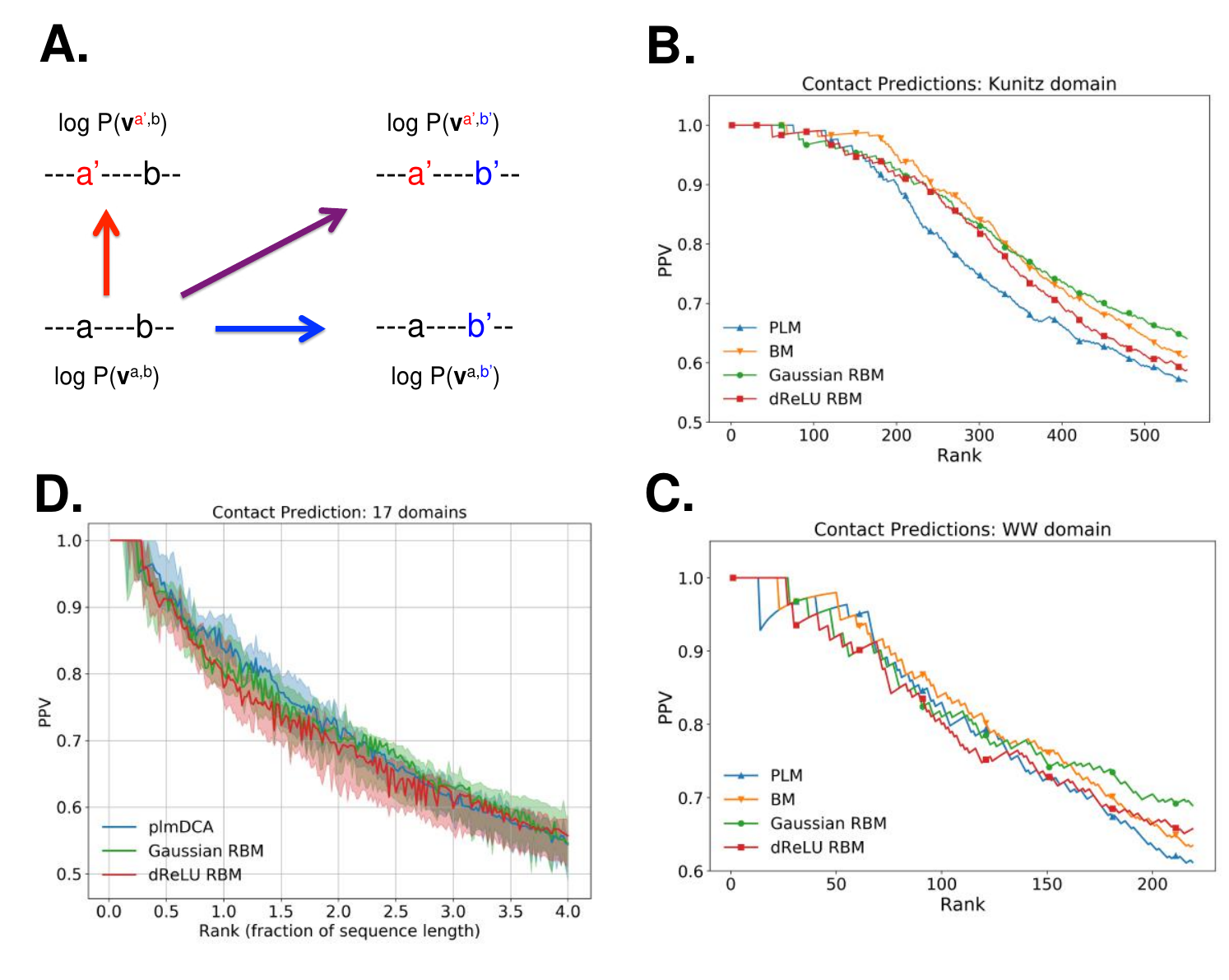}
\caption{{\bf Contact predictions using RBM. }
{\bf A.} Sketch of the derivation of effective epistatic interactions between residues with RBM. The change in log probability resulting from a double mutation (purple arrow) is compared to the sum of the changes accompanying the single mutations (blue and red arrows), see text and Methods, Eqns.~(10,11).
{\bf B}. Positive Predictive Value (PPV) vs. pairs $(i,j)$ of residues, ranked according to their scores for the Kunitz domain. RBM predictions with quadratic (Gaussian RBM) and dReLU potentials are compared to direct coupling-based methods -- Pseudo-Likelihood Method (plmDCA) \cite{ekeberg2014fast}, and Boltzmann Machine (BM) learning \cite{Sutto13567}. 
{\bf C.} Same as panel {\bf B} for the WW domain. 
{\bf D.} \rev{Distant contact predictions for the 17 protein domains used for benchmarking plmDCA in  \cite{ekeberg2014fast} obtained using fixed regularization $\lambda_1^2=0.1$ and $M = 0.3 \times N \times 20$. Positive Predictive Value for contacts between residues separated by at least 5 sites along the protein backbone vs. ranks of the corresponding couplings, expressed as fractions of the protein length $N$; solid lines indicate the median PPV and colored areas the corresponding 1/3 to 2/3 quantiles.} }
\label{fig5}
\end{fullwidth}
\end{figure}
\clearpage

\subsection*{Benchmarking on Lattice Proteins}

Lattice protein (LP) models were introduced in the $90's$ to study protein folding and design \cite{mirny2001}. In one of those models \cite{shakhnovich1990enumeration}, a `protein' of $N=27$ amino acids may fold into $\sim 10^5$ distinct structures on a  $3\times 3\times 3$ cubic lattice, with probabilities depending on its sequence (Methods and Figs.~7A\&B). LP sequence data were used to benchmark the Direct-Coupling Analysis in  \cite{jacquin2016benchmarking}, and we here follow the same approach to assess the performances of RBM in a case where the ground truth is known. We first generate a MSA containing sequences having large probabilities ($p_{nat}>0.99$) of folding into one structure shown in Fig.~7A \cite{jacquin2016benchmarking}.  A RBM with $M=100$ dReLU hidden units is then learned, see Appendix 1 for details about regularization and cross-validation. 

Various structural LP features are encoded by the weights as in real proteins, including complex negative-design related modes, see Figs.~7C\&D and the remaining weights in Supporting Information. Performances in terms of contact predictions are comparable to state-of-the art methods on LP, see Appendix 1, Fig.~11.

The capability of RBM to design new sequences with desired features and high values of fitness, exactly computable in LP as the probability of folding into the native structure in Fig.~7A, can be quantitatively assessed. Conditional sampling allows us to design sequences with specific hidden-unit activity levels, or combinations of features not found in the MSA (Fig.~7E). These designed sequences are diverse and have large fitnesses, comparable to the MSA sequences and even higher when generated by duplicated RBM  (Fig.~7F), and  well correlated with the RBM probabilities $P({\bf v})$ (Appendix 1, Fig.~6).

\begin{figure}
\begin{fullwidth}
\centering
\includegraphics[scale=0.8,angle=90]{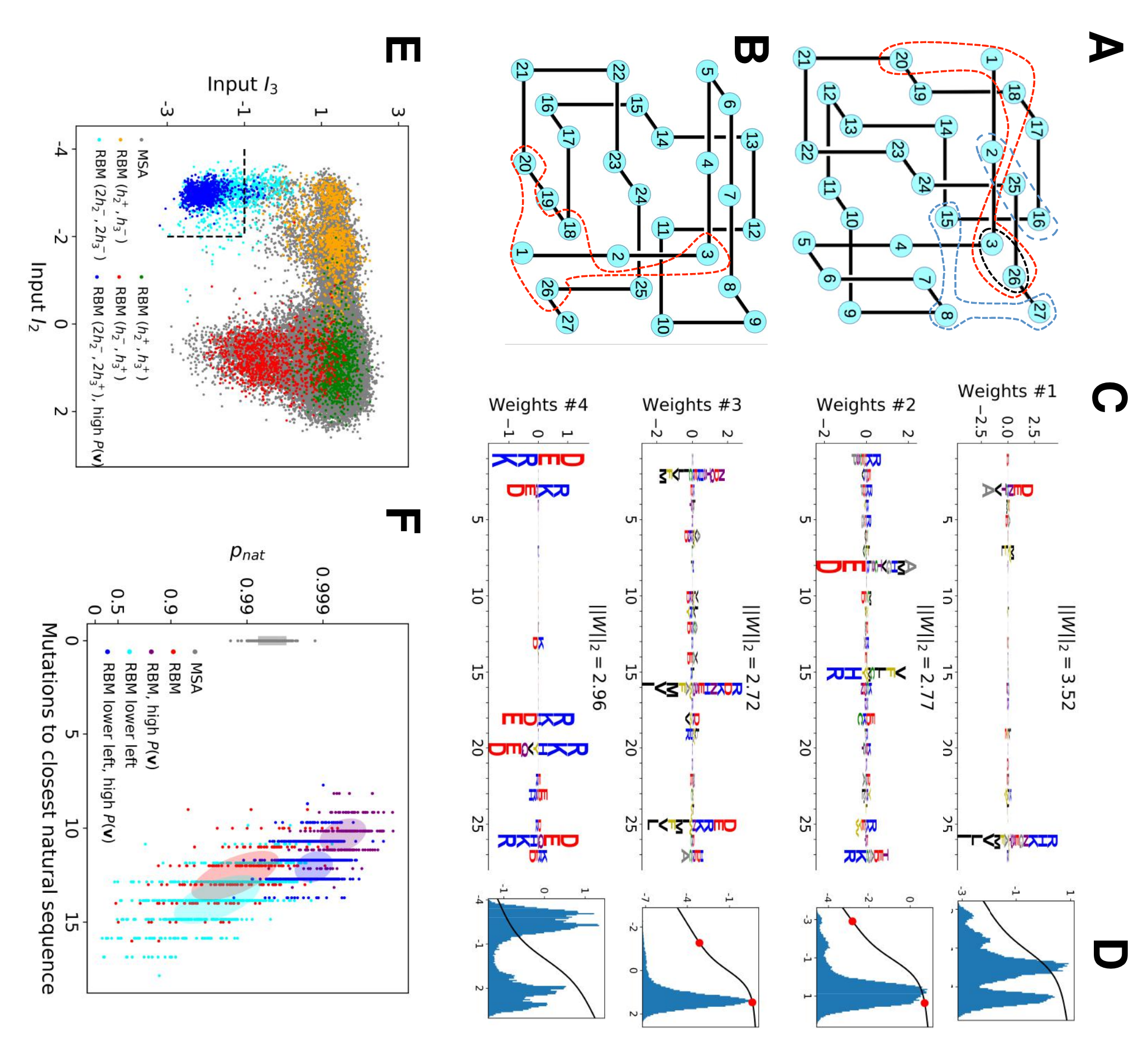}
\caption{{\bf Benchmarking RBM with lattice proteins.} {\bf A.} $S_A$, one of  the $103,406$ distinct structures that a 27-mer can adopt on the cubic lattice \cite{shakhnovich1990enumeration}. Circled sites are related to the features shown in panel 6C. {\bf B.} $S_G$, another fold with a contact map (set of neighbouring sites) close to $S_A$ \cite{jacquin2016benchmarking}.  {\bf C.} Four weight logos for a RBM inferred from sequences folding into $S_A$, see Supporting Information for the remaining 96 weights. Weight 1 corresponds to the contact between sites 3 and 26, see black dashed contour in panel A; the contact can be realized by amino acids of opposite (-+) charges ($I_1>0$), or by hydrophobic residues ($I_1<0$). Weights 2 and 3 are related to, respectively, the triplets of amino acids 8-15-27 and 2-16-25, each realizing two overlapping contacts on $S_A$ (blue dashed contours). Weight 4 codes for electrostatic contacts between 3-26, 1-18 and 1-20, and imposes that the  charges of amino acids 1 and 26 have same sign. The latter constraint is not due to the native fold (1 and 26 are `far away' on $S_A$) but impedes folding in the `competing' structure, $S_G$ (Fig.~7B and Methods), in which sites 1 and 26 are neighbours \cite{jacquin2016benchmarking}. {\bf D.}  Distributions of inputs $I$ and average activities (full line, left scale). All features are activated across the entire sequence space (not shown). {\bf E.} Conditional sampling with activities $(h_2,h_3)$ fixed to $(h_2^\pm, h_3^\pm)$, see red dots in panel D. Designed sequences occupy specific clusters in the sequence space, corresponding to different realizations of the overlapping contacts encoded by weights 2 and 3 (panel 6C). Conditioning to $(h_2^-, h_3^+)$ makes possible to generate sequences combining features not found together in the MSA, see bottom left corner, even with very high probabilities (Methods). {\bf F.} Scatter plot of the number of mutations to the closest natural sequence vs. the probability $p_{nat}$ of folding into structure $S_A$ (see \cite{jacquin2016benchmarking} for a precise definition) for natural (gray) and artificial (colored) sequences. Note the large diversity and the existence of sequences with higher $p_{nat}$ than in the training sample. }
\label{fig7}
\end{fullwidth}
\end{figure}
\clearpage

\subsection{Cross-validation of the model and interpretability of the representations}

Each RBM was trained on a randomly chosen subset of 80\% of the sequences in the MSA, while the remaining 20\% (called test set) were used for validation of its predictive power. In practice, we compute the average log-probability of the test set to assess the performances of the RBM for various values of the number $M$ of hidden units, of the regularization strength $\lambda_1^2$ and for different hidden-unit potentials. Results for the WW and Kunitz domains and for Lattice Proteins are reported  in Fig.~8 and in Appendix~2 (Model Selection). The dReLU potential, which include quadratic and Bernoulli (another popular choice for RBM) potentials as special cases, is consistently better than the latters. As expected, increasing $M$ allows RBM to capture more features in the data distribution and, therefore, improves performances up to some level, after which overfitting starts to occur. 

The impact of the regularization strength $\lambda_1^2$ favoring weight sparsity (see definition in Methods Eqn.~(8)) is twofold, see Fig.~8A for the WW domain. In the absence of regularization ($\lambda_1^2=0$) weights have components on all sites and residues, and the RBM overfit the data, as found from the large difference between the log-probabilities of the training and test sets. \rev{Overfitting notably results in generated sequences that are close to the natural ones and not very diverse, as seen from the entropy of the sequence distribution (Appendix 1 Fig.~8)}. Imposing mild regularization allows the RBM  to avoid overfitting and maximize the log-probability of the test set ($\lambda_1^2=0.03$ in Fig.~8A), but most sites and residues carry non-zero weights. Interestingly, imposing stronger regularizations has low impact on the generalization abilities of RBM (weak decrease of test set log-probability), while making weights much sparser  ($\lambda_1^2=0.25$ in Fig.~3). For too large regularizations, too few non-zero weights remain available and the RBM is not powerful enough to adequately model the data (drop in log-probability of the test set).

Favoring sparser weights in exchange for a small loss in log-probability has a deep impact on the nature of the representation of the sequence space by the RBM, see Fig.~8B. Good representations are expected to capture invariant properties of sequences across evolutionary divergent organisms, rather than idiosyncratic features attached to a limited set of sequences (mixture model in Fig.~8C). For sparse enough weights, the RBM is driven into the compositional representation regime (see \cite{tubiana2017emergence}) of Fig.~8E, in which each hidden unit encodes a limited portion of a sequence and the representation of a sequence is defined by the set of hidden units with strong inputs. Hence, the same hidden unit (e.g. weights 1 and 2 coding for the realizations of  contacts in the Kunitz domain in Fig.~2B) can be recruited in many parts of the sequence space corresponding to very diverse organisms (see bottom histograms attached to weights 1 and 2 in Fig.~2C, showing that the sequences corresponding to strong inputs are scattered all over the sequence space). In addition, silencing or activating one hidden unit affects only a limited number of residues (contrary to the entangled regime of Fig.~8D), and a large diversity of sequences can be generated through combinatorial choices of the activity states of the hidden units, which guarantees efficient sequence design.

\rev{In addition, inferring sparse weights makes their comparison across many different protein families easier. We show in Figs.~9\&10 some representative weights obtained after training RBMs with the MSAs of the 16 families considered in }\cite{ekeberg2014fast} \rev{(the 17th family, the Kunitz domain, is shown in Fig.~2), chosen to illustrate the broad classes of encountered motifs; see Supporting Information for the other top weights of the 16 families. We find that weights may code for a variety of structural properties}
\begin{itemize}
\item \rev{Pairwise contacts on the corresponding structures, realized by various types of residue-residue  physico-chemical interactions, see Figs.~9A\&B. These motifs are similar to weights 2 of the Kunitz domain (Fig.~2B) and weight 1 of the WW domain (Fig.~3B).}
\item \rev{Structural triplets, carrying residues in proximity either on the tertiary or on the secondary structure, see Figs.~9C,D,E\&F. Many such triplets arise from electrostatic interactions and carry amino acids with alternating charges (Figs.~9C,D\&E); they are often found in} $\alpha$\rev{-helices and reflect their} $\sim 4$\rev{-site periodicity (Fig.~9E and last two sites in Fig.~9D), in agreement with weight 1 of the Kunitz domain (Fig.~2B). Triplets may also involve residues with non-electrostatic interactions (Fig.~9F). }
\item \rev{Other structural motifs involving four or more residues, \textit{e.g.} between} $\beta$\rev{-strands, see Fig.~9G. Such motifs were also found in the WW domain, see weight 2 in Fig.~3B.}
\end{itemize}
\rev{In addition, weights may also reflect non-structural properties, such as:}
\begin{itemize}
\item \rev{Stretches of gaps at the extremities of the sequences, indicating the presence of subfamilies containing shorter proteins, see Fig.~10A\&B.}
\item \rev{Stretches of gaps in regions corresponding to internal loops of the proteins, see Figs.~10C\&D. These motifs control the length of these loops, similarly to weight 1 of HSP70, see Fig.~4C.}
\item \rev{Contiguous residue motifs on loops (Figs.~10E\&F) and }$\beta-$\rev{strands (Fig.~10G). These motifs could be involved in binding specificity, as found in the Kunitz and WW domains (weights 4 in Fig.~2B\&3B).}
\item \rev{Phylogenetic properties shared by a subset of evolutionary close sequences, see bottom histograms Figs.~19H\&I, contrary to the motifs listed above. These motifs are generally less sparse and scattered over the protein sequence, as weight 5 of the Kunitz domain in Fig.~2B.}
\end{itemize}
\rev{For all those motifs, the top histograms of the inputs on the corresponding hidden units indicate how the protein families cluster into distinct subfamilies with respect to the features.}

\begin{figure}
\begin{fullwidth}
\centering
\includegraphics[scale=0.7,angle=0]{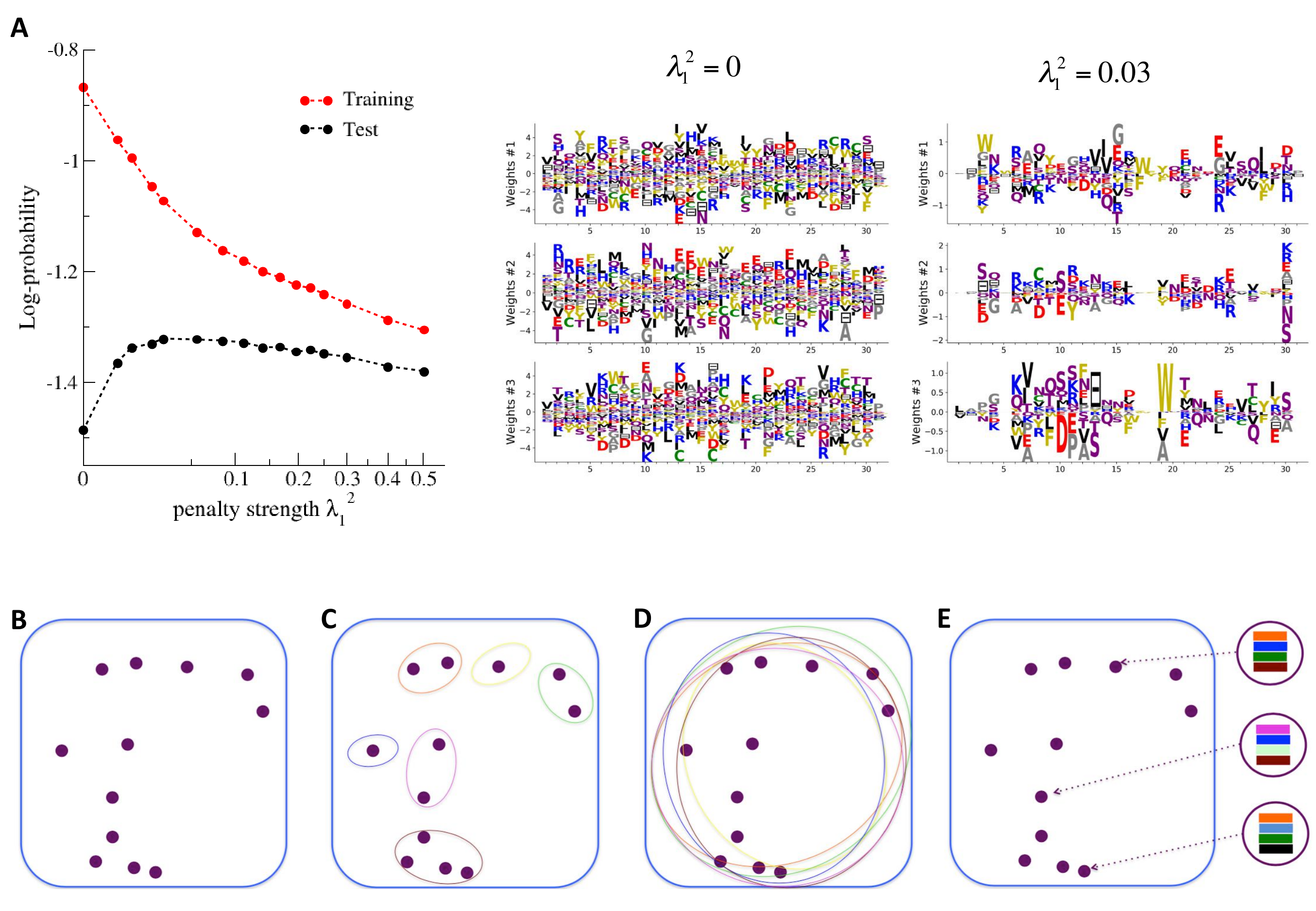}
\caption{ {\bf Nature of representations built by RBM and interpretability of weights.} 
{\bf A.} \textit{Effect of sparsifying regularization.}  Left: log-probability (Methods, Eqn.~(8)) as a function of the regularization strength $\lambda_1^2$  (square root scale) for RBM with $M=100$ hidden units trained on WW domain sequence data. Right: Weights attached to three representative hidden units are shown for $\lambda_1^2=0$ (no regularization) and 0.03 (optimal log-likelihood for the test set, see left panel); weights shown in Fig.~3 were obtained at higher regularization  $\lambda_1^2=0.25$. For larger regularization, too many weights vanish, and the log-likelihood diminishes. 
{\bf B.} Sequences (purple dots) in the MSA attached to a protein family define a highly sparse subset of the sequence space (symbolized by the blue square), from which a RBM model is inferred. The RBM then defines a distribution over the entire sequence space, with high scores for natural sequences and over many more other sequences putatively belonging to the protein family. The representations of the sequence space by RBM can be of different types, three examples of which are sketched in the following panels. 
{\bf C.} \textit{Mixture model:} each hidden unit focuses on a specific region in sequence space (color ellipses, different colors correspond to different units), and the attached weights form a template for this region. The representation of a sequence thus involves one (or a few) strongly activated hidden units, while all remaining units are inactive. 
{\bf D.} \textit{Entangled model:} all hidden units are moderatly active across the sequence space. The pattern of activities vary from one sequence to another in a complex manner.
{\bf E.} \textit{Compositional model:} a moderate number of hidden units are activated for each protein sequence, each recognizing one of the motifs (shown by colors) in the sequence and controlling one of the protein biological properties. Composing the different motifs in various way (right circled compositions) generates a large diversity of sequences.
} 
\label{fig8}
\end{fullwidth}
\end{figure}

\clearpage
\begin{figure}
\begin{fullwidth}
\centering
\includegraphics[scale=0.15,angle=0]{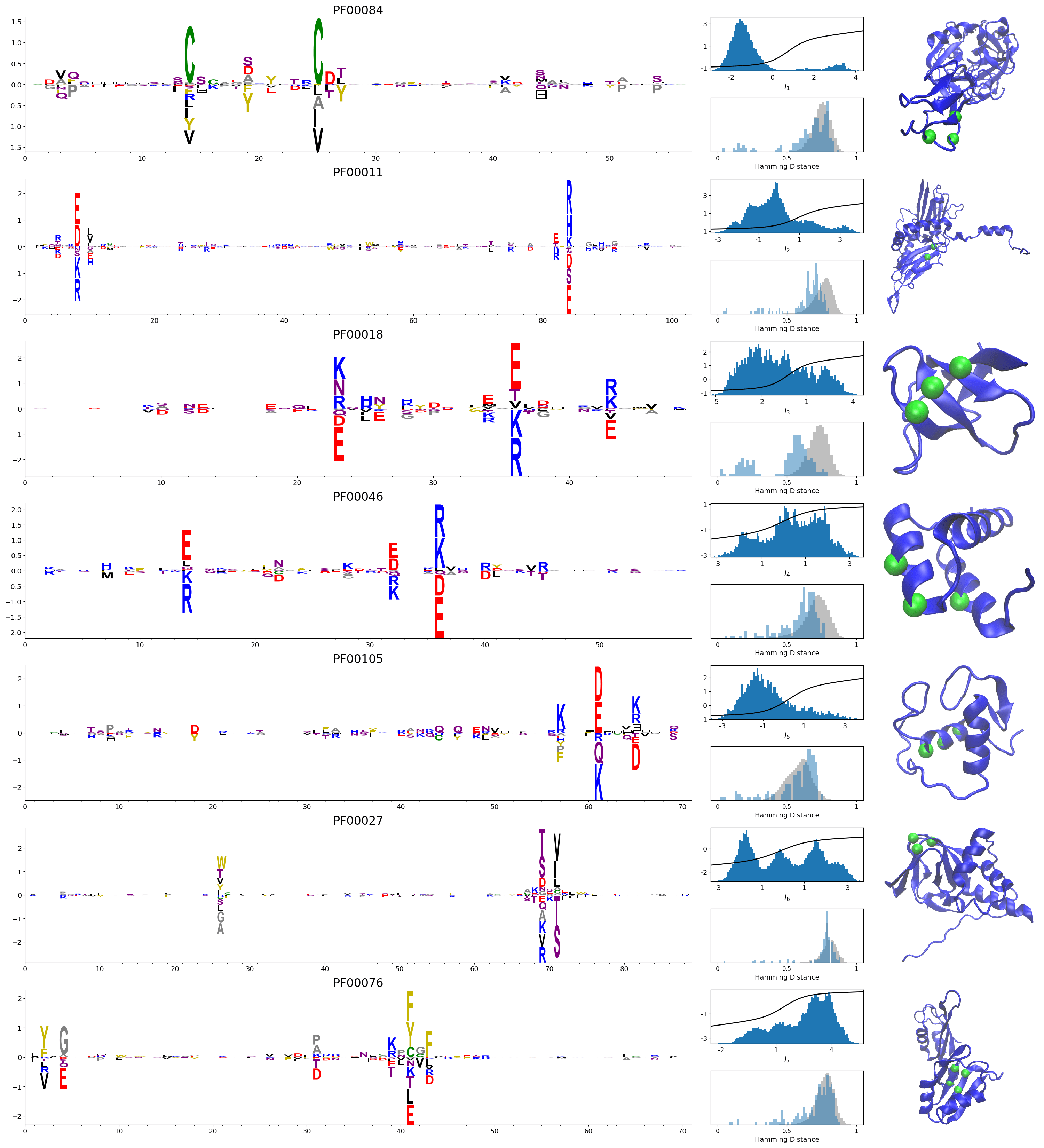}
\caption{ {\bf Representative weights of the protein families selected in}  \cite{ekeberg2014fast} \ref{coding for structural properties}. \rev{RBM parameters: $\lambda_1^2=0.25$, $M=0.05 \times N \times 20$. Same format as Figs.~2B, 3B and 4B. Weights are ordered by similarity, from top to bottom: Sushi domain (PF00084), Heat shock protein Hsp20 (PF00011), SH3 Domain (PF00018),  Homeodomain protein (PF00046), Zinc finger--C4 type (PF00105), Cyclic nucleotide-binding domain (PF00027), RNA recognition motif (PF00076). Green spheres show the sites carrying largest weights on the 3D folds (in order, PDB: 1elv,2bol,2hda,2vi6,1gdc,3fhi,1g2e). The ten weights with largest norms in each family are shown in Supporting Files 5-6.}
}
\label{fig9}
\end{fullwidth}
\end{figure}
\clearpage

\clearpage
\begin{figure}
\vspace*{-2.0cm}
\begin{fullwidth}
\centering
\includegraphics[scale=0.15,angle=0]{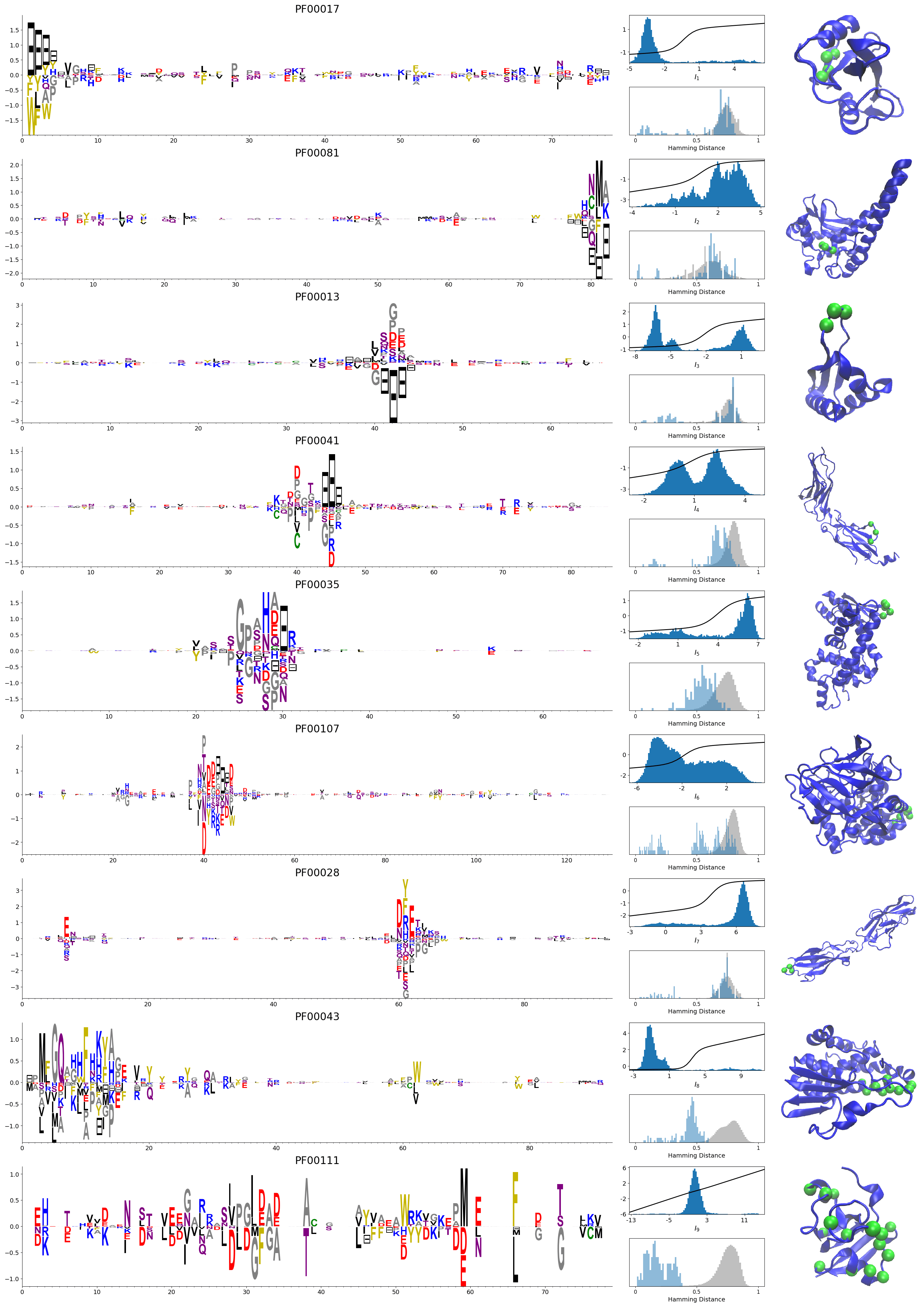}
\caption{ {\bf Representative weights of the protein families selected in}  \cite{ekeberg2014fast} \ref{coding for non-structural properties}. \rev{RBM parameters: $\lambda_1^2=0.25$, $M=0.05 \times N \times 20$. Same format as Figs.~2B, 3B and 4B. Weights are ordered by similarity, from top to bottom:  SH2 domain (PF00017), Superoxide dismutase (PF00081), K Homology domain (PF00013), Fibronectin type III domain (PF00041), Double-stranded RNA binding motif (PF00035), Zinc-binding dehydrogenase (PF00107), Cadherin (PF00028), Glutathione S-transferase, C-terminal domain (PF00043), 2Fe-2S iron-sulfur cluster binding domain (PF00111). Green spheres show the sites carrying largest weights on the 3D folds (in order, PDB: 1o47,3bfr,1wvn,1bqu,1o0w 1a71,2o72,6gsu,1a70). The ten weights with largest norms in each family are shown in Supporting Files 5-6.}
} 
\label{fig10}
\end{fullwidth}
\end{figure}
\clearpage

\section*{Discussion}

In summary, we have shown that RBM are a promising, versatile, and unifying method for modeling and generating protein sequences. RBM, when trained on protein sequence data, reveal a wealth of structural, functional and evolutionary features. To our knowledge, no other method has been able to extract such detailed information in a unique framework so far. In addition, RBM can be used to design new sequences: hidden units can be seen as representation-controlling knobs, tunable at will to sample specific portions of the sequence space corresponding to desired functionalities. A major and appealing advantage of RBM is that the two-layer architecture of the model embodies the very concept of genotype-phenotype mapping (Fig. 1C). Codes for learning and visualizing RBM are attached to this publication (Methods).

From a machine-learning point of view, the values of RBM defining parameters (class of potentials and number $M$ of hidden units, regularization penalties) were selected based on the log-probability of a test set of natural sequences not used for training and on the interpretability of the model. The dReLU potentials we have introduced in this work (Eqn.~(6)) consistently outperform other potentials for generative purposes. As expected, increasing $M$ improves likelihood up to some level, after which overfitting starts to occur. Adding  sparsifying regularization not only prevents overfitting but also facilitates the biological  interpretation of weights (Fig.~8A). It is thus an effective way to enhance the correspondence between representation and phenotypic spaces (Fig.~1C). It also allows us to drive the RBM operation point in which most features can be activated across many regions of the sequence space (Fig.~8E); examples are provided by hidden units 1 and 2 for the Kunitz domain in Figs.~2B\&C and hidden unit 3 for the WW domain in Figs.~3B\&C. Combining these features allows us to generate a variety of new sequences with high probabilities, such as those shown in Fig.~5.  
Note that some inferred features, such as hidden unit 5 in Figs.~2C\&D and, to a lesser extent, hidden unit 2 in Figs.~3B\&C, are, on the contrary, activated by evolutionary close sequences. Our inferred RBMs thus share some partial similarity with the mixture models of Fig.~8C. Interestingly, the identification of specific sequence motifs with structural, functional or evolutionary meaning does not seem to be restricted to a few protein domains or proteins, but could be a generic property as suggested by our study of 16 more families (Figs.~9\&10).

Despite the algorithmic improvements developed in the present work (Methods), training RBM is challenging as it requires intensive sampling. Generative models alternative to RBM and not requiring Markov Chain sampling exist in machine learning, such as Generative Adversarial Networks \cite{goodfellow2014generative} and Variational Auto--encoders (VAE) \cite{kingma2013auto}. VAE were recently applied to protein sequence data for fitness prediction \cite{novak,marks}. Our work differs in several important points: our RBM is an extension of direct-based coupling approaches, requires much less hidden units (about 10 to 50 times less than \cite{novak} and \cite{marks}), has a simple architecture with two layers carrying sequences and representations, infers interpretable weights with biological relevance, and can be easily tweaked to design sequences with desired statistical properties. We have shown that RBM can successfully model small domains (with a few tens of amino acids) as well as much longer proteins (with several hundreds of residues). The reason is that, even for very large proteins, the computational effort can be controlled through the number $M$ of hidden units, see Methods for discussion about the running time of our learning algorithm. Choosing moderate values of $M$ makes the number of parameters to be learned reasonable and avoids overfitting, yet allowing for the discovery of important functional and structural features. It is, however, unclear how $M$ should scale with $N$ to unveil `all'  the functional features of very complex and rich proteins (such as Hsp70).

From a computational biology point of view, RBM unifies and extends previous approaches in the context of protein coevolutionary analysis. From the one hand, the features extracted by RBM identify `collective modes' controlling the biological functionalities of the protein, in a similar way to the so-called sectors extracted by statistical coupling analysis \cite{halabi2009protein}. However, contrary to sectors, the collective modes are not disjoint: a site may participate to different features, depending on the value of the residue it carries. On the other hand, RBM coincide with direct-coupling analysis \cite{morcos2011direct} when the potential ${\cal U}(h)$ is quadratic in $h$. For non-quadratic potentials ${\cal U}$, couplings to all orders between the visible units are present. The presence of high-order interactions allows for a significantly better description of gap modes \cite{feinauer2014improving}, of multiple long-range couplings due to ligand binding, and of outliers sequences (Appendix 1, Fig.~5). Our dReLU RBM model offers an efficient way to go beyond pairwise coupling models, without an explosion in the number of interaction parameters to be inferred, as all high-order interactions (whose number, $~q^N$, is exponentially large in $N$) are effectively generated from the same $M\times N\times q$ weights $w_{i\mu}(v)$. \rev{RBM also outperforms the Hopfield-Potts framework} \cite{cocco2013principal}, \rev{an approach previously introduced to capture both collective and localized structural modes. Hopfield-Potts 'patterns' were derived with no sparsity regularization and within the mean-field approximation, which made the Hopfield-Potts model not sufficiently accurate for sequence design, see Appendix 1,  Figs.~14-18. }

\rev{The weights shown in Figs.~2B, 3B and 4B are stable with respect to subsampling (Appendix 1, Fig.~13)} and could be unambiguously interpreted and related to existing literature. However, the biological significance of some of the inferred features remains unclear, and would require experimental investigation. Similarly, the capability of RBM to design new functional sequences need experimental validation besides the comparison with past design experiments (Fig.~5E) and the benchmarking on \textit{in silico} proteins (Fig.~7). While recombining different parts of natural proteins sequences from different organisms is a well recognized procedure for protein design \cite{stemmer1994rapid,khersonsky2016reinvent}, RBM innovates in a crucial aspect. Traditional approaches cut sequences into fragments at fixed positions based on secondary structure considerations, but such parts are learnt and need not be contiguous along the primary sequence in RBM models. We believe protein design with detailed computational modeling methods, such as Rosetta \cite{simons1997assembly,khersonsky2016reinvent}, could be efficiently guided by our RBM-based approach, in much the same way protein folding greatly benefited from the inclusion of long-range contacts found by direct-coupling analysis \cite{marks2011protein,hopf2012three}.

Future projects include developing systematic methods for identifying function-determining sites, and analyzing more protein families. \rev{As suggested by the analysis of the 16 families shown in Figs.~9\&10, such a study could help establish a general classification of motifs into broad classes with structural or functional relevance, shared by distinct proteins}. In addition, it would be very interesting to use RBM to determine evolutionary paths between two, or more, protein sequences in the same family, but with distinct phenotypes. In principle, RBM could reveal how functionalities continuously change along the paths, and provide a measure of viability of intermediary sequences.

\section{Materials and Methods}

\subsection{Data preprocessing}
We use the PFAM sequence alignments of the V31.0 release (March 2017) for both Kunitz (PF00014) and WW (PF00397) domains. All columns with insertions are discarded, after which duplicate sequences are removed. We are left with, respectively, $N=53$ sites and $B= 8062$ unique sequences for Kunitz, and $N=31$ and $B=7503$ for WW; each site can carry $q=21$ different symbols. To correct for the heterogeneous sampling of the sequence space, a reweighting procedure is applied: each sequence ${\bf v}^\ell$ with $ \ell=1,...,B$ is assigned a weight $w_\ell$ equal to the inverse of the number of sequences with more than $90\%$ amino-acid identity (including itself). In all that follows, the average over the sequence data of a function $f$ is defined as
\begin{equation}
\langle f({\bf v}) \rangle_{MSA}=\left(\sum_{\ell=1}^B w_\ell \; f({\bf v^\ell})\right)\bigg/\left(\sum_{\ell=1}^B w_\ell \right)\ .
\end{equation}

\subsection{Learning procedure}
\subsubsection{Objective function and gradients}
Training is performed by maximizing, through stochastic gradient ascent, the difference  between the log-probability of the sequences in the MSA and the regularization costs, 
\begin{equation} \label{cost_total}
\langle  \log P({\bf v})\rangle_{MSA} - \frac{\lambda_f}{2} \sum_{i,v} g_i(v)^2 - \frac{\lambda_1^2}{2 q N} \sum_\mu \left( \sum_{i,v} | w_{i\mu}(v)| \right)^2\ ,
\end{equation}
Regularization termes include a standard $L_2$ penalty for the potentials acting on the visible units, and a custom $L_2/L_1$ penalty for the weights. The latter penalty corresponds to an effective $L_1$ regularization with an adaptive strength increasing with the weights, thus promoting homogeneity among hidden units \footnote{This can be seen from the gradient of the regularization term, which reads $\lambda_1^2 \left(\sum_{i,v'} | w_{i\mu}(v')|/qN  \right) \text{sign}(w_{i\mu}(v) )$}. 
Besides, it prevents hidden units from ending up entirely disconnected ($w_{i\mu}(v) = 0 \; \forall i, v$), and makes the determination of the penalty strength $\lambda_1^2$ more robust, see Appendix 1, Fig.~2.

According to Eqn.~(5), the probability of a sequence $\bf v$ can be written as, 
\begin{equation}\label{pdev}
P ({\bf v}) = e^{-E_\text{eff}({\bf v})} \bigg/ \bigg(\sum _{{\bf v}'}e^{-E_\text{eff}({\bf v}')}\bigg)\ , \quad \text{where}\quad E_\text{eff}({\bf v}) = - \sum_{i=1}^N g_i(v_i) - \sum _{\mu=1}^M \Gamma \big( I_\mu({\bf v})\big)\end{equation}
is the effective `energy' of the sequence, which depends on all the model parameters. The gradient of $\langle  \log P({\bf v})\rangle_{MSA}$ over one of these parameters, denoted generically by $\psi$, is therefore
\begin{equation}\label{grad56}
\frac{\partial }{\partial\psi} \langle  \log P({\bf v})\rangle_{MSA} =\sum_{\bf v} P({\bf v})  \frac{\partial E_\text{eff}}{\partial\psi}   ({\bf v})- \bigg\langle \frac{\partial E_\text{eff}}{\partial\psi}  ({\bf v})\bigg\rangle_{MSA} \ .
\end{equation}
Hence, the gradient is the difference between the average values of the derivative of $E_{eff}$ with respect to $\psi$ over the model and the data distributions.

\subsubsection{Moment evaluation}
\rev{Several methods have been developped to evaluate the model average in the gradient, see Eqn.~}(\ref{grad56}) \cite{fischer2012introduction}. \rev{The naive approach is to run for each gradient iteration a full Markov Chain Monte Carlo (MCMC) simulation of the RBM until the samples reach equilibrium, then use these samples to compute the model average }\cite{ackley1987learning}. \rev{A more efficient approach is the Persistent Constrastive Divergence} \cite{tieleman2008training}\rev{: the samples obtained from the previous simulation are used to initialize for the next MCMC simulation, and only a small number of Gibbs updates }($N_{MC} \sim 10$) \rev{is performed between each gradient evaluation. If the model parameters evolve slowly, the samples are always at equilibrium, and we obtain the same accuracy as the naive approach at a fraction of the computational cost. In practice, PCD successes if the mixing rate of the Markov Chain - which depends on the nature and dimension of data, and model parameters - is fast enough. In our training sessions, PCD proved sufficient to learn relevant features and good generative models for small proteins and regularized RBM. For larger  proteins, to speed up mixing, we use Parallel Tempering techniques }\cite{parallel tempering,future_algo}.

\subsubsection{Stochastic Gradient Ascent}
\rev{The optimization is carried out by Stochastic Gradient Ascent. At each step, the gradient is evaluated using a mini-batch of the data, as well as a small number of Markov Chain Monte Carlo configurations. In most of our training sessions, we used the same batch size }($=100$) \rev{for both sets. The model is initialized as follows:}

\begin{itemize}
\item Weights $w_{i\mu}(v)$, are randomly and independently drawn from a Gaussian distribution with zero mean and variance equal to $ \frac{0.1}{N}$. The scaling factor $\frac{1}{N}$ ensures that the initial input distribution has variance of the order of $1$.
\item The potentials $g_i(v)$ are given their values in the independent-site model: $g_i(v) = \log \esp{\delta_{v_i,v}}_{\text{MSA}}$, where $\delta$ denotes the Kronecker function.
\item For all hidden-unit potentials, we set $\gamma_+=\gamma_-=1$, $\theta_+=\theta_-=0$.
\end{itemize}

\rev{The learning rate is initially set to }$0.1$\rev{, and decays exponentially after a fraction of the total training time (e.g. 50\%) until it reaches a final, small value, e.g. }$10^{-4}$.

\subsubsection{Dynamic reparametrization}
\rev{For Gaussian and dReLU potentials, there is a redundancy between the slope of the hidden unit average activity and the global amplitude of the weight vector. Indeed, for the Gaussian potential, the model distribution is invariant under rescaling transformations }$\gamma_\mu \rightarrow \lambda^2 \gamma_\mu$, $w_{i\mu} \rightarrow \lambda w_{i\mu}$, $\theta_\mu \rightarrow \lambda \theta_\mu$ \rev{and offset transformation }$\theta_\mu \rightarrow \theta_\mu + K_\mu$, $g_i \rightarrow g_i - \sum_\mu w_{i\mu} \frac{K_\mu}{\gamma_\mu}$. \rev{Though we can set }$\gamma_\mu =1, \; \theta_\mu=0 \; \forall \mu$ \rev{without loss of generality, it can lead either to numerical instability (at high learning rate) or slow learning (at low learning rate). A significantly better choice is to dynamically adjust the slope and offset so that }$\esp{h_\mu} \sim 0$ and $\text{Var}(h_\mu) \sim 1$ \rev{at all time. This new approach, reminiscent of batch normalization for deep networks, is implemented in the training algorithm released with this work and is benchmarked in} \cite{future_algo}.

\subsubsection{Gauge choice}
\rev{Since the conditional probability Eqn.~\ref{condv} is normalized, the transformations }$g_i(v) \rightarrow g_i(v) + \lambda_i$ and $w_{i\mu}(v) \rightarrow w_{i\mu}(v) + K_{i\mu}$ \rev{leave the conditional probability invariant. We choose the zero-sum gauges, defined by }$\sum_v g_i(v) = 0$, $\sum_v w_{i\mu}(v) = 0$. \rev{Since the regularization penalties over the fields and weight depend on the gauge choice, the gauge must be enforced throughout all training and not only at the end. The updates on the fields leave the gauge invariant, so the transformation }$g_i(v) \rightarrow g_i(v) - \frac{1}{q} \sum_{v'} g_i(v')$ \rev{can be used only once, after initialization. On the other hand, it is not the case for the updates on the weights, so the transformation }$w_{i\mu}(v) - \frac{1}{q} \sum_{v'} w_{i\mu}(v')$ \rev{must be applied after each gradient update.}

\subsubsection{Evaluating the partition function}\label{evaluate_Z}
\rev{Evaluating }$P({\bf v})$ \rev{requires knowledge of the partition function} $Z = \displaystyle{\sum_{\bf v} \exp \left( -E_\text{eff}({\bf v}) \right)}$, \rev{see denominator in }Eqn.~(\ref{pdev}). \rev{The later expression, which involves summing over }$q^N$ \rev{terms is not tractable. Instead, we estimate} $Z$ \rev{using the Annealed Importance Sampling algorithm (AIS)} \cite{neal2001annealed,salakhutdinov2008quantitative}. \rev{Briefly, the idea is to estimate partition function ratios. Let }$P_1({\bf v}) = \frac{P_1^*({\bf v})}{Z_1}$, $P_0 = \frac{P_0^*({\bf v})}{Z_0}$ \rev{be  two probability distributions with partition functions }$Z_1$, $Z_0$\rev{. Then:}
\begin{equation}
\esp{\frac{P_1^*({\bf v})}{P_0^*({\bf v})} }_{ {\bf v} \sim P_0 } = \sum_{ {\bf v}} \frac{P_1^*({\bf v})}{P_0^*({\bf v})} \frac{P_0^*({\bf v})}{Z_0} = \frac{1}{Z_0} \sum_{ {\bf v}} P_1^*({\bf v}) = \frac{Z_1}{Z_0} 
\end{equation}
\rev{Therefore, provided that }$Z_0$ \rev{is known (e.g. if }$P_0$\rev{ is an independent model with no couplings), one can in principle estimate }$Z_1$ \rev{through Monte Carlo sampling. The difficulty lies in the variance of the estimator: if }$P_1$, $P_0$ \rev{are very different from one another, some configurations can be very likely for} $P_1$ \rev{and have very low probability with} $P_0$\rev{; these configurations appear almost never in the Monte Carlo estimate of }$\esp{.}$\rev{, but the probability ratio can be exponentially large. In Annealed Importance Sampling, we address this problem by constructing a continuous path of interpolating distributions} $P_\beta ({\bf v})= P_1({\bf v})^\beta\; P_0({\bf v})^{1-\beta}$, \rev{and estimate} $Z_1$ \rev{as a product of ratios of partition functions:}
\begin{equation}
Z_1 = \frac{Z_1}{Z_{\beta_{l_{max}}}} \frac{Z_{\beta_{l_{max}-1}}}{Z_{\beta_{l_{max}-2}}} ... \frac{Z_{\beta_1}}{Z_0} \times Z_0\ ,
\end{equation}
\rev{where we choose a linear set of interpolating inverse temperatures of the form }$\beta_l = \frac{l}{l_{\text{max}}}$. \rev{To evaluate the successive expectations, we use a fixed number }$C$ \rev{of samples initially drawn from} $P_0$\rev{, and gradually anneal them from }$P_0$ \rev{to }$P_1$\rev{ by successive applications of Gibbs sampling at }$P_\beta$\rev{. Moreover, all computations are done in logarithmic scales for numerical stability purposes: we estimate }$\log \frac{Z_1}{Z_0} \approx \esp{ \log \frac{P_1^*({\bf v})}{P_0^*({\bf v})} }_{ {\bf v} \sim P_0 }$\rev{, which is justified if }$P_1$\rev{ and }$P_0$\rev{ are close. In practice, we used }$C=20$ \rev{chains, }$n_\beta = 5\times 10^4$ \rev{steps. For the initial distribution }$P_0$\rev{, we take the closest (in terms of KL divergence) independent model to the data distribution} $P_{MSA}$. \rev{The visible layer fields are the ones of the independent model inferred from the MSA, and the weights are} ${\bf w}^{\beta =0} = 0$. \rev{For the hidden potential values, we infer the parameters from the statistics of the hidden layer activity conditioned to the data.} 

\subsubsection{Explicit formula for sampling and training RBM}
\rev{Training, sampling and computing the probability of sequences with RBM requires: (1) Sampling from }$P({\bf v}|{\bf h})$, \rev{(2) Sampling from} $P({\bf h}|{\bf v})$\rev{, and (3) Evaluating the effective energy} $E_{\text{eff}}({\bf v})$ \rev{and its derivatives. This is done as follows:}

\begin{enumerate}
\item \rev{Each sequence site} $i$ \rev{is encoded as a categorical variable taking integer values }$v_i \in [0,20]$\rev{, with each integer corresponding to one of the 20 amino-acids + 1 gap. Similarly, the fields and weights are encoded as respectively a }$N \times 21$ \rev{matrix, and a }$M \times N \times 21$ \rev{tensor. Owe to the bipartite structure of the graph, }$P({\bf v} | {\bf h}) = \prod_i P({\bf v_i} | {\bf h})$\rev{, see Eqn.~(4). Therefore, sampling from} $P({\bf v} | {\bf h})$ \rev{is done in three steps: compute the inputs received from the hidden layer, then the conditional probabilities} $P(v_i |{\bf h})$ \rev{given the inputs, and sample each visible unit independently from the others  from the corresponding conditional distributions.}

\item \rev{The conditional probability} $P({\bf h} | {\bf v} )$ \rev{factorizes. Given a visible configuration} $\bf v$\rev{, each hidden unit is sampled independently from the others via }$P(h_\mu | {\bf v})$\rev{, see Eqn.~(3). For a quadratic potential} ${\cal U}(h)=\frac12 \gamma h^2 + \theta h$, \rev{this conditional distribution is Gaussian. For the dReLU potential} $\mathcal{U}(h)$ \rev{in Eqn.~(6), we introduce first} $\Phi(x) = \exp(\frac{x^2}{2}) \left[1- \text{erf}(\frac{x}{\sqrt{2}} ) \right] \sqrt{\frac{\pi}{2}}$

\rev{Some useful properties of }$\Phi$ \rev{are:}

\begin{itemize}
\item $\Phi(x) \sim_{x \rightarrow -\infty} \exp(\frac{x^2}{2}) \sqrt{2\pi} $
\item $\Phi(x) \sim_{x \rightarrow \infty} \frac{1}{x} - \frac{1}{x^3} + \frac{3}{x^5} + \mathcal{O}(\frac{1}{x^7})$
\item $\Phi'(x) = x \Phi(x) - 1$
\end{itemize}

\rev{To avoid numerical issues, }$\Phi$ \rev{is computed in practice with its definition for }$x<5$ \rev{and with its asymptotic expansion otherwise. We also write }$\mathcal{TN}(\mu,\sigma^2,a,b)$ \rev{the truncated Gaussian distribution of mode} $\mu$\rev{, width }$\sigma$ \rev{and support }$[a,b]$.

\rev{Then,} $P(h|I)$ \rev{is given by a mixture of two truncated Gaussians:}
\begin{equation}
P(h| I) = p^+ \mathcal{TN} \left(\frac{I - \theta^+}{\gamma_+},\frac{1}{\gamma_+}, 0,+\infty \right) + p^- \mathcal{TN} \left(\mu = \frac{I - \theta^-}{\gamma^-},\sigma^2 = \frac{1}{\gamma^-},- \infty,0 \right)
\end{equation}
where $Z^\pm = \Phi \left( \frac{\mp (I - \theta^\pm)}{\sqrt{\gamma^\pm}} \right) \frac{1}{\sqrt{\gamma^\pm}}$, and $p^\pm = \frac{Z^\pm}{Z^++Z^-}$.

\item \rev{Evaluating} $E_{\text{eff}}$ \rev{and its derivatives requires an explicit expression for the cumulant--generating function} $\Gamma(I)$. \rev{For quadratic potentials} $\Gamma(I)$ \rev{is quadratic too. For dReLU potentials, we have }$\Gamma(I) = \log (Z^+ + Z^-)$ \rev{where }$Z^\pm$ \rev{are defined above}. 
\end{enumerate}
\subsubsection{Computational complexity}
The computational complexity is of the order of $M \times N \times B$, with more accurate variants taking more time. The algorithm scales reasonably to large protein sizes, and was tested successfully for $N$ up to $\sim 700$, taking of the order of 1-2 days on an Intel Xeon Phi processor with 2 $\times$ 28 cores.

\subsection{Sampling procedure}
\label{sectraining}

Sampling from $P$ in Eqn.~(5) is done with Markov Chain Monte Carlo methods, with the standard alternate Gibbs sampler described in the main text and in \cite{fischer2012introduction}. 
Conditional sampling, \textit{i.e.} sampling from $P({\bf v} | h_\mu = h_\mu^c)$ is straightforward with RBM: it is achieved by the same Gibbs sampler while keeping $h_\mu$ fixed.

The RBM architecture can be modified to generate sequences with high probabilities, such as in Figs.~5E\&F. The trick is to duplicate the hidden units, the weights, and the local potentials acting on the visible units, as shown in Methods, Fig.~1. By doing so, the sequences $\bf v$ are distributed according to
\begin{equation}
P_2({\bf v}) \propto \int \prod_{\mu} dh_{\mu1}\,dh_{\mu2} \; P({\bf v} | {\bf h}_1)\, P({\bf v} | {\bf h}_2)  = P({\bf v})^2 \ .
\end{equation}
Hence, with the duplicated RBM, sequences with high probabilities in the original RBM model are given a boost compared to low-probability sequences.
Note that more subtle biases can be introduced by duplicating some (but not all) of the hidden units in order to give more importance in the sampling to the associated statistical features.

\begin{center}
\includegraphics[width=.6\columnwidth,angle=0]{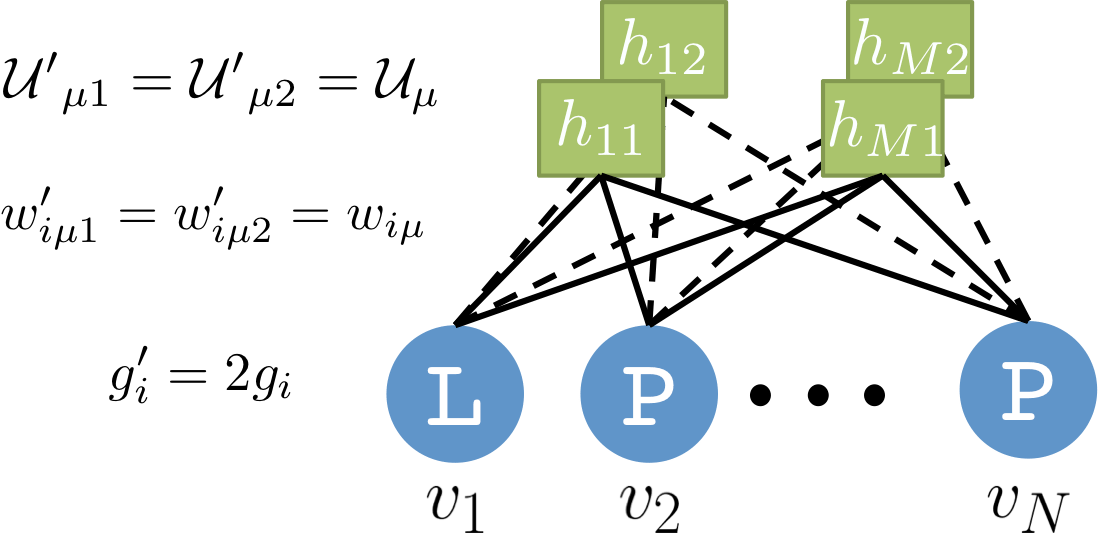}
\justify
\captionof{figure}{Duplicate RBM for biasing sampling toward high-probability sequences. Visible-unit configurations $\bf v$ are sampled from $P_2({\bf v})\propto P({\bf v})^2$.}
\label{figMethods2}
\end{center}

\subsection{Contact map estimation}

RBM can be used for contact prediction in a manner similar to pairwise coupling models, after derivation of an effective coupling matrix $J_{ij}^{\text{eff}}(a,b)$. Consider a sequence $\bf v$, and two sites $i,j$. Define the set of mutated sequences ${\bf v}^{a,b}$ with amino acid content: $v^{a,b}_k = v_k$ if $k \neq i,j$, $a$ if $k = i$, $b$ if $k=j$ (Fig.~6A). 
The differential likelihood ratio 
\begin{equation} \label{epistasis}
\Delta \Delta R_{ij}({\bf v}; a,a',b,b') \equiv \log \left[ \frac{P( {\bf v}^{a,b}) \, P({\bf v}^{a',b'})}{P({\bf v}^{a',b}) \, P({\bf v}^{a,b'})} \right] \ ,
\end{equation}
where $P$ is the marginal distribution in Eqn.~(\ref{marginal}), measures epistatic contributions to the double mutation $a \rightarrow a'$ and $b \rightarrow b'$ on, respectively, sites $i$ and $j$ in the background defined by sequence $\bf v$, see Fig.~6A. The effective coupling matrix is then defined as
\begin{equation} \label{couplings_from_epistasis}
J_{ij}^{\text{eff}}(a,b) = \esp{\frac{1}{q^2}\sum_{a',b'} \Delta \Delta R_{ij}({\bf v}; a,a',b,b')}_{MSA} \ ,
\end{equation}
where the average is taken over the sequences $\bf v$ in the MSA.
For a pairwise model, $\Delta \Delta R_{ij}$ does not depend on the background sequence ${\bf v}$, and Eqn.~(\ref{couplings_from_epistasis}) coincides with the true coupling in the zero-sum gauge. Contact estimators are based on the Frobenius norms of $J^{\text{eff}}$, with the Average Product Correction, see \cite{cocco2018inverse}.

\subsection{Code availability} 
The Python 2.7 package for training and visualizing RBMs, used to obtained the results reported in this work, is available at \url{https://github.com/jertubiana/ProteinMotifRBM}. In addition, Jupyter notebooks are provided for reproducing most figures of the article.

\section{Acknowledgments}
We thank D. Chatenay for useful comments on the manuscript and L. Posani for his help on lattice proteins. This work was partly funded by the ANR project RBMPro CE30-0021-01.

\section{Supporting files}
\begin{itemize}
\item Supporting file 1 (pdf): Weight logo for all hidden units inferred from the Kunitz domain MSA.
\item Supporting file 2: Weight logo for all hidden units inferred from the WW domain MSA.
\item Supporting file 3: Weight logo for all hidden units inferred from the LP MSA.
\item Supporting file 4: Weight logo of 12 Hopfield-Potts pattern inferred from the Hsp70 protein MSA. Same format as Appendix 1 Figures 14-16.
\item Supporting file 5: Weight logo and associated structures of the 10 weights with highest norms, excluding the gap modes for each of the 16 additional domains shown in Fig.~9.
\item Supporting file 6: Weight logo and associated structures of the 10 sparse (i.e. within the 30\% most sparse weights of the RBM) weights with highest norms, excluding the gap modes for each of the 16 additional domains shown in Fig.~9.
\end{itemize}

\bibliography{bibliography}

\appendix
\begin{appendixbox}
\section{Supporting Methods and Figures}
\vskip .5cm

\section{Lattice-protein synthetic sequences}
LP models have been introduced in the $'90$ to investigate the uniqueness of folding shared by the majority of real proteins \cite{shakhnovich1990enumeration}, and have been more recently used to benchmark graphical models inferred from sequence data \cite{jacquin2016benchmarking}. 
There are  ${\cal N}=103,406$ possible folds, \textit{i.e.} self-avoiding path of the 27 amino-acid-long chains, on a $3 \times 3\times 3$ lattice cube  \cite{shakhnovich1990enumeration}. The probability that the protein sequence ${\bf v}=(v_1,v_2,...,v_{27})$ folds in one of these, say, $S$, is 
\begin{equation}
 P_{nat}({\bf v} ; S)=\frac{\displaystyle e^{-{\cal E}({\bf v} ; S)}}{\displaystyle \sum_{S'=1}^{\cal N} e^{-{\cal E}({\bf v}; S')}} \ ,
\label{pnat}
\end{equation}
where the energy of sequence ${\bf v}$ in structure $S$ is given by
\begin{equation}
{\cal E}({\bf v};S) = \sum_{i<j} c_{ij}^{(S)}\; E(v_i,v_j)\ .
\label{energy}
\end{equation}
\noindent 
In the formula above, $c^{(S)}$ is the contact map: $c^{(S)}_{ij}=1$ if the pair of sites ${ij}$ is in contact, \textit{i.e.} $i$ and $j$ are nearest neighbors on the lattice and zero otherwise. The pairwise energy $E(v_i,v_j)$ represents the amino-acid physico-chemical interactions, given by the the Miyazawa-Jernigan (MJ) knowledge-based potential \cite{miyazawa1996residue}.

A collection of 36,000 sequences that specifically fold on structure $S_A$ (Fig.~7A, Main Text) with high probability $P_{nat}({\bf v};S_A)>0.995$ were generated by Monte Carlo simulations as described in \cite{jacquin2016benchmarking}. As real MSA, Lattice Protein data feature short- and long-range correlations between amino-acid on different sites, as well as high-order interactions that arise from competition between folds \cite{jacquin2016benchmarking}.

\section{Model selection}

We discuss here the choice of parameters (strength of regularization, number of hidden units, shape of hidden-unit potentials, ...) for the RBM used in main text. Our goal is to achieve good generative performances and to learn biologically interpretable representations. We estimate the accuracy of the fit to the data distribution using the average log-likelihood, divided by the number of visible units $\frac{1}{N} \langle \log P({\bf v}) \rangle _{MSA}$. For visible-unit variables with $q=21$ possible values (20 amino acids + gap symbol), this number typically ranges from $-\log 21\simeq -3.04$ (uniform distribution) to 0. Evaluating $P({\bf v})$ (Methods Eqn.~1) requires knowledge of the partition function $Z = \sum_{\bf v} \exp \left( \sum_{i=1}^N g_i(v_i) + \sum_{\mu=1}^M \Gamma_\mu(I_\mu({\bf v})) \right)$, see Section \ref{evaluate_Z}.

\subsection{Number of hidden units}
The number of hidden units is critical for the generative performance. We trained RBMs on the Lattice Protein data set for various potentials (Bernoulli, quadratic and dReLU), number of hidden units (1-400) and regularizations ($\lambda_1^2 =0$, $\lambda_1^2=0.025$). The likelihood estimation shows that, as expected, the larger $M$, the better the ability to fit the training data  (Appendix 1, Fig. 1). Overfitting \textit{i.e.} a decrease in test set performance may occur for large $M$. For the regularized case, the likelihood saturates with about 100 hidden units. Similar results are obtained on WW, see Appendix 1, Fig.~2. 

\begin{center}
\includegraphics[scale=0.35]{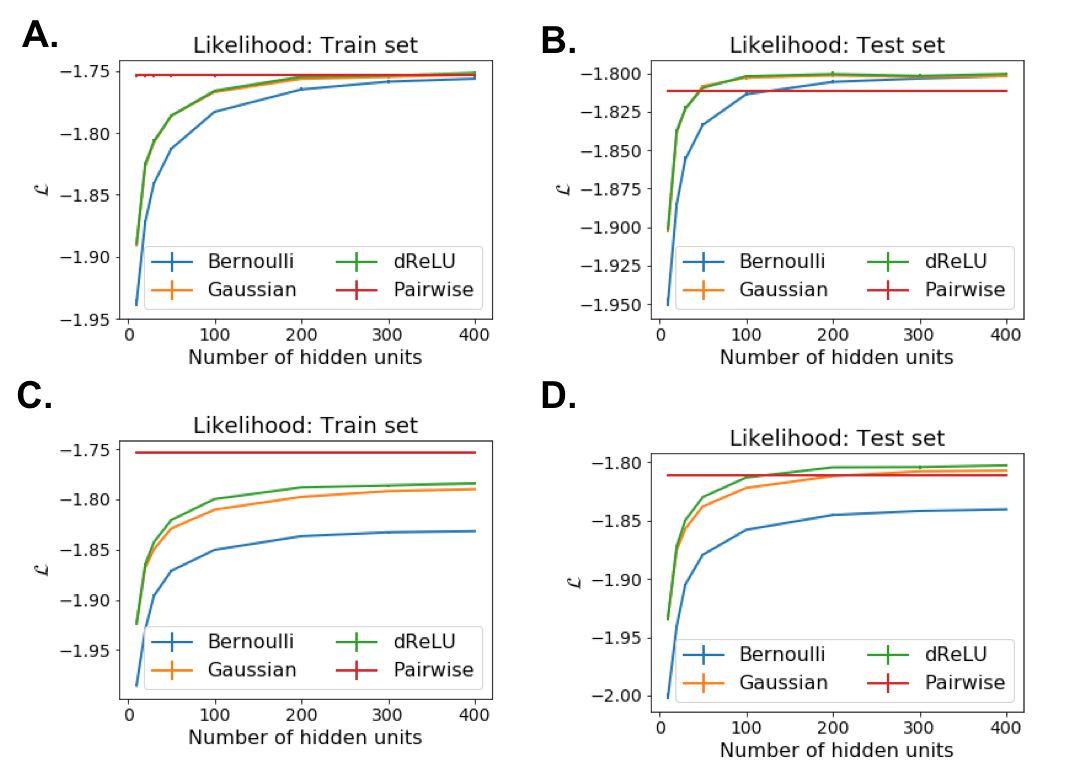}
\captionof{figure}{Model selection for RBM trained on the Lattice Proteins MSA. Likelihood estimates for various potentials and number of hidden units, evaluated on train and held out test set. Top row: without regularization ($\lambda_1^2 = 0$). Bottom row: with regularization ($\lambda_1^2 = 0.025$).}
\label{model_selection_LP}
\end{center}
\begin{center}
\includegraphics[scale=0.35]{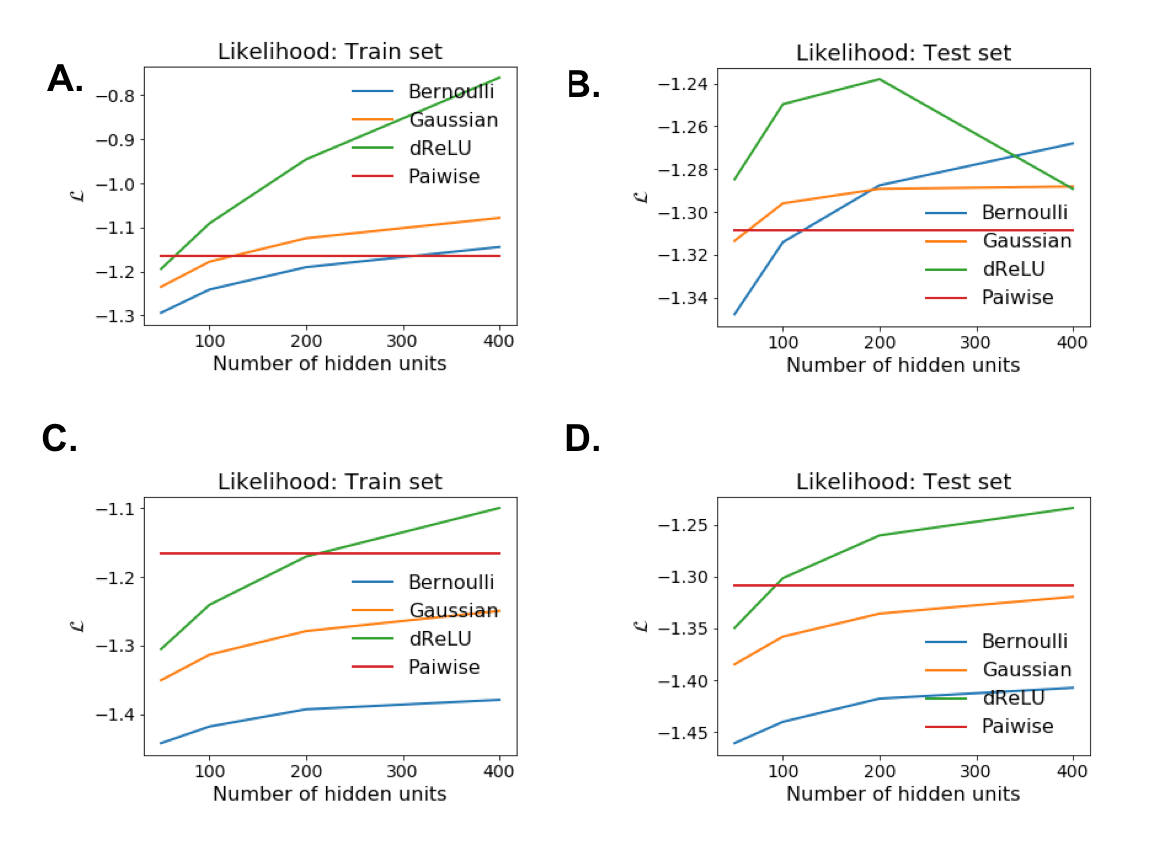}
\captionof{figure}{Model selection for RBM trained on the WW domain MSA. Likelihood estimates for various potentials and number of hidden units, evaluated on train and held out test set. Top row: without regularization ($\lambda_1^2 = 0$). Bottom row: with regularization ($\lambda_1^2 = 0.25$).}
\label{model_selection_WW}
\end{center}
Besides generative performance, the representation also changes as $M$ increases. For very low values of $M$, each hidden unit tries to explain as much covariation as possible and its corresponding weight vector is extended, similarly to PCA. For larger numbers of hidden units, weights tend to become more sparse; they stabilize at some point, after which new hidden units simply duplicate previous ones.
\subsection{Sparse regularization}
We first investigate the importance of the sparsifying penalty term. Our study shows that, unlike the case of MNIST digit data \cite{tubiana2017emergence}, sparsity does not arise naturally from training RBM on protein sequences but requires the introduction of a specific sparsifying regularization, see Fig.~8, Main Text. On the one hand, sparse weights, as the ones shown in Fig.~2, 3, 4 and 7 of the Main Text, are easier to interpret, but, on the other hand, regularization generally leads to a decrease in the generative performance. We show below that the choice of regularization strength used in main text is a good compromise between sparsity and generative performance. 

\begin{center}
\includegraphics[scale=0.4]{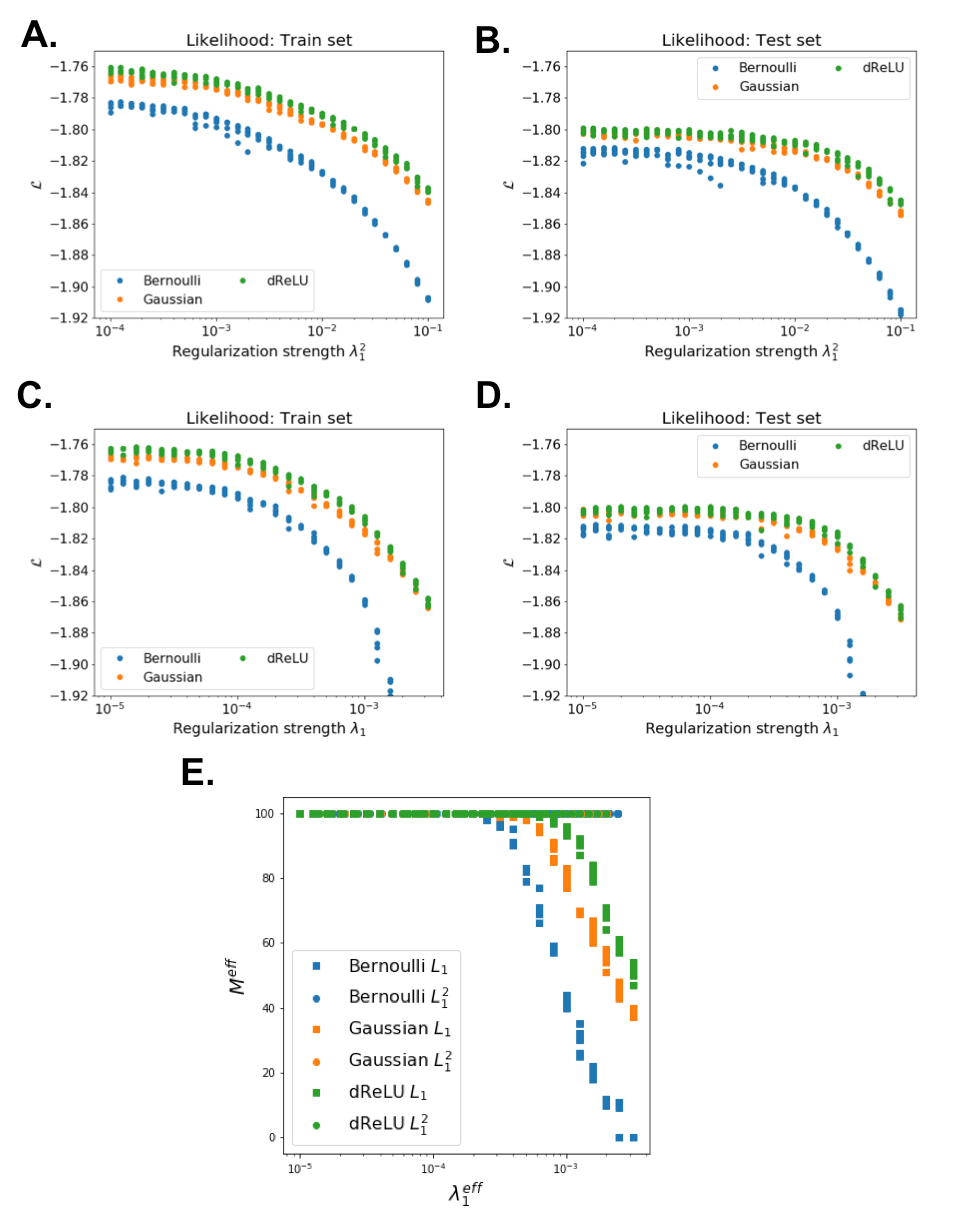}
\captionof{figure}{Sparsity-generative performance trade-off for RBM trained on the MSA of the Lattice Protein $S_A$. Panels A-D: Likelihood as function of regularization strength, for $L_1^2$ (top) and $L_1$ (bottom) sparse penalties, on train (left) and test (middle) sets. E: Number $M_{eff}$ of connected hidden units (such that $\max_{i,v} |w_{i\mu}(v)| > 0$) against effective strength penalty, for $L_1$ and $L_1^2$ penalties. For $L_1$ penalty $\lambda_1^{eff} = \lambda_1$; for $L_1^2$, $\lambda_1^{eff} = \lambda_1^2 \frac{1}{NMq}\sum_{\mu,i,v} |w_{\mu i}(v)|$.}
\label{sparsity_performance}
\end{center}
We train several RBM on the Lattice Proteins MSA, with a fixed number of hidden units ($M=100$), fixed potential, and varying strength of the sparse penalty $\lambda_1^2$ (defined in Methods, Eqn.~(8)), and evaluate their likelihoods. We repeat the same procedure using the standard $L_1$ regularization ($\lambda_1 \sum_{i,v,\mu} |w_{i\mu}(v)|$) instead of $L_1^2$. Results are shown in Appendix 1, Fig. 3. In both cases, the likelihood on test set decreases mildly with the regularization strength. However, for $L_1$ regularization, several hidden units become disconnected (\textit{i.e.} $w_{i\mu}(v) = 0$ for all $i,v$) as we increase the penalty strength. The $L_1^2$ penalty achieves sparse weights without disconnecting hidden units when the penalty is too large, hence it is more robust and requires less fine tuning.

\subsection{Hidden-unit potentials}
Lastly, we discuss the choice of the hidden-unit potentials. A priori, the major difference between Bernoulli, quadratic and dReLU potentials are that (i) Bernoulli hidden unit take discrete values whereas quadratic and dReLU take continuous ones and (ii) After marginalization, quadratic potentials create pairwise effective interactions whereas Bernoulli and dReLU create non-pairwise ones. It was shown in the context of image processing and text mining that non-pairwise models are more efficient in practice, and theoretical arguments also highlight the importance of high-order interactions \cite{tubiana2017emergence}. 

In terms of generative performance, our results on Lattice Proteins and WW domain MSAs show that, for the same number of parameters, dReLU RBM perform better than Gaussian and Bernoulli RBM. Similar results, not shown, were obtained for the Kunitz domain MSA. Although RBM with Bernoulli hidden units are known to be universal approximators as $M \rightarrow \infty$ \cite{le2008representational}, they require more hidden units than the other types; hence more data. This can be intuitively explained by the fact that Bernoulli units cannot naturally express modulation in the degree of presence of a feature. To overcome this issue, one needs more than one hidden unit to encode each feature, as in \cite{nair2010rectified}. This is consistent with the heavier distribution of hidden units correlations observed in all data sets, see Appendix 1, Fig.~4. For example, for RBM  for Bernoulli potentials, 51 out of 100 hidden units encode gap stretches, as opposed to 23 for quadratic and 15 for dReLU potentials; on WW, the numbers are respectively 18, 15 and 9. For both data sets, dReLU encode more efficiently the gap modes.

\begin{center}
\includegraphics[scale=0.15]{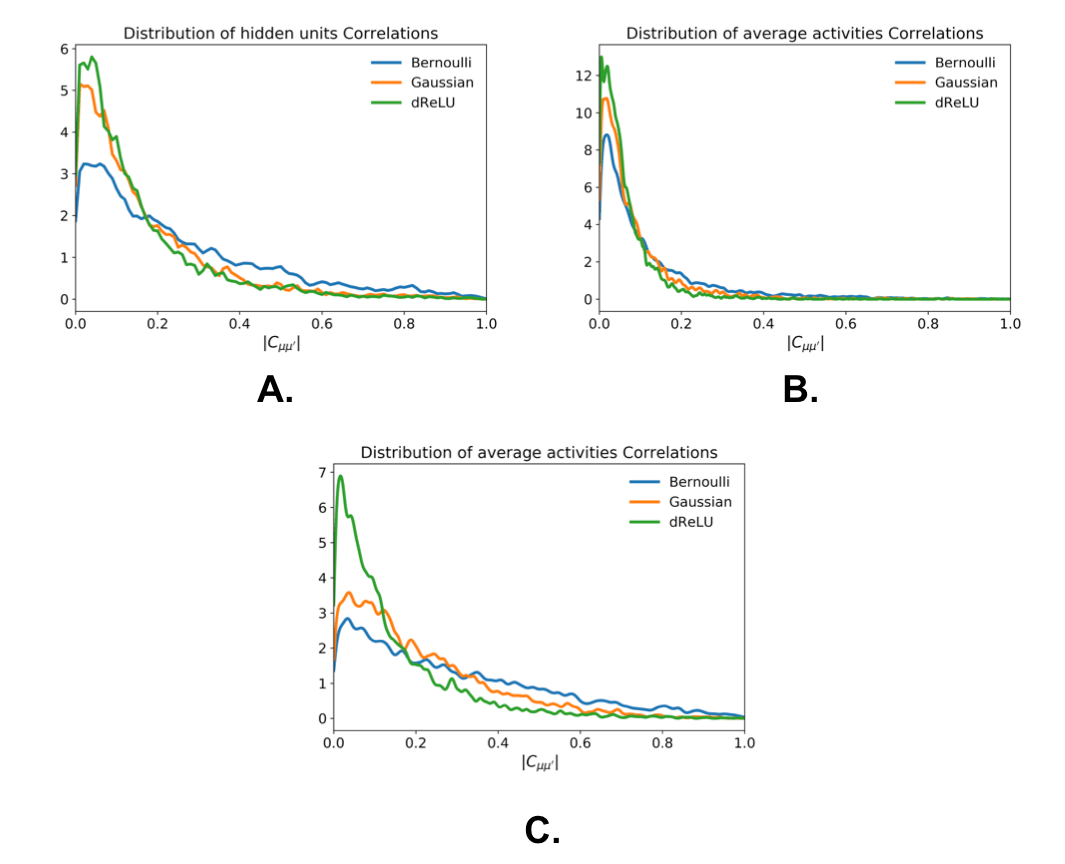}
\captionof{figure}{Hidden layer representation redundancy as function of the hidden-unit potentials. Distribution of Pearson correlations coeffcients between hidden-unit average activities, for RBM trained with $M=100$, on (a) Lattice Proteins MSA, (b) Kunitz domain MSA, (c) WW domain MSA. Bernoulli RBM feature higher correlations}
\label{correlation_hiddens}
\end{center}
One of the key aspect that explains the difference of performance between dReLU and Gaussian RBM is the ability of the former to better model 'outlier' sequences, with rare extended features such as Bikunin-AMBP (Weight 5 in Main Text, Fig.~2) or non-aromatic W28-substitution feature (Weight 3 in Main Text, Fig.~3). Indeed, thanks to the thresholding effect of the average activity, dReLU can account for outliers, without altering the distribution for the bulk of other sequences - unlike quadratic potentials. To illustrate this property, we compare in Appendix 1, Fig.~5 the likelihoods for all sequences of two RBMs trained with quadratic (resp. dReLU) potentials, $M=100$, $\lambda_1^2=0.25$ on the Kunitz domain MSA. The color code measures the degree of anomaly of the sequence, which is obtained as follows:

\begin{enumerate}
\item Compute average activity $h_\mu^l$ of dReLU RBM for all data sequences ${\bf v}^l$,
\item Normalize (z-score) each dimension: $\hat{h}_\mu = \frac{h_\mu - \esp{h_\mu}_{MSA} }{\sqrt{\var{h_\mu}_{MSA}}}$,
\item Define: \begin{equation} \label{color_code}
c^l = \arg \max_\mu |\hat{h}_\mu^l|
\end{equation}
\end{enumerate}

\noindent For instance, a sequence ${\bf v}^l$ with $c^l = 10$ has at least one hidden-unit average activity that is 10 standard deviations away from the mean. Clearly, most sequences have very similar likelihood but the outlier sequences are better modeled by dReLU potentials.

\begin{center}
\includegraphics[scale=0.06]{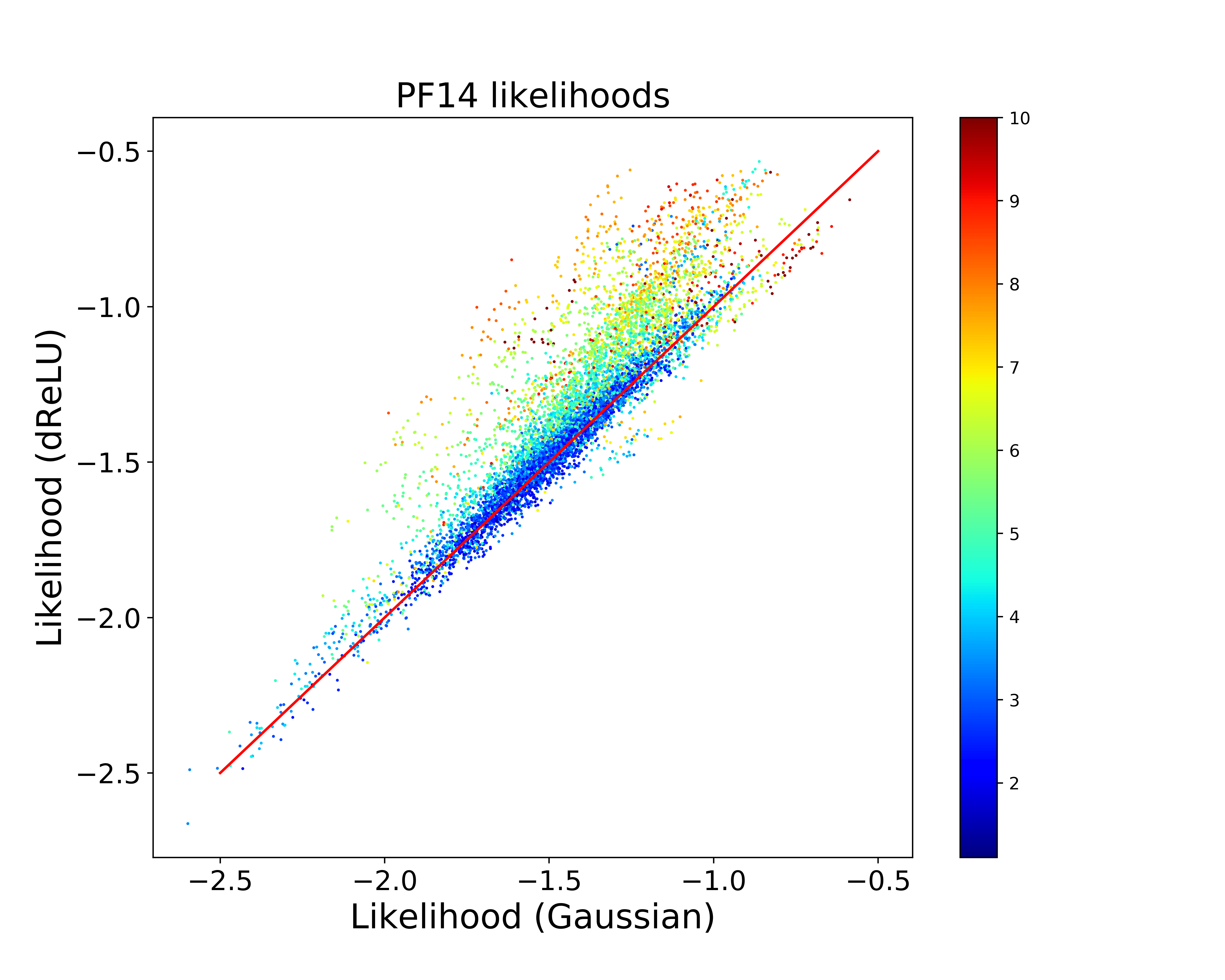}
\captionof{figure}{Comparison of Gaussian and dReLU RBM with $M=100$ trained on the Kunitz domain MSA. Scatter plot of likelihoods for each model, where each point represents a sequence of the MSA. The color code is defined in Eqn.~\ref{color_code}; hot colors indicate 'outlier' sequences}
\label{dReLU_vs_Gaussian}
\end{center}

The features extracted are fairly robust with respect to the choice of potential when regularization is used. Clearly, the nature of the potentials does not matter for finding contacts features because for any potential, a hidden unit connected to only two sites will create only pairwise effective interaction. For larger collective modes, some difference arise. As discussed above, Bernoulli features are more redundant, and Gaussian RBM tend to miss outlier features.

\subsection{Summary}
To summarize, the systematic study suggests that:
\begin{itemize}
\item More general potentials like dReLU perform better than the simpler quadratic and Bernoulli potentials;
\item There exist values of sparsity regularization penalties allowing for both good generative performance and interpretability;
\item As the number of hidden units increases, more features are captured and generative performance improve. Beyond some point, increasing $M$ simply adds duplicate hidden units and does not enhance performance.
\end{itemize}

\section{Sequence generation}

We use Lattice Proteins to check that our RBM is a good generative model, {\em i.e.} is able to generate sequences that have both high fitness and high diversity (far away from one another and from the sequences provided in the training data set), as was done for Boltzmann Machines \cite{jacquin2016benchmarking}. 
Various RBM are trained, sequences are generated for each RBM and scored using the ground truth $p_{nat}$, see Appendix 1, Fig.~6. We find that (i) RBMs with low likelihood (Bernoulli and/or small $M$) generate low quality sequences; (ii) Unregularized BMs and RBMs, which tend to overfit, generate sequences with higher fitness but low diversity; (iii) The true fitness function is well predicted by the inferred log probability. Moreover, conditional sampling also generates high-quality sequences, even when conditioning on unseen combination of features.

\begin{center}
\includegraphics[scale=0.33]{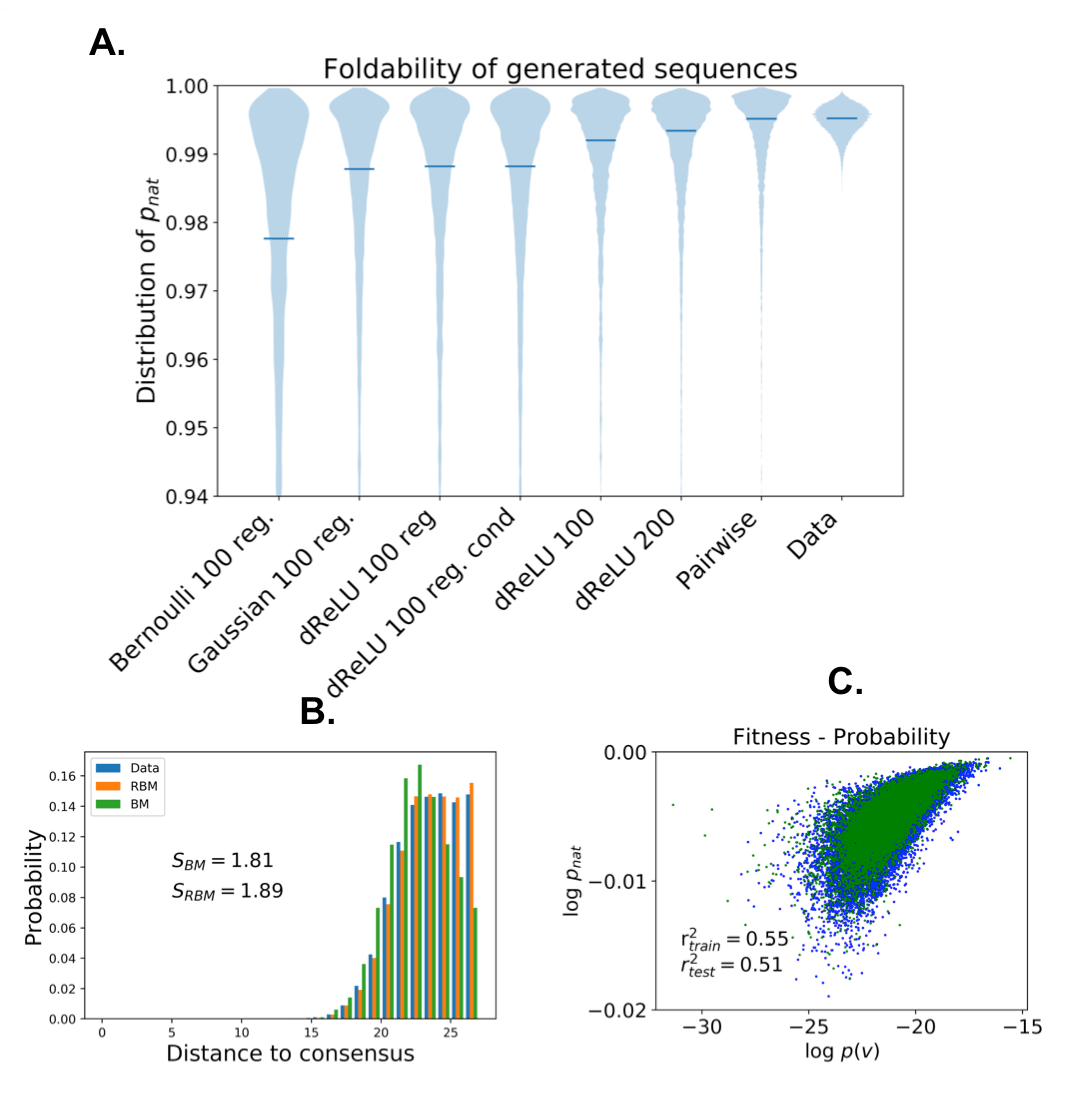}
\captionof{figure}{Quantitative quality assessment of sequences generated by RBM trained on the Lattice Protein MSA. (a) Distributions of the probability $p_{nat}$ of folding into the native structure $S_A$ (Eqn.~(14) in Methods), for sequences generated by various models. The horizontal bars locate the average values of $p_{{\text nat}}$. Models with higher capacity (more parameters, less regularization) generate sequences with higher quality but lower diversity. (b) Distribution of distances from a randomly selected wild type. The unregularized BM samples have lower diversity, whereas the regularized RBM samples better reproduce the data distribution. (c) log-probability of dReLU RBM $M=100$ shown in Main Text Fig.~7 vs true fitness evaluated on sequences from the MSA used (train) or not (test) for training.}
\label{generative_LP}
\end{center}

For RBMs trained on real proteins sequences, no ground-truth fitness is available and sequence quality cannot be assessed numerically. Appendix 1, Fig.~7 shows nonetheless that the generated sequences, including the ones with recombined features that do not appear in nature are consistent with a pairwise model trained on the same data.

\begin{center}
\includegraphics[scale=.25]{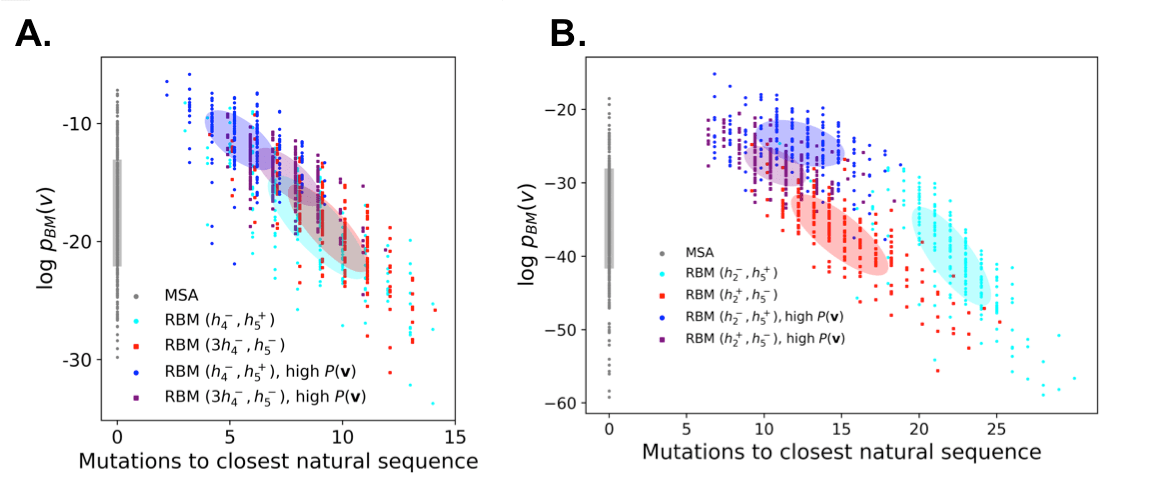}
\captionof{figure}{Quality assessment of sequences generated by RBM trained on (a) the Kunitz domain MSA and (b) the WW domain MSA. Scatter plot of the number of mutations to the closest natural sequence vs log-probability of a BM trained on the same data, for natural (gray) and RBM-generated (colored) WW domain sequences. Same color code as Main Text Fig.~5A. Note similar likelihoods values for RBM-generated sequences, and natural ones - including the unseen $(h_4^-,h_5^+)$ combinations}
\label{generative_natural}
\end{center}
\rev{Finally, we show in Appendix 1, Fig.~8 the role of regularization and sequence reweighting on sequence generation. Sequences drawn from unregularized model are closer to the ones of the training data, and the corresponding sequence distribution has significantly lower entropy} $S = - \sum_{ {\bf v} } P({\bf v}) \log P({\bf v})$ \rev{(i.e. the average negative log-probability of the generated sequences). There are respectively about} $e^{S} \sim 10^{12}$ \rev{and }$10^{18}$ \rev{distinct sequences for unregularized and regularized model. We find that sequence reweighting plays a similar role as regularization: with reweighting, sequences are slightly farther away from the training set and the model has higher entropy.}

\begin{center}
\includegraphics[scale=.15]{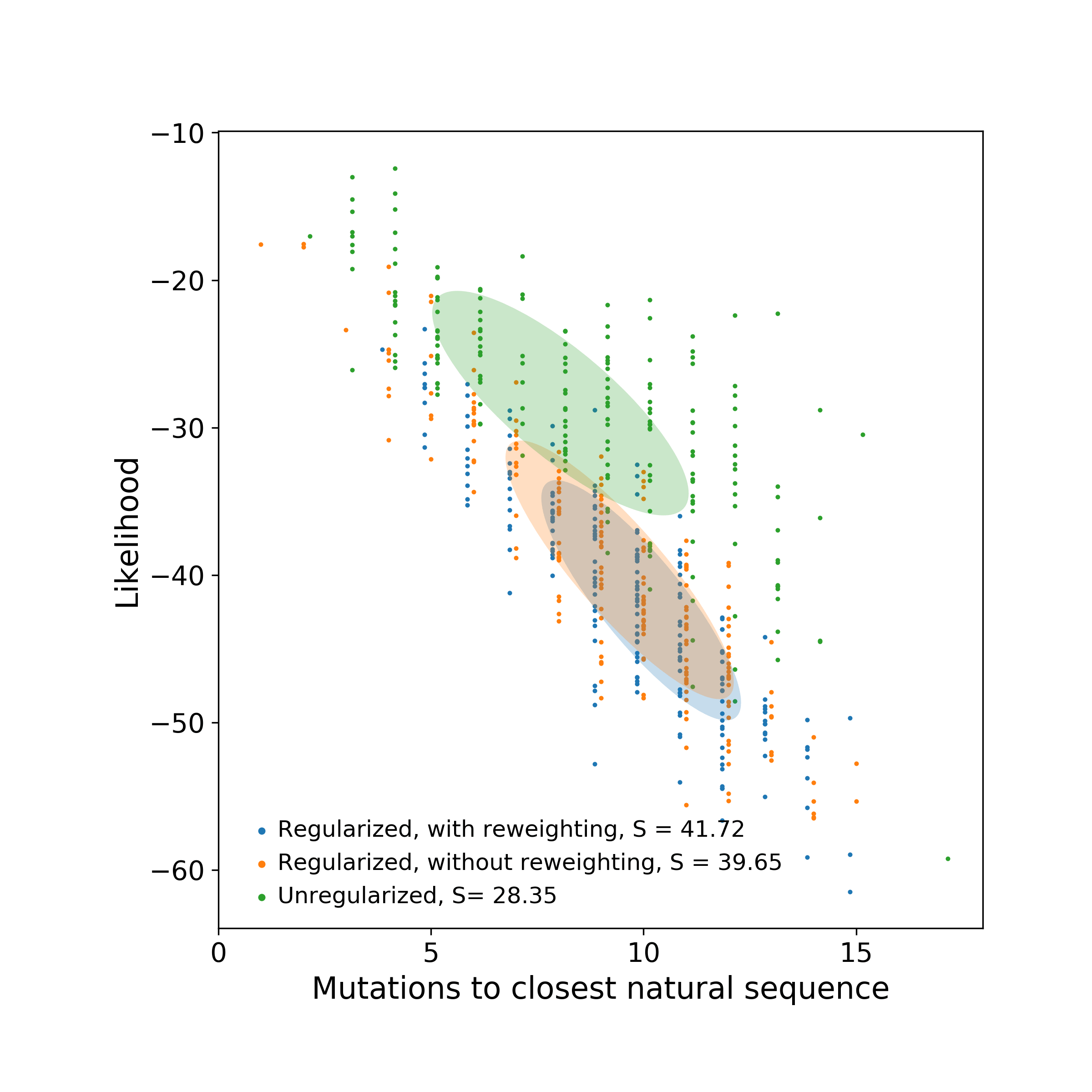}
\captionof{figure}{\rev{Evaluating the role of regularization and sequence reweighting on generated sequence diversity for the WW domain. The y-axis indicates the log-likelihood of the data generated by the model; entropy is the negative average log-likelihood}}
\label{generative_regularization}
\end{center}

\section{Contact predictions}

Since RBMs learn a full energy landscape, they can predict epistatic interactions, see Methods, and therefore contacts, as shown in Main Text Fig.~6. The effective couplings derived with RBM are consistent with the ones inferred from a pairwise model, see Appendix 1, Fig.~9. Predictions for distant contacts in the Kunitz domain are shown in Appendix 1, Fig.~10, and are slightly worse than with DCA. 

We discuss briefly the best set of parameters for contact prediction. As seen from Appendix 1, Fig.~11, all RBMs can predict more or less accurately contacts maps on Lattice Proteins. As for the likelihood and generative performance, increasing the number of hidden units significantly improves contact prediction. The best hidden unit potentials for predicting contacts are dReLU and quadratic. 

We have also studied how constraints on the sparsity of weights, tuned by the regularization penalty $\lambda_1^2$, influenced the performance. Since weights are never exactly zero, proxies are required for an appropriate definition of sparsity. In order to
avoid arbitrary thresholds, we use Participation Ratios. The Participation Ratio $(PR_e)$ of a vector ${\bf x}=\{x_i \}$ is
\begin{equation}
PR_e(\bf x) =  \frac{(\sum_{i} |x_i|^e)^2}{\sum_{i} |x_i|^{2e} }
\end{equation}
If $\bf x$ has $K$ nonzero and equal (in modulus) components
PR is equal to $K$ for any $e$. In practice we use the values $e=2$
and 3: the higher $e$ is, the more small components are discounted
against strong components in $\bf x$. Note also that it is invariant
under rescaling of ${\bf x}$. We then define the weight sparsity $p_\mu$ of a hidden
unit, through
\begin{equation}\label{sparsity}
p_\mu = \frac{1}{N} PR_3( \bf{x_\mu}) \quad \text{with}\quad ({\bf x}_{\mu})_i \equiv \sqrt{\sum_v w_{i\mu}(v)^2} \\
\end{equation}
and average it over $\mu$ to get a unique estimator of weight sparsity across the RBM. Results are reported in Appendix 1, Fig.~12, and shows that performance strongly worsen when increasing sparsity, both in Lattice Proteins and in real families. 

\begin{center}
\includegraphics[scale=0.10]{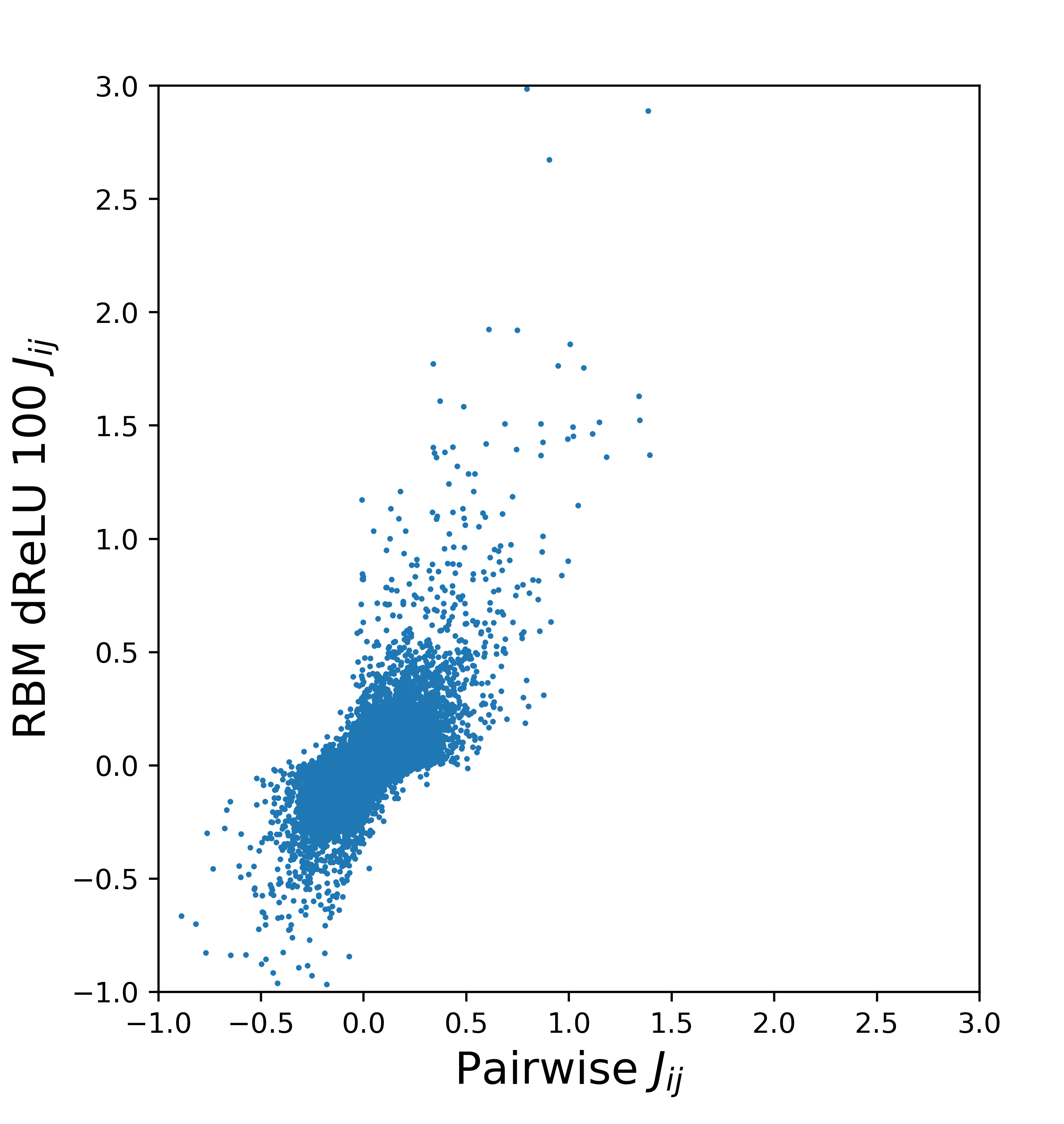}
\captionof{figure}{ {\bf Pairwise Couplings learnt from Kunitz domain MSA}. Scatter plot of inferred pairwise direct couplings learnt by BM vs effective pairwise couplings computed from the RBM through Eqn.~(15) in Methods.}
\label{scatter_JJ}
\end{center}
\begin{center}
\includegraphics[scale=0.3]{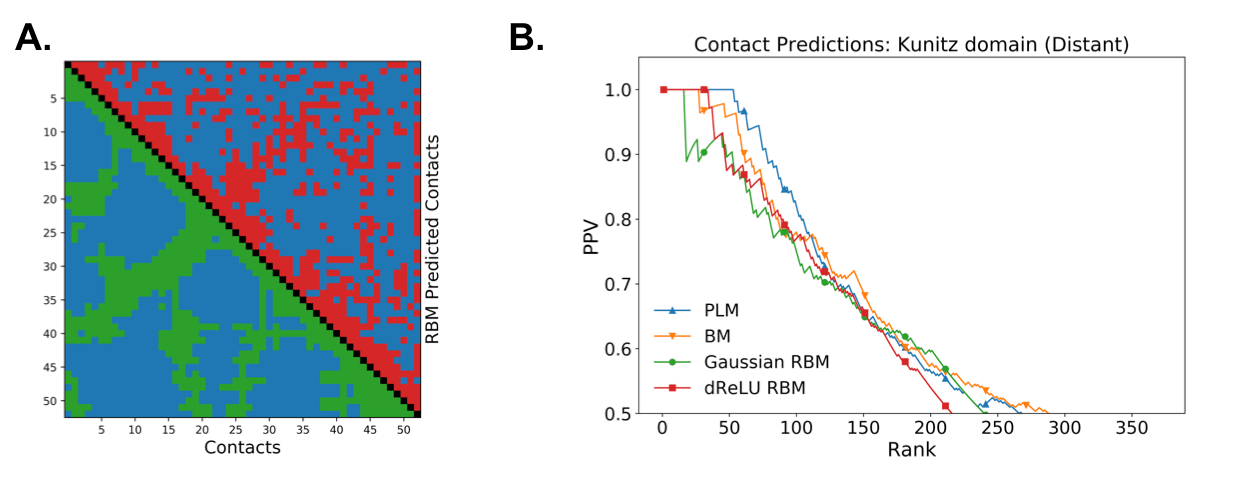}
\captionof{figure}{Contact map and contact predictions for Kunitz domain. (a) Lower diagonal: the 551 pairs of residues at $D<0.8$~nm in the structure. Upper diagonal: top 551 contacts predicted by dReLU RBM with $M=100$, shown in Main Text Fig.~2. (b) Positive Predicted Value vs rank for distant contacts $|i-j|>4$ for RBM ($M=100$) and pairwise models. Distant contacts are well predicted, including the ones involved in the secondary structure}
\label{Contact_map_PF14}
\end{center}

\begin{center}
\includegraphics[scale=0.4]{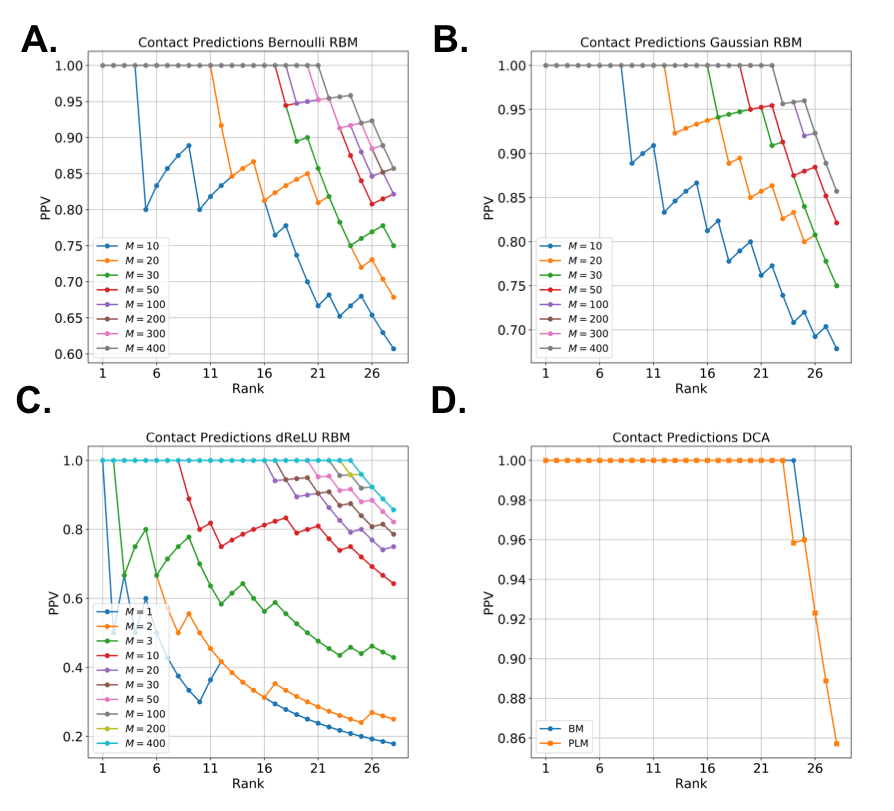}
\captionof{figure}{ {\bf Contact Predictions for Lattice Proteins}, with (a) Bernoulli (b) Gaussian (c) dReLU RBM and (d) BM. Models with quadratic/dReLU potentials and large number of hidden units typically perform the same as pairwise models, trained either with Monte Carlo or Pseudo-likelihood Maximization.}
\label{Contacts_Predictions_LP}
\end{center}
\begin{center}
\includegraphics[scale=0.3]{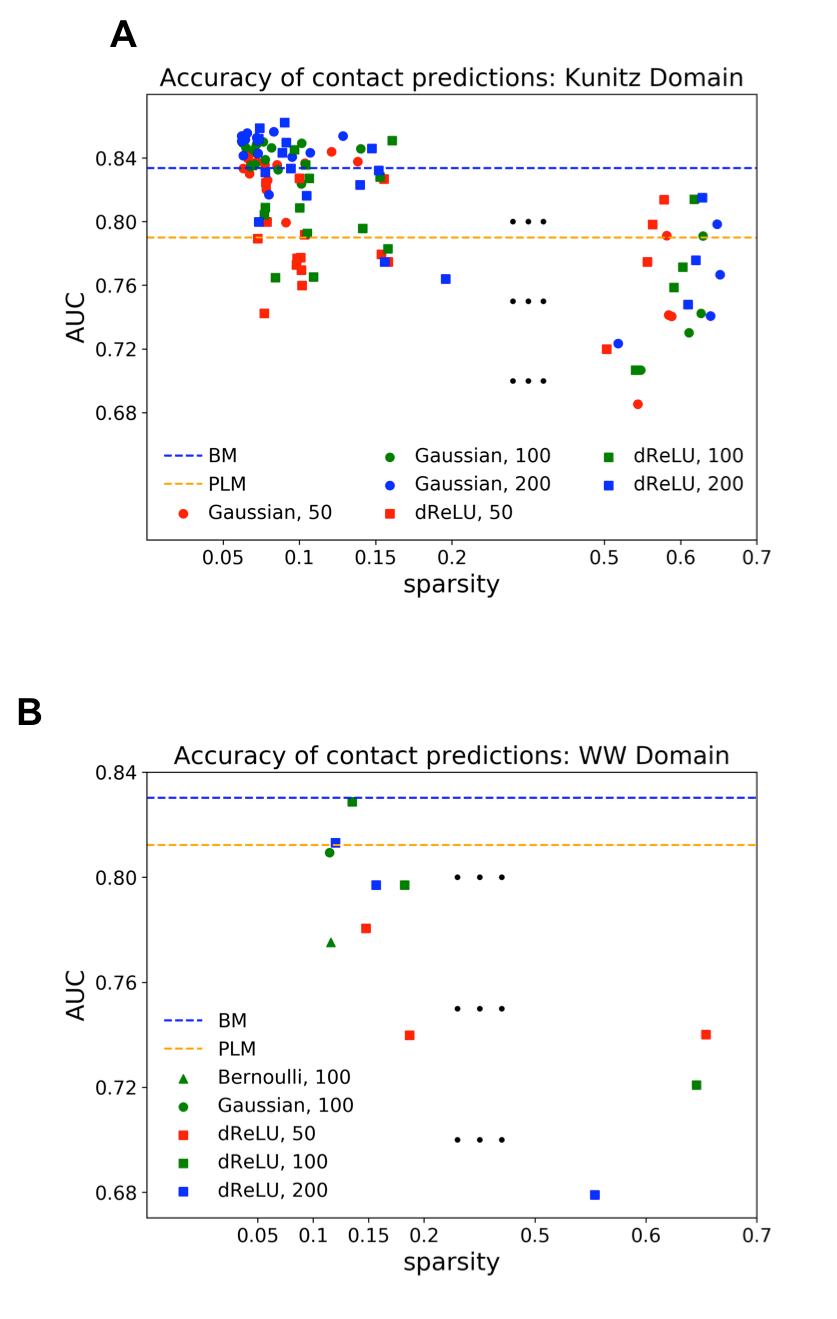}
\captionof{figure}{{\bf Contact Predictions as function of RBM parameters} for (a) Kunitz and (b) WW domains. Both panels shows the area under curve metric (integrated up to the true number of contacts) for various trainings, with different training parameters, regularization choice and hidden units number/potentials, against the weight sparsity. In both case, large sparse regularization and high number of hidden units reproduce the performance of pairwise models.}
\label{Contacts_Predictions_others}
\end{center}
\section{Feature robustness}

\rev{To assess feature robustness, we repeat the training on WW using only one of the two half of the sequences data, and look for the closest features to the ones shown in Main. The closest features, shown below, are quite similar to the original ones.}

\begin{center}
\includegraphics[scale=0.17]{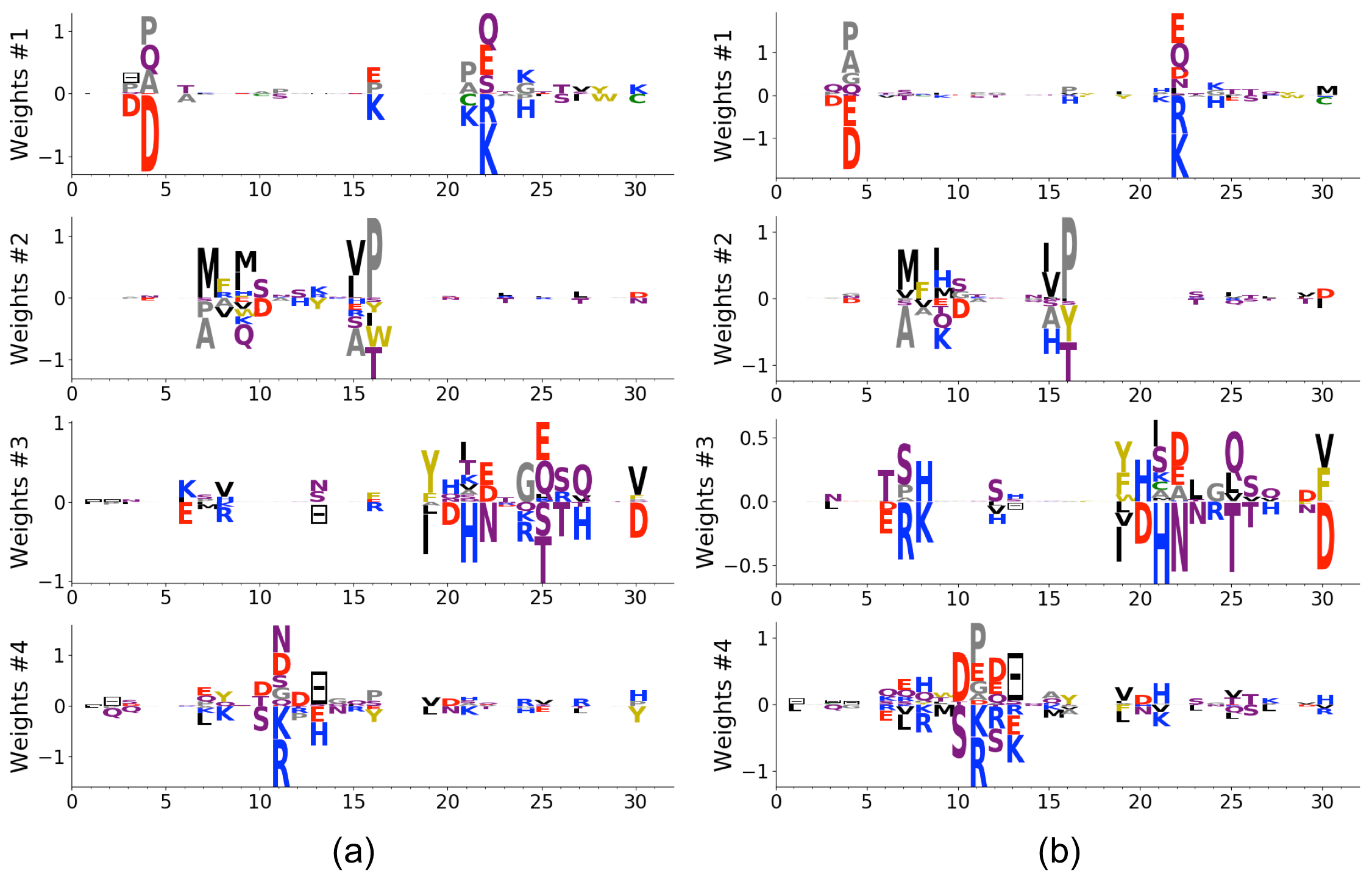}
\captionof{figure}{Features inferred using the first and second half of the sequences}
\label{WW_half_training1}
\end{center}

\newpage
\section{Comparison with the Hopfield-Potts model}

\rev{The Hopfield-Potts model is a special case of RBM with i) quadratic potentials for hidden units, ii) No regularization but orthogonality constraints on the weights, iii) Mean-field inference rather than PCD Monte Carlo learning. The consequences are that: i)	We cannot model high-order interactions, ii) We do not observe a compositional regime in which the weights are sparse and typical configurations are obtained by combinations of these weights. Instead, the representation is entangled and the weights attached to high eigenvalues are extended over most sites of the protein. iii) The model is not generative, i.e. does not reproduce the data moments and cannot generate a diverse set of sequences. To illustrate this fact, we show:}
\begin{itemize}
\item \rev{Examples of weights inferred from the the Kunitz and WW domains, and for Lattice Proteins; Weights corresponding to Hsp70 can be found in a Supporting Information file. Low-eigenvalue weights are sparse, as reported in} \cite{cocco2013principal}, \rev{but high eigenvalue weights that encode collective modes are extended, and therefore hard to interpret and to relate to function.}

\item  \rev{Contact predictions with Hopfield-Potts, showing worse performance than RBM or plmDCA.}

\item \rev{Benchmarking of generated sequences with Hopfield Potts on Lattice Proteins, similar to Fig.~7F. Using a small pseudo-count, sequences are very bad (have very low folding probability). Using a larger pseudo-count, sequences have reasonable fitness} $p_{\text{nat}}$,\rev{ though lower than high}-$P({\bf v})$ \rev{RBM, but quite low diversity. This phenomenon is characteristic of sequences generated with mean-field models, see Fig.~3A in} \cite{jacquin2016benchmarking}.\rev{ We also note that the Lattice Protein benchmark is actually optimistic for Hopfield-Potts model, as the pseudo-count trick does not work as well whenever a sequence has many conserved sites.}

\end{itemize}

\begin{center}
\includegraphics[scale=0.1]{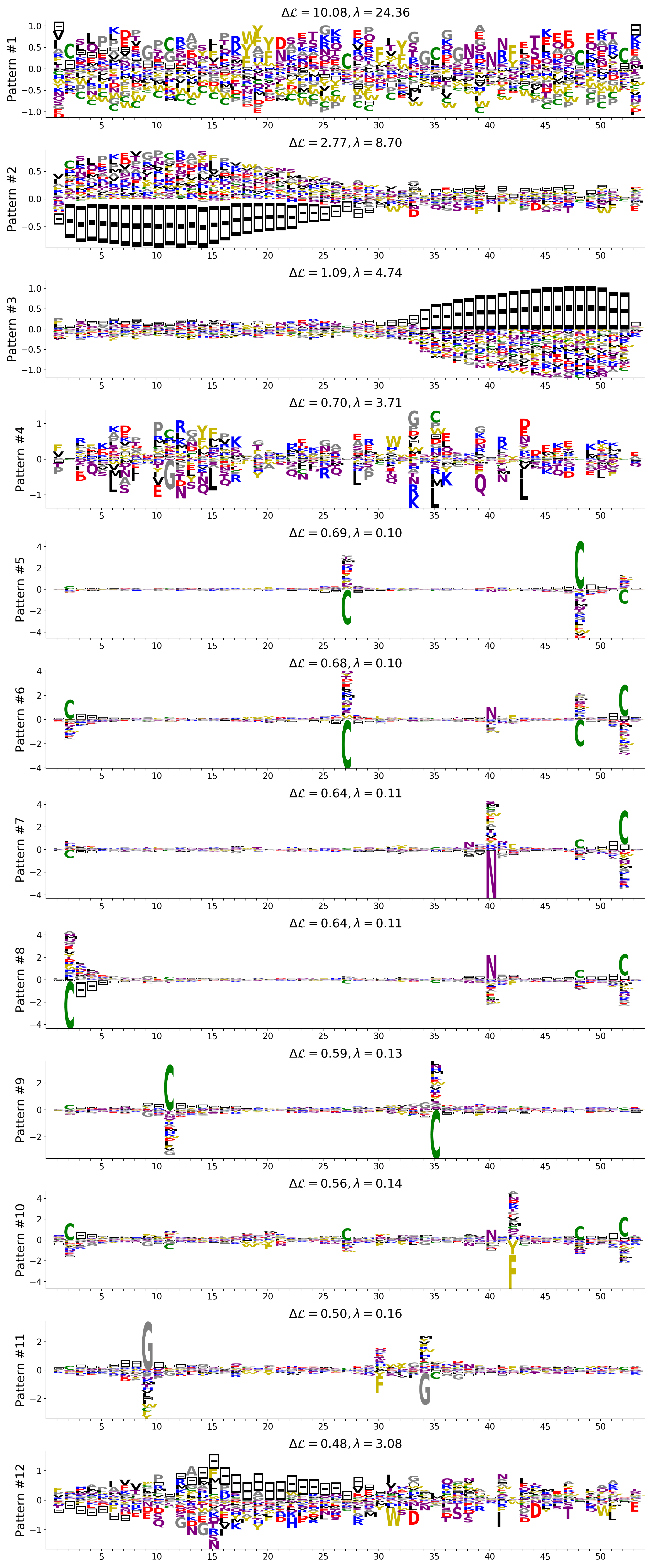}
\captionof{figure}{\rev{Top 12 patterns with highest contributions to the log-probability, see eqn (23) in} \cite{cocco2013principal}, \rev{inferred by the Hopfield-Potts model on the Kunitz domain. }}
\label{HP_Kunitz}
\end{center}
\begin{center}
\includegraphics[scale=0.1]{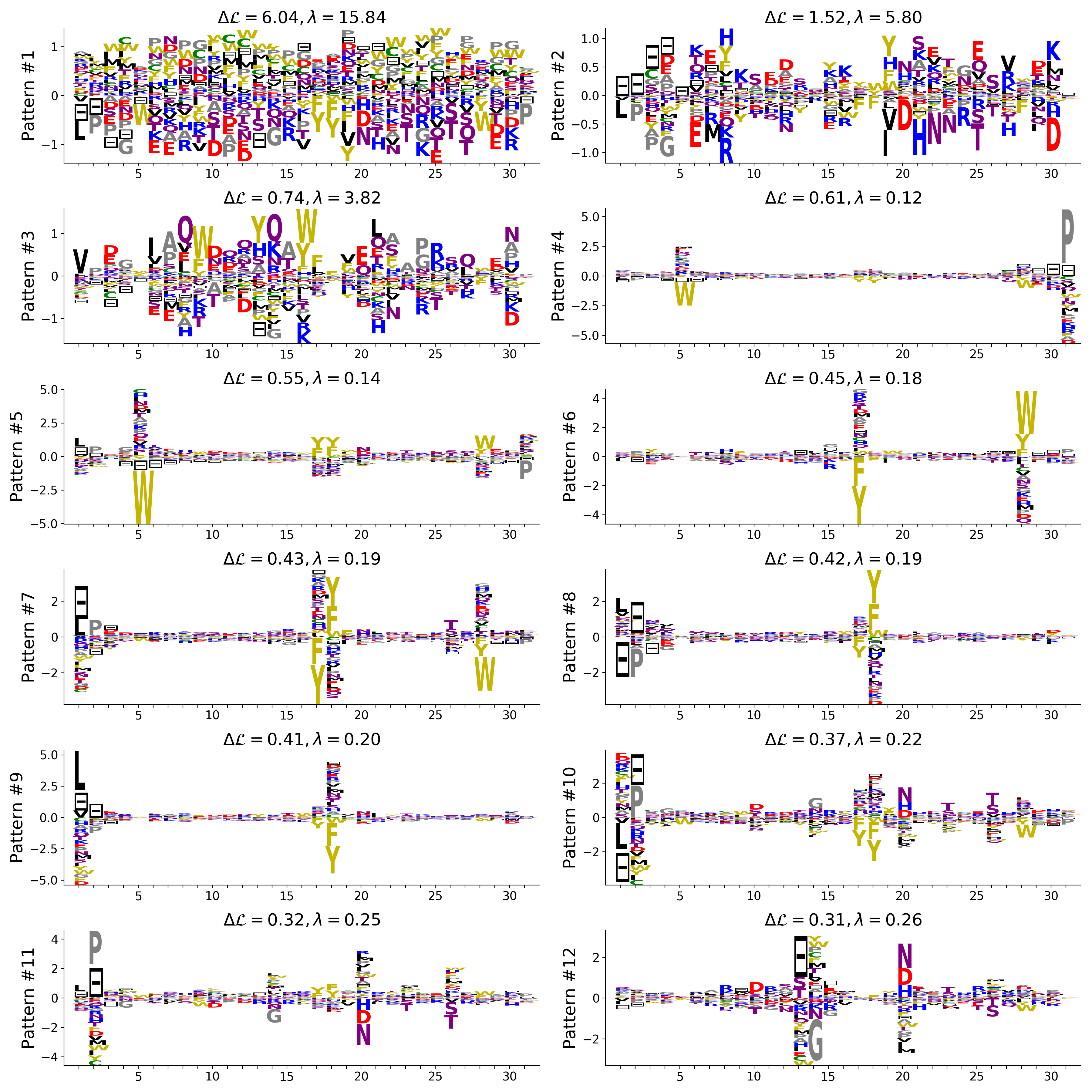}
\captionof{figure}{\rev{Top 12 patterns with highest contributions to the log-probability, see eqn (23) in} \cite{cocco2013principal}, \rev{inferred by the Hopfield-Potts model on the WW domain}}
\label{HP_WW}
\end{center}
\begin{center}
\includegraphics[scale=0.4]{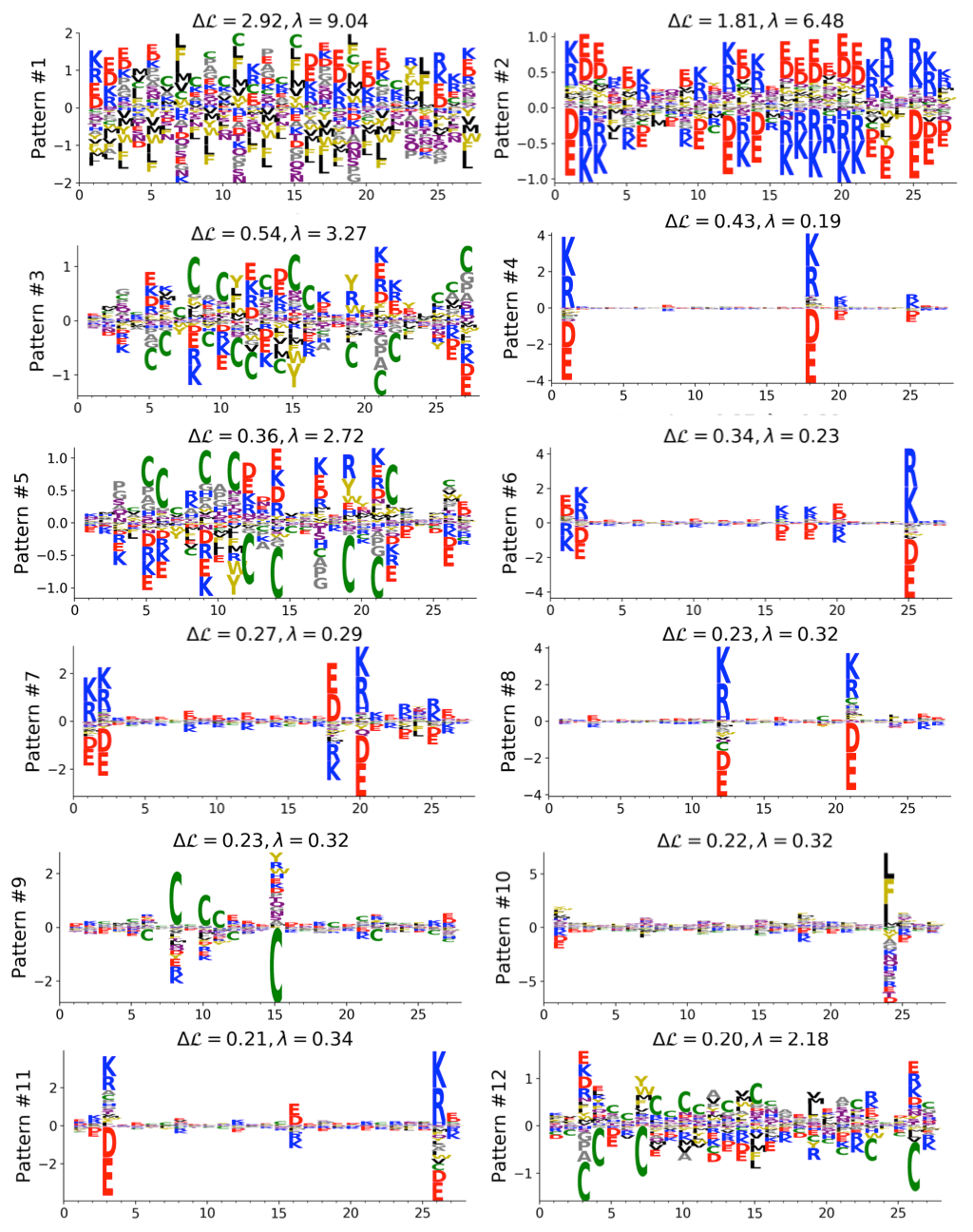}
\captionof{figure}{\rev{Top 12 patterns with highest contributions to the log-probability, see eqn (23) in} \cite{cocco2013principal},\rev{ inferred by the Hopfield-Potts model on the Lattice Proteins data}}
\label{HP_LP}
\end{center}

\begin{center}
\includegraphics[scale=0.25]{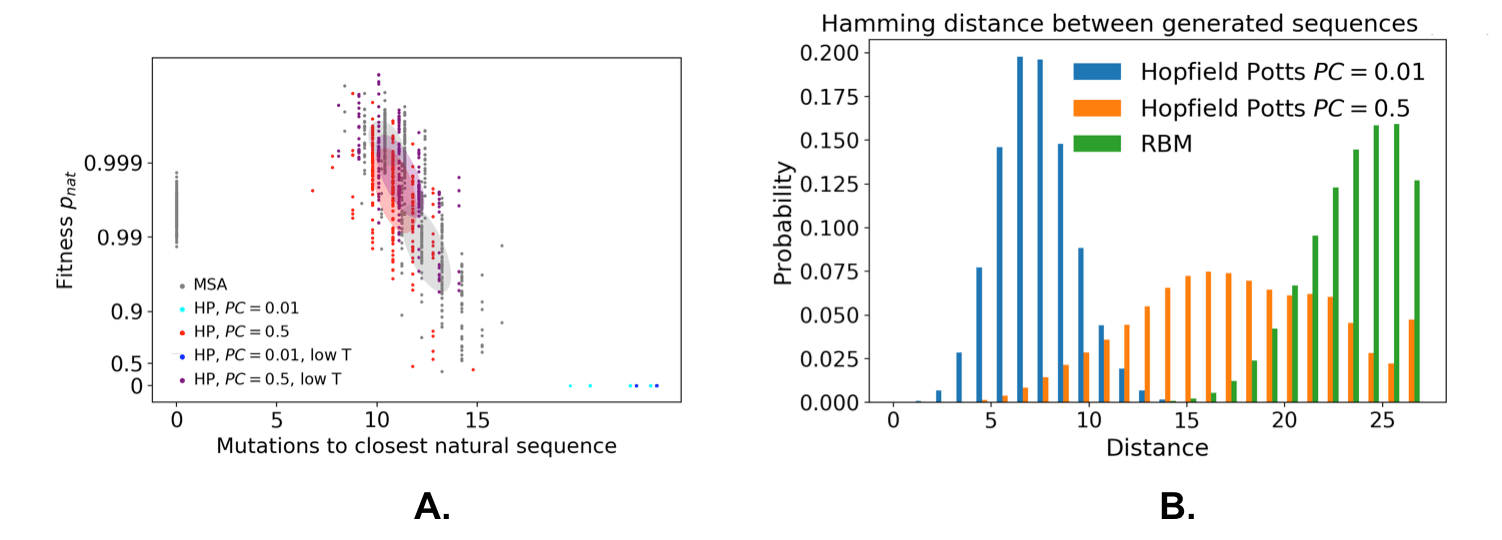}
\captionof{figure}{ {\bf Hopfield-Potts model for sequence generation} \rev{A. Fitness} $p_{\text{nat}}$ \rev{against distance to closest sequence for Hopfield-Potts model with pseudo-count 0.01 or 0.5, sampled with or without the high }$P({\bf v})$ \rev{bias. Gray ellipses denote the corresponding values for the RBM. B. Distribution of distances between generated sequences.}}
\label{HP_LP2}
\end{center}

\begin{center}
\includegraphics[scale=0.05]{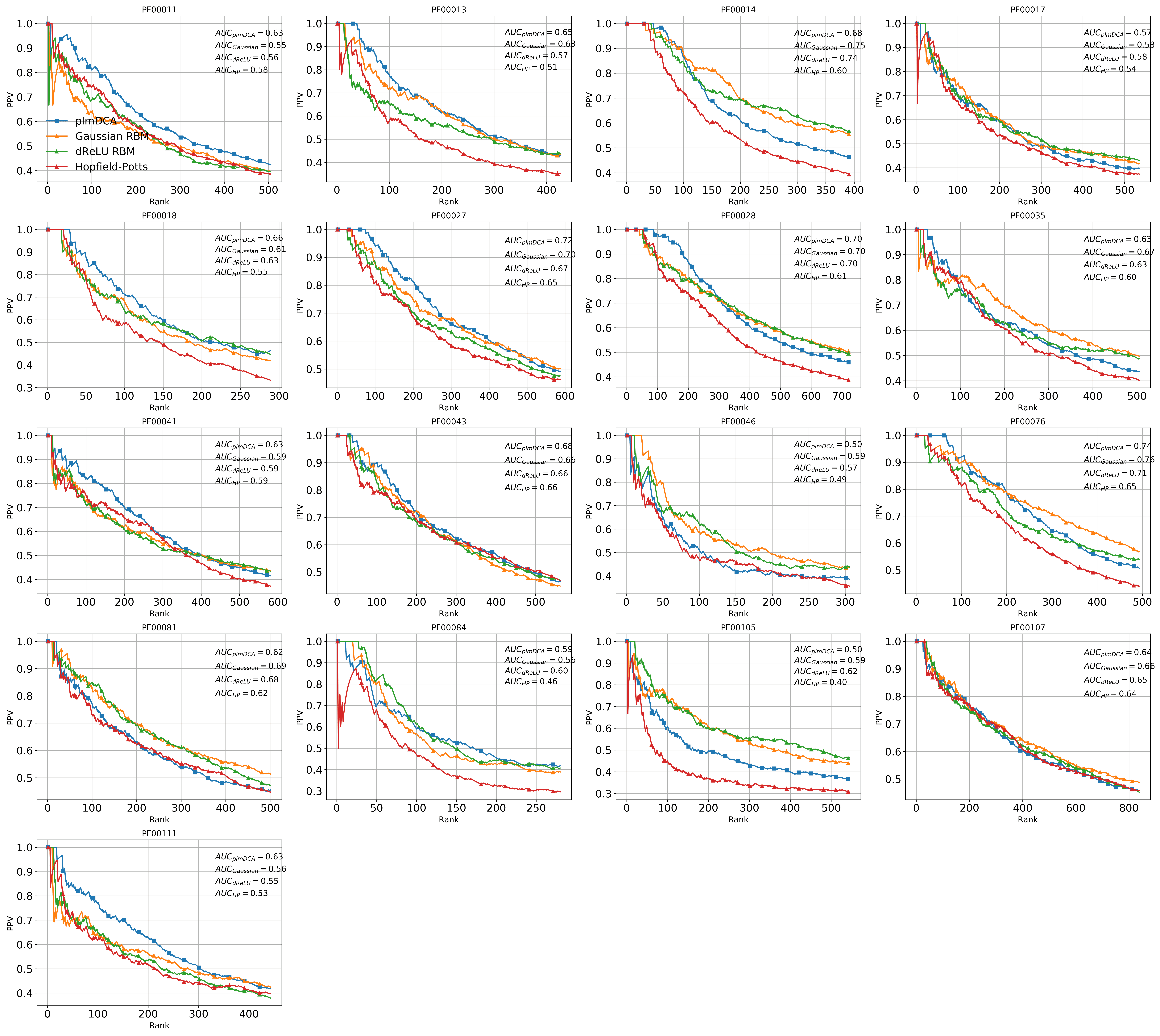}
\captionof{figure}{\rev{Contact prediction for 17 protein families; same figure as in main text but with the Hopfield-Potts model results.} }
\label{HP_contacts}
\end{center}

\section{Additional figure:  hidden-input distribution for Kunitz domain, separated by phylogenetic identity and genes}
\begin{center}
\includegraphics[scale=0.4]{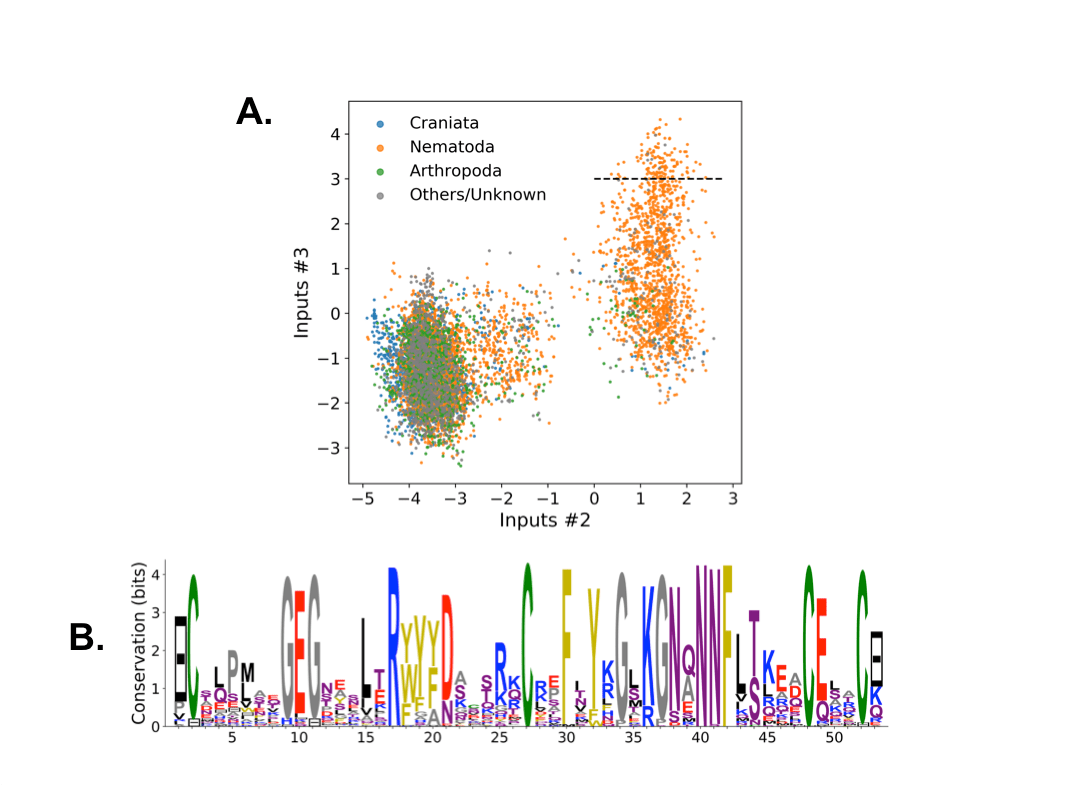}
\captionof{figure}{Phylogenetic identity of feature-activating Kunitz sequences with the RBM shown in Main Text, Fig.~2. {\bf A.} Scatter plot of inputs of hidden units 2 and 3; color depicts organism position in the phylogenic tree of species. Most of the sequences that lack the disulfide bridge come from nematodes. {\bf B.} Sequence logo of the 137 sequences above the dashed line ($I_3>3$), showing the electrostatic triangle that putatively replaces the disulfide bridge.}
\label{additional_figures_PF14}
\end{center}

\begin{center}
\includegraphics[scale=0.13]{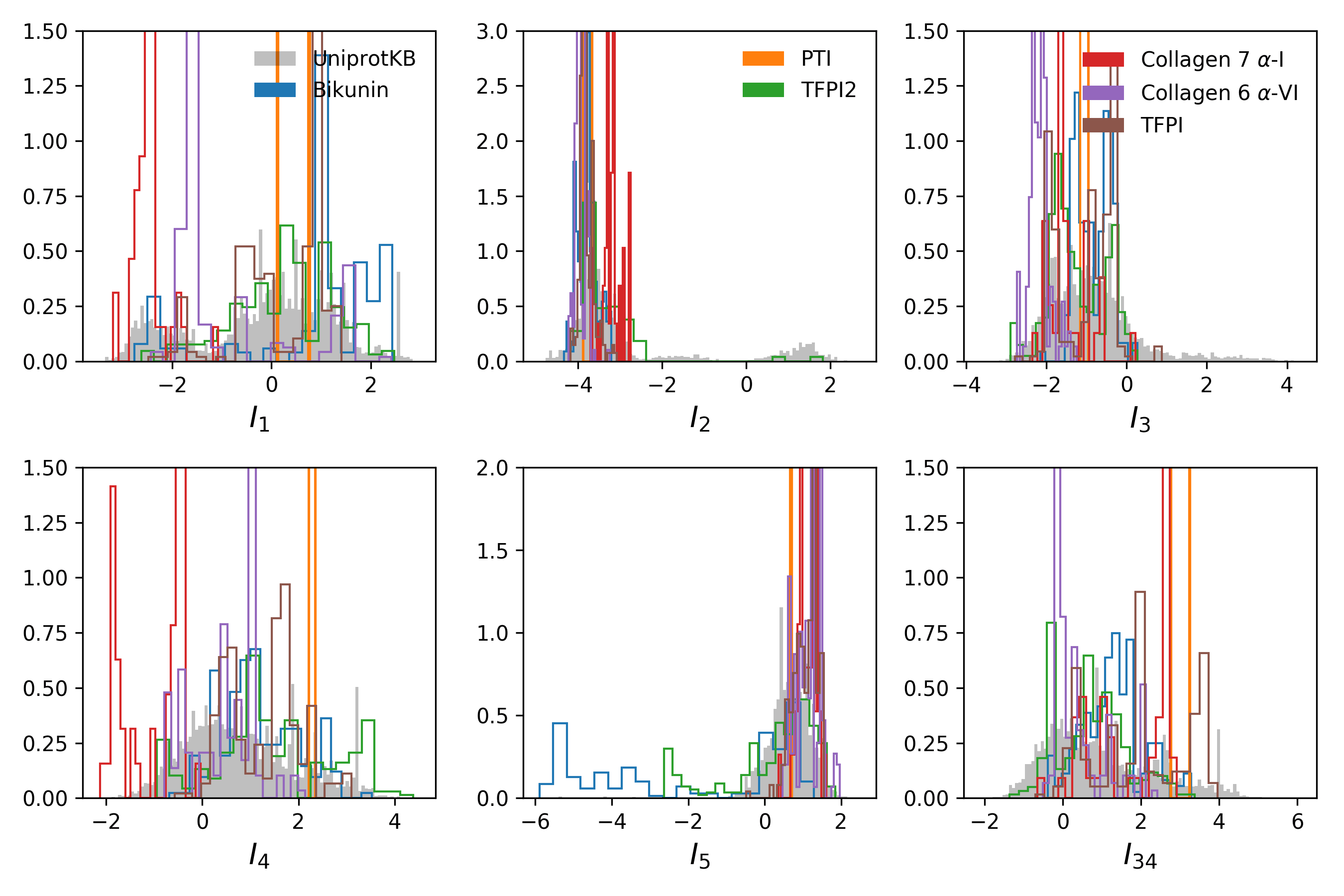}
\captionof{figure}{Distribution of inputs for the five features shown in main text plus hidden unit 34. Distributions of inputs for Kunitz domains belonging to specific genes are shown.}
\label{additional_figures_PF14_bis}
\end{center}

\section{Additional figure:  Weight logos, 3D visualizations, input distributions of 10 hidden units for Hsp70}
Hidden unit numbering: 
1 = short vs long loop
2 = function feature on SBD.
3 = LID/SBD interdomain.
4 = NBD/SBD interdomain and non-allosteric specific.
5 = Unstructured tail
6 = short/long vs very short loop
7 = Long loop variant
8 = ER-specific
9 = second non-allosteric specific
10 = Dimer contacts
\end{appendixbox}

\begin{figure}
\begin{fullwidth}
\centering
\includegraphics[scale=0.45,angle=90]{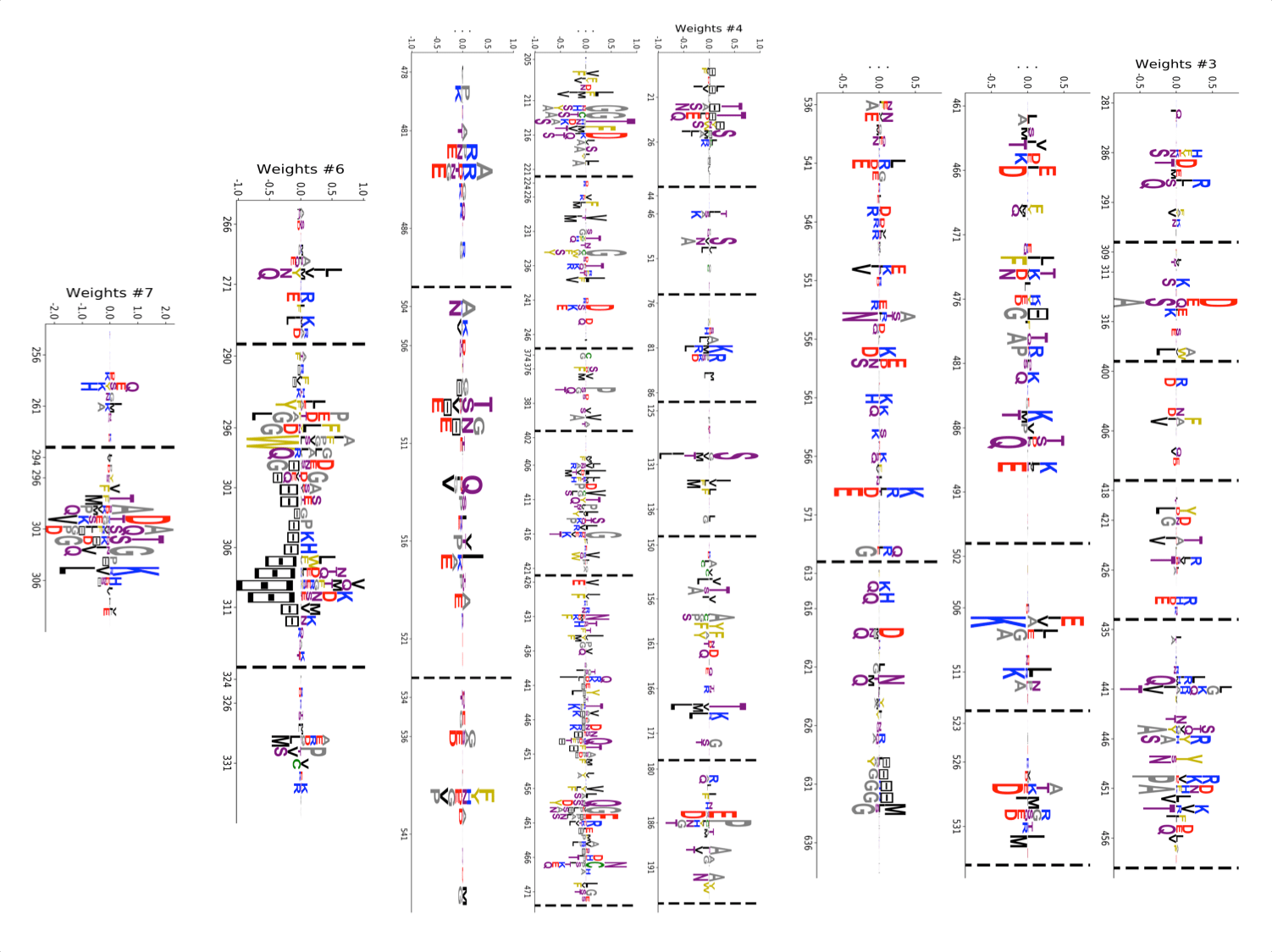}
\caption{Truncated Weight logo of 10 selected HSP70 hidden units (1/2)}
\end{fullwidth}
\end{figure}
\begin{figure}
\begin{fullwidth}
\centering
\includegraphics[scale=0.45,angle=90]{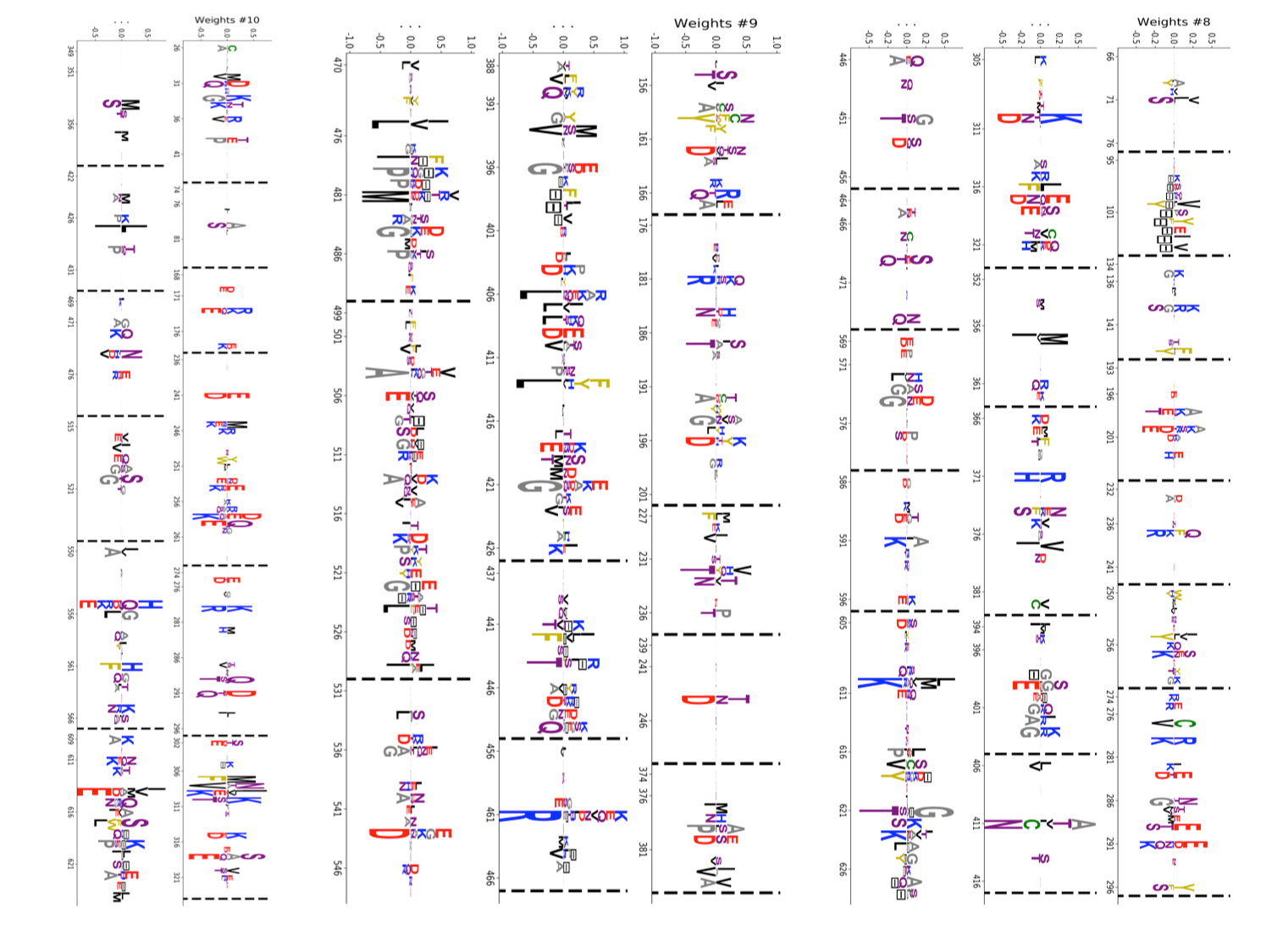}
\caption{Truncated Weight logo of 10 selected HSP70 hidden units (2/2)}
\label{additional_figures_HSP70_1}
\end{fullwidth}
\end{figure}

\begin{figure}
\begin{fullwidth}
\centering
\includegraphics[scale=0.45,angle=90]{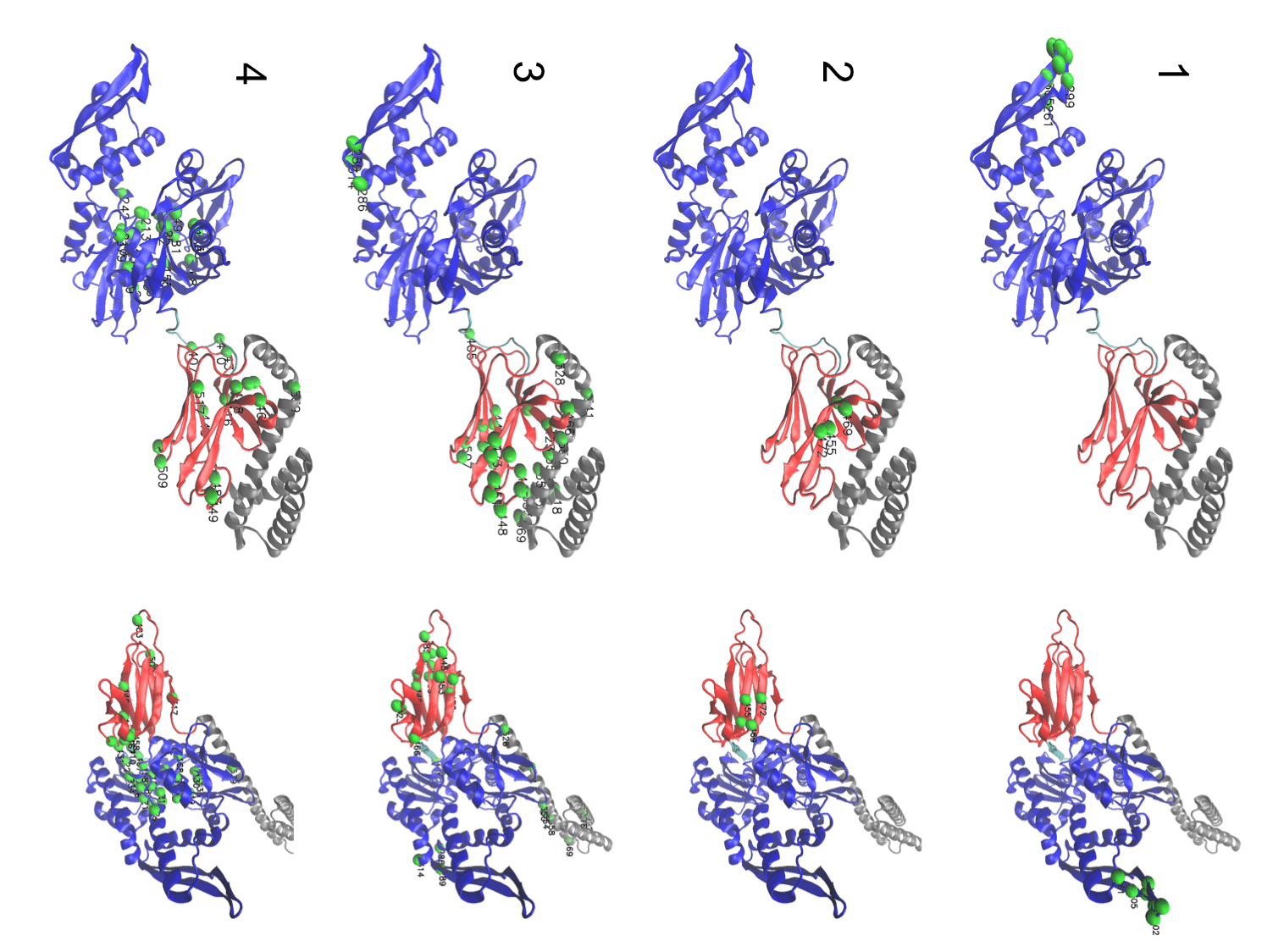}
\caption{Corresponding structures (1/3). Left: ADP-bound conformation (PDB: 2kho). Right: ATP-bound conformation (PDB: 4jne). For the last hidden unit, we show the structure of the dimer Hsp70/Hsp70 in ATP conformation (PDB: 4JNE), highlighting dimeric contacts.}
\end{fullwidth}
\end{figure}

\begin{figure}
\begin{fullwidth}
\centering
\includegraphics[scale=0.45,angle=90]{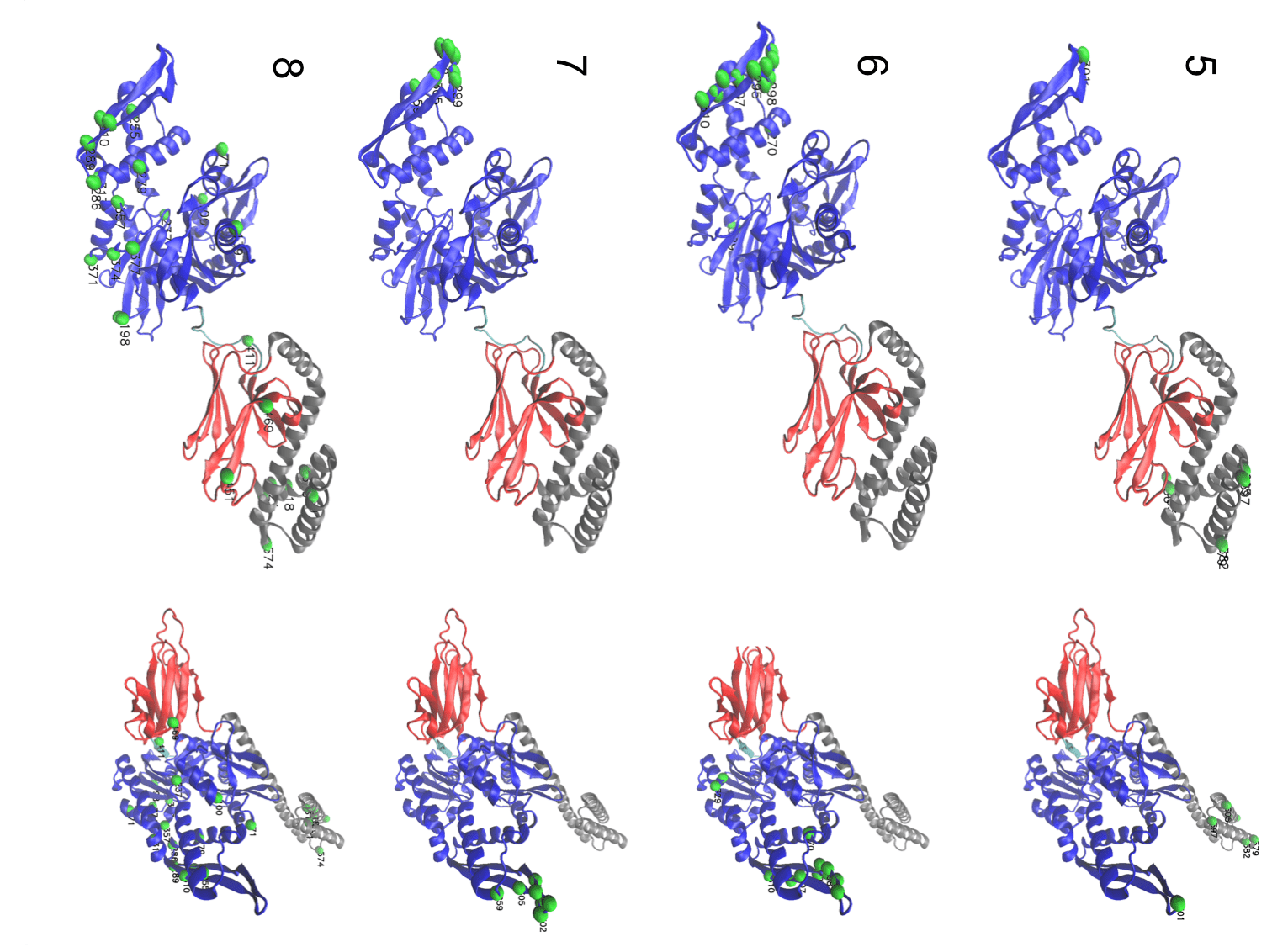}
\caption{Corresponding structures (2/3). Left: ADP-bound conformation (PDB: 2kho). Right: ATP-bound conformation (PDB: 4jne). For the last hidden unit, we show the structure of the dimer Hsp70/Hsp70 in ATP conformation (PDB: 4JNE), highlighting dimeric contacts.}
\end{fullwidth}
\end{figure}

\begin{figure}
\begin{fullwidth}
\centering
\includegraphics[scale=0.45,angle=90]{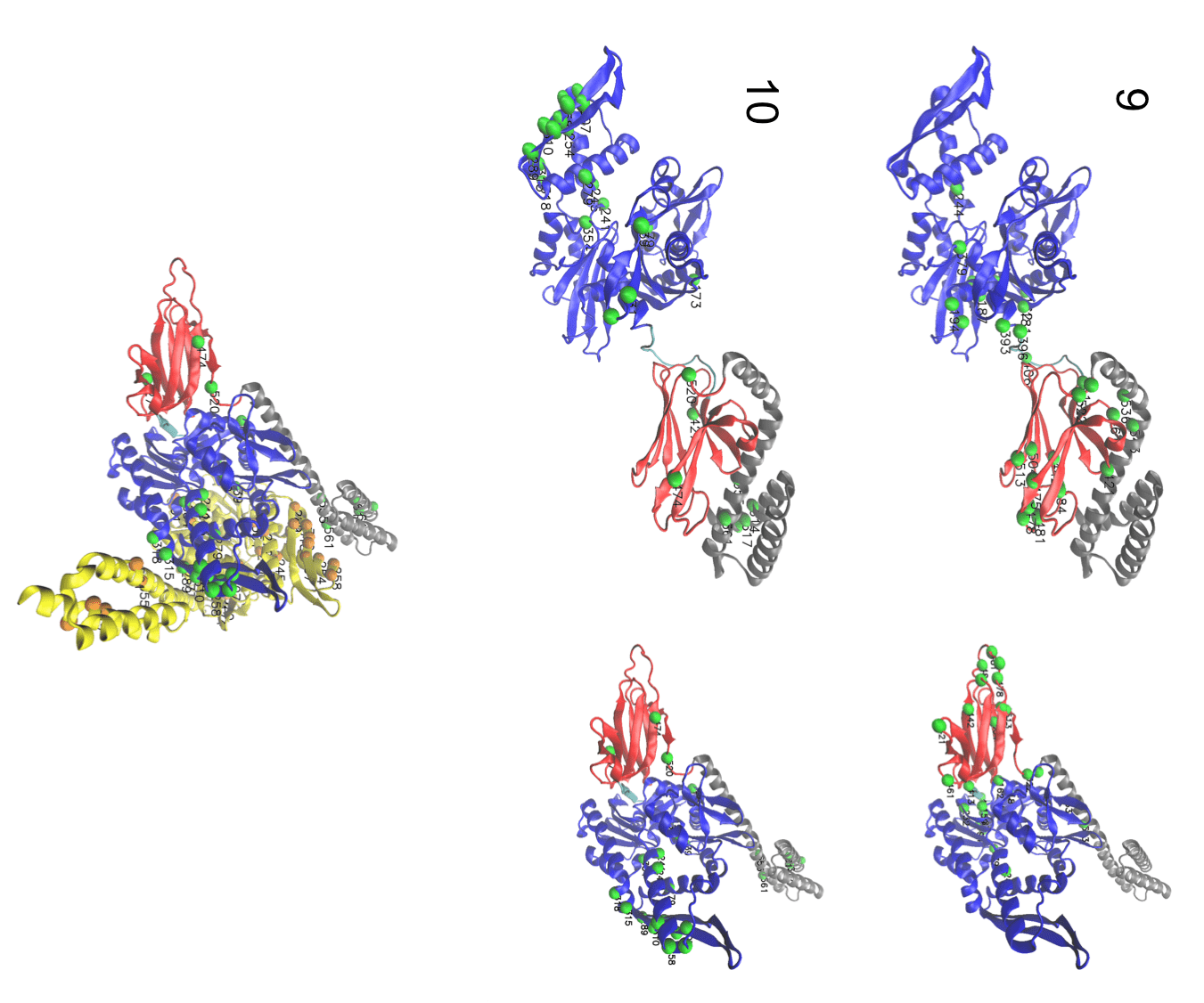}
\caption{Corresponding structures (3/3). Left: ADP-bound conformation (PDB: 2kho). Right: ATP-bound conformation (PDB: 4jne). For the last hidden unit, we show the structure of the dimer Hsp70/Hsp70 in ATP conformation (PDB: 4JNE), highlighting dimeric contacts.}
\label{additional_figures_HSP70_2}
\end{fullwidth}
\end{figure}

\begin{figure}
\begin{fullwidth}
\centering
\includegraphics[scale=0.17]{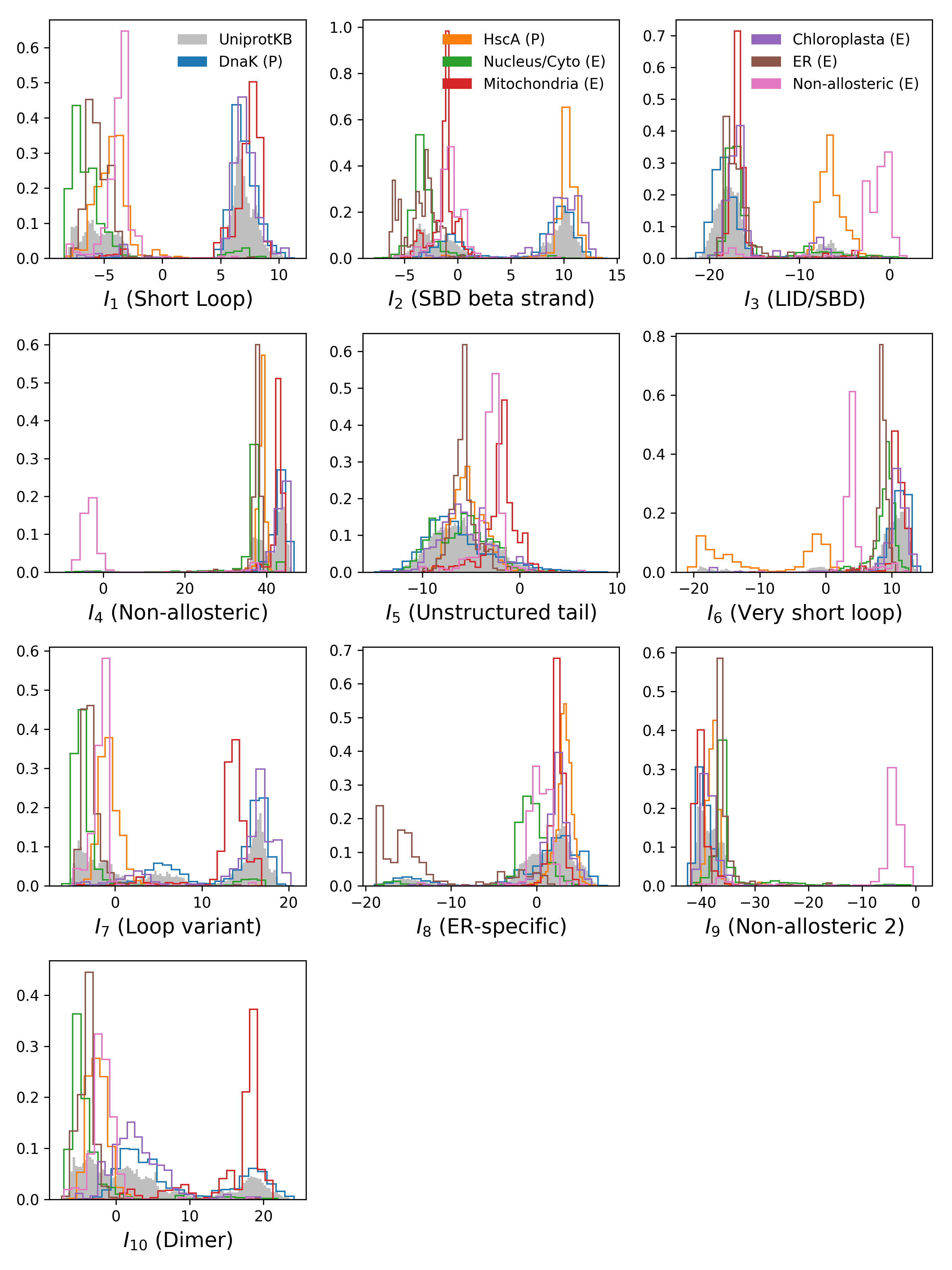}
\caption{Corresponding input distributions. Note that both hidden unit 4 and 9 discriminate the non-allosteric subfamily from the rest; and that hidden unit 8 discriminates Eukaryotic Hsp expressed in the Endoplasmic Reticulum from the rest.}
\label{additional_figures_HSP70_3}
\end{fullwidth}
\end{figure}
\begin{figure}
\begin{fullwidth}
\centering
\includegraphics[scale=0.11]{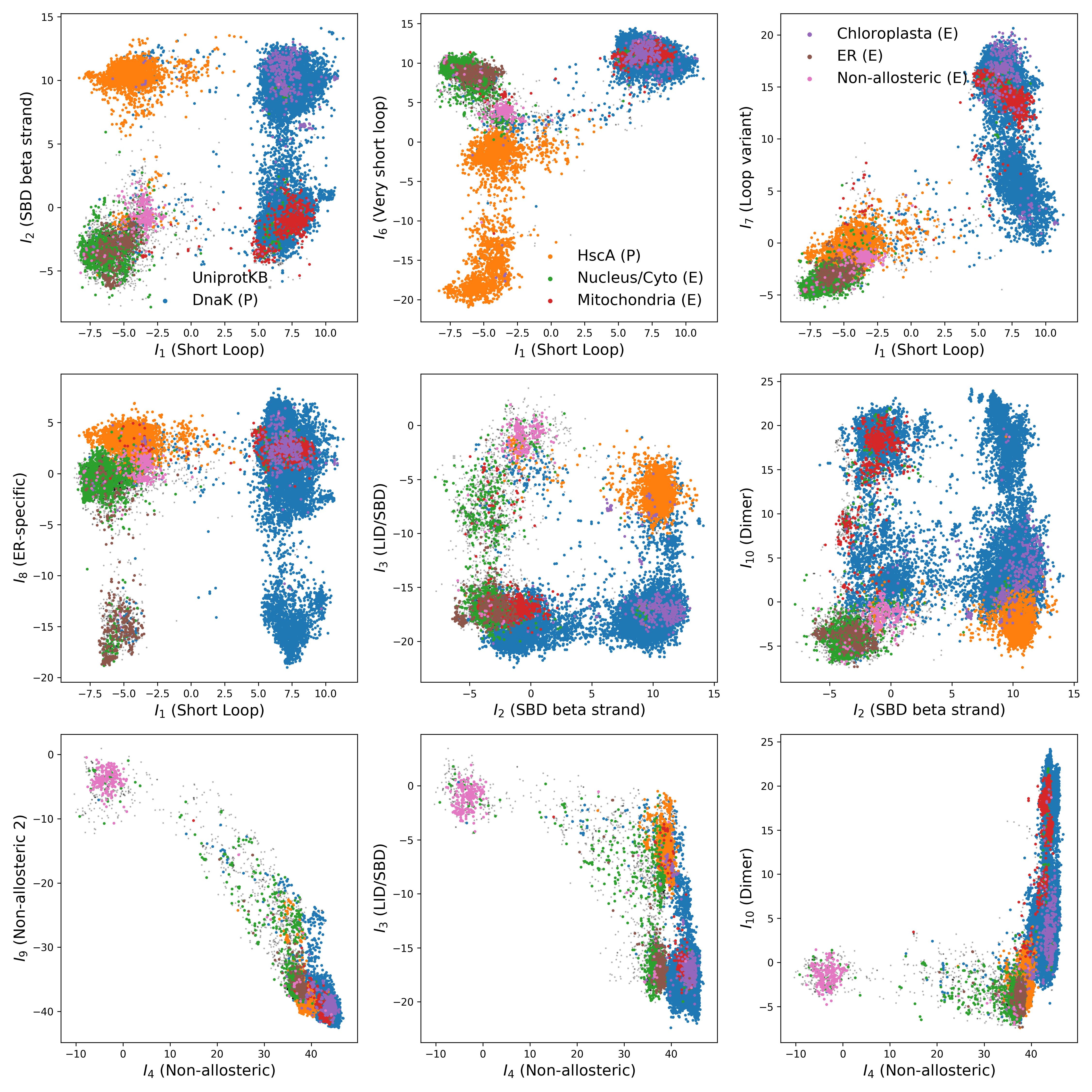}
\caption{Some scatter plots of inputs for the 10 hidden units shown.}
\label{additional_figures_HSP70_4}
\end{fullwidth}
\end{figure}

\begin{figure}
\begin{fullwidth}
\centering
\includegraphics[scale=0.13]{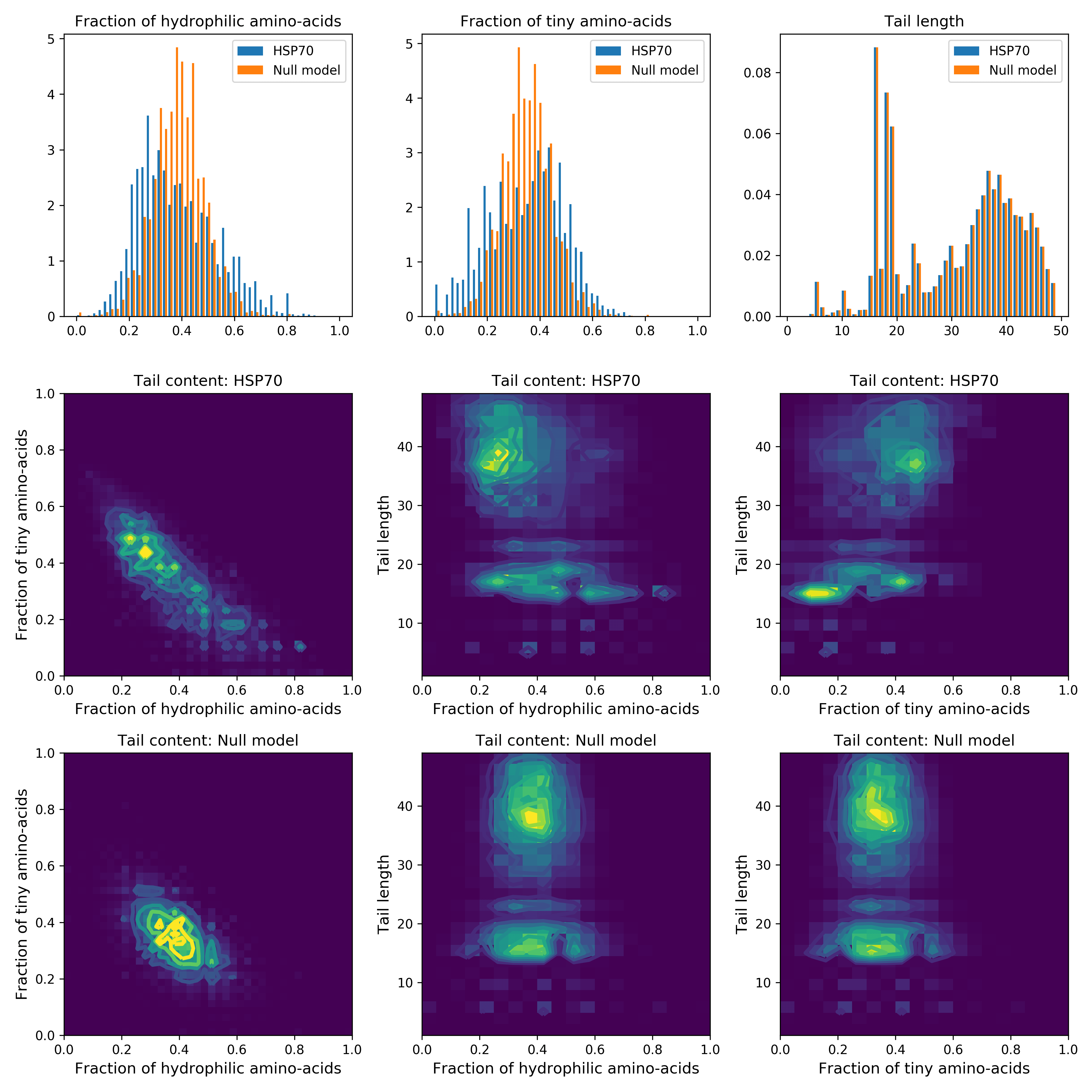}
\caption{{\bf Statistics of length and amino-acid content of the unstructured tail of Hsp70}. Hidden unit 5 defines a set of sites, mostly located on the unstructured tail of Hsp70; its sequence logo and input distribution suggests that for a given sequence, the tail can be enriched either in tiny (A,G) or hydrophilic amino-acids (E,D,K,R,T,S,N,Q). This is qualitatively confirmed by the non-gaussian statistics of the distributions of fraction of tiny and hydrophilic amino-acids in the tail (blue histograms and top left contour plots). This effect could however be due to the variable length of the loop (bottom histogram). To assess this enrichment, we build a null model where the tail size is random (same statistics as Hsp70), and each amino-acid is drawn randomly, independent from the others, using the same amino-acid frequency as in the tail of Hsp70. The null model statistics (orange histograms and lower left contour plots) are clearly different, validating the collective mode.}
\label{additional_figures_HSP70_5}
\end{fullwidth}
\end{figure}

\end{document}